\documentclass[12pt]{article}
\usepackage{graphicx}
\usepackage{natbib}
\newcommand{\blind}{0}

\usepackage[utf8]{inputenc}
\usepackage{bbm}
\usepackage{amsmath}
\usepackage{bm}
\usepackage{algorithm}
\usepackage{algpseudocode}
\usepackage{caption}
\usepackage{subcaption}
\usepackage{hyperref}
\usepackage{amsfonts}
\usepackage{booktabs}

\usepackage{amsthm}
\usepackage{amssymb}
\usepackage{color}

\DeclareMathOperator*{\argmin}{arg\,min}
\DeclareMathOperator*{\var}{var}
\DeclareMathOperator*{\tr}{tr}
\newtheorem{theorem}{Theorem}[section]

\newtheorem{lemma}[theorem]{Lemma}
\usepackage[dvipsnames]{xcolor}
\newtheorem{proposition}{Proposition}

\usepackage{authblk}
\title{Bibliography management: \texttt{natbib} package}
\numberwithin{equation}{section}

\begin{document}
\def\spacingset#1{\renewcommand{\baselinestretch}%
{#1}\small\normalsize} \spacingset{1}

\if0\blind
{
  \title{\bf Assessment of Case Influence in the Lasso with a Case-weight Adjusted Solution Path}
  \author{Zhenbang Jiao\thanks{Email: \texttt{jiao.180@osu.edu}}\hspace{.2cm}\vspace{-.2cm}\\
    and\\
    Yoonkyung Lee\thanks{Email: \texttt{yklee@stat.osu.edu}. This research was supported in part by the National Science Foundation Grant DMS-20-15490.} \\
    Department of Statistics, The Ohio State University}
    \date{}
  \maketitle
} \fi

\if1\blind
{
  \bigskip
  \bigskip
  \bigskip
  \begin{center}
    {\LARGE\bf Title}
\end{center}
  \medskip
} \fi

\bigskip

\begin{abstract}
We study case influence in the Lasso regression using Cook's distance which measures overall change in the fitted values when one observation is deleted. 
Unlike in ordinary least squares regression, the estimated coefficients in the Lasso do not have a closed form due to the nondifferentiability of the $\ell_1$ penalty, and neither does Cook's distance. 
To find the case-deleted Lasso solution without refitting the model, we approach it from the full data solution by introducing a weight parameter ranging from 1 to 0 and generating a solution path indexed by this parameter. 
We show that the solution path is piecewise linear with respect to a simple function of the weight parameter under a fixed penalty. 
The resulting case influence is a function of the penalty and weight, and it becomes Cook's distance when the weight is 0. 
As the penalty parameter changes, selected variables change, and
 the magnitude of Cook's distance for the same data point may vary with the subset of variables selected. 
In addition, we introduce a case influence graph to visualize how the contribution of each data point changes with the penalty parameter. 
From the graph, we can identify influential points at different penalty levels and make modeling decisions accordingly. 
Moreover, we find that case influence graphs exhibit different patterns between underfitting and overfitting phases, which can provide additional information for model selection. 
\end{abstract}

\noindent%
\vfill

\newpage

\section{Introduction} \label{intro}
The Lasso regression, introduced by \cite{tibshirani1996regression}, has become a staple choice for regression modeling in the field of statistical learning for its effectiveness in feature selection and regularization. By imposing an $\ell_1$ penalty on the regression coefficients, Lasso not only helps improve prediction accuracy but also enhances model interpretability by selecting only a subset of relevant predictors. This has led to its widespread adoption across various domains, including economics, engineering, finance, and genomics.

While considerable progress has been made in the application of Lasso over the years in terms of efficient computation \citep{lars}, statistical inference \citep{lee2016exact, taylor2018post}, and extension to generalized linear models \citep{park2008penalized, ivanoff2016adaptive} and Bayesian regression \citep{hans2009bayesian}, there are still some gaps to be filled for the Lasso as a modeling procedure in comparison with the ordinary least squares (OLS) regression, unarguably the primary method for modeling. Model diagnostics is an area far less developed for the Lasso, and this paper focuses specifically on the assessment of case influence in the Lasso.

Identification of influential points has long been a substantial part of regression model diagnostics \citep{belsley1980regression}. 
The presence of contaminated or unusual observations either in the response or predictor space, known as influential points, can severely distort the analysis result. 
It is arguably an even more compelling issue in Lasso for two reasons.
First, influential points in Lasso do not only alter the regression coefficients as in other regression models but also affect the model selection decision. Second, when the number of predictors $p$ is far greater than the number of observations $n$ ($p \gg n$), any changes in the observations can lead to a substantial change in the selected predictors. As a result, the Lasso model can be extremely sensitive to individual observations. 

Extensive research has been devoted to the influence of individual observations in the context of OLS regression and beyond. 
As a primary example, Cook's distance \citep{COOKDIS} in OLS regression evaluates the influence of an observation by considering the overall difference in the fitted values between the full data model and the model with the observation removed (leave-one-out model).
Since the seminal work of Cook's, researchers have developed different measures of case influence and generalized Cook's distance for detecting influential data to hierarchical linear models, mixed effect models, and generalized linear models \citep{belsley1980regression, pregibon1981logistic, pourahmadi2002dynamic, roy2004regression, pena2005new, zhu2012perturbation, loy2014hlmdiag, loy2017model}. 
Other similar case deletion statistics such as DFBETA and DFFITS \citep{welsch1977linear,belsley1980regression} have also been studied under these contexts. Moreover, these statistics have been extended to ridge regression \citep{walker1988influence, jahufer2009assessing}. 
Since the estimated coefficients in ridge regression can be calculated explicitly similar to OLS estimates, these influence measures can still be obtained without refitting the leave-one-out models.
However, this is not the case for Lasso.

To get around the computational issue in evaluating case influence on the Lasso model,
\cite{Kim2015} utilized a ridge-like closed-form approximation as a substitute for the Lasso solution and directly applied it to Cook's distance formula for influence diagnostics. \cite{jang2017influence} proposed an influence plot that includes all versions of Lasso solution path from the full data and leave-one-out data, and used this graphical tool to identify influential observations.
\cite{influencediag} proposed four new metrics that measure the 
change in various aspects (e.g., active set and optimal $\lambda$) between the models fitted to the full data and leave-one-out data. All metrics require fitting Lasso models $(n+1)$ times (once with the full data and $n$ times with each observation removed). 
More generally, for high-dimensional settings ($p\approx n$ or $p\gg n$), \cite{zhao2013high} defined the influence of an observation by the overall difference in marginal correlations between each of the features and the response using the full data and leave-one-out data 
 in place of Cook's distance and applied the measure to high dimensional regression methods including Lasso.

Inspired by the general approach to assessing the influence of individual observations through a case influence graph in \cite{pregibon1981logistic, cook1986assessment, tu2019case},
we consider a continuous weight attached to each case and define the Lasso solution path indexed by the case weight parameter, given a fixed value of penalty. We call this a case-weight adjusted Lasso model path. When the weight is 1 for a given case, the model corresponds to the full data fit, and when the weight is 0 while those for the rest are kept at 1, it corresponds to the leave-one-out fit. Using this model path, we can define a generalized Cook's distance viewed as a function of the case weight parameter.  This approach has both statistical and computational advantages. 

Statistically, it allows for a much broader and more comprehensive view of case influence, encompassing Cook's distance as a case-deletion statistic at the one end, and enabling us to trace the extent of change in the fitted model due to a single case on a continuum. Moreover, using the case influence graph and focusing on its local behavior around the weight equal to 1, we can also examine the sensitivity of the Lasso model to an infinitesimal change in each case and define the notion of local influence differently from the case-deletion statistic.
Computationally, linking the full data problem to the leave-one-out problem continuously with the weight parameter permits efficiency in obtaining the leave-one-out solution with the full data solution as a warm start. This computational strategy is proven to be very effective in getting the exact Lasso model when an observation is deleted, thereby allowing for the exact evaluation of Cook's distance in the Lasso. To the best of our knowledge, this homotopic approach to case influence diagnostics in the Lasso is new.

To characterize the case-weight adjusted Lasso model path analytically, we derive the optimality conditions for the model parameters and describe how to update the active set of non-zero regression coefficients, in particular, based on the conditions as the case weight reduces from 1 to 0. Further, we devise a new algorithm implementing this idea to offer a solution path of the coefficient estimates as a function of the weight for any fixed penalty parameter. This algorithm works for both cases when $p \le n$  and $p>n$. As a result, we can calculate the exact Cook's distance for the Lasso and perform leave-one-out cross-validation without refitting models. We also provide an appropriate procedure to identify influential points based on Cook's distance for the Lasso.

Since the estimates of coefficients and the active set in the Lasso model vary with penalty, case influence changes in response to the penalty parameter and resulting estimates. 
Influential points when the penalty is small may become less influential or not influential at all when a larger penalty is applied, and vice versa. 
For example, an influential data point with abnormal values in some features could become much less influential in the Lasso if those features are dropped from the active set with a large penalty. 
To monitor the influence dynamics of each observation as the penalty changes, 
we adopt the case influence graph as a visualization tool by indexing Cook's distance for the Lasso by
the penalty parameter rather than the case weight as in the original graph.
Empirically, we observe that case influence graphs exhibit different patterns between underfitting and overfitting models. 
Such differences can provide extra insights for model selection.

The rest of the paper is organized as follows. Section~\ref{cwasp} formally defines the case-weight adjusted Lasso model and presents a solution path algorithm. We show that the solution path is piecewise linear with respect to a simple function of the weight parameter. Section~\ref{caseinf} presents Cook's distance for the Lasso and a statistical procedure for identifying influential observations using the distance.
We also propose local influence as an alternative measure of case influence.
The case influence graph is introduced and explained in detail in Section~\ref{cig}. 
 We use synthetic and real datasets to illustrate the effectiveness of Cook's distance in detecting influential points in Section~\ref{simu}. We further demonstrate that the removal of such points generally improves consistency in model selection and parameter estimates. 
We also compare the exact Cook's distance with other proposed measures numerically.
In Section~\ref{conclusion}, we provide concluding remarks and suggest future research directions.

\section{Case-weight Adjusted Lasso}\label{cwasp}
In this section, we formulate a case-weight adjusted Lasso problem (Section~\ref{settings}) and provide the optimality conditions for the solution to the problem (Section~\ref{opt_cond}). Using the conditions, we show how the solution changes as a case-weight parameter varies. To describe piecewise changes in the solution as a function of the case-weight parameter, we discuss how to identify the breakpoints for the parameter (Section~\ref{break_deter}). Finally, we present a solution path algorithm based on the optimality conditions (Section~\ref{algor}). The detailed derivations for this section are presented in Appendix~\ref{app_cwasp}.

\subsection{Case-weight adjusted Lasso model} \label{settings}
We consider the standard linear regression setting with $p$ predictors. Let $X=[x_{ij}]\in \mathbb{R}^{n\times p}$ be the design matrix with $n$ observations $\bm{x}_i$ on the predictors and 
$\bm{y}=(y_1,\dots,y_n)^\top\in \mathbb{R}^n$ be the response vector. 
Let $\bm{x}_i^\top$ be the $i$th row of $X$ for $i \in [n]$, where $[n]=\{1,\dots,n\}$.
Without loss of generality, we also assume that each column of $X$ is centered with mean 0. With an unknown regression coefficient vector $\bm{\beta}^* = (\beta_1^*, \dots, \beta_p^*)^\top$, an intercept $\beta_0^*$, and error variance $\sigma^2$, we assume that  
\begin{align*}
    \bm{y}\sim N(\bm{1}_n\beta_0^* + X\bm{\beta}^*, \sigma^2 I_n).
\end{align*}

To assess the influence of each case in Lasso, we introduce a case weight parameter $\omega \in [0,1]$. Let $k\in [n]$ be the index for the case of interest whose weight $\omega$ changes from 1 to 0.
 Then the case-weight adjusted Lasso problem given a specified value of $\lambda$ is defined as follows:
\begin{equation}
\begin{aligned} 
    \min_{\beta_0, \bm{\beta}}\quad \frac{1}{2}\sum_{i\neq k} (y_i - \beta_0 - \bm{x}_i^\top\bm{\beta})^2 + \frac{1}{2}\omega(y_k - \beta_0 - \bm{x}_k^\top\bm{\beta})^2 +\lambda \sum_{j=1}^p |\beta_j|. \label{org_formula}
\end{aligned}
\end{equation}
We examine how the solution to \eqref{org_formula}, denoted by $\hat{\beta}_0^\omega$ and $\hat{\bm{\beta}}^\omega$, changes as a function of the case weight parameter $\omega$ while $\lambda$ is fixed. {Note that the dependence of $\hat{\beta}_0^\omega$ and $\hat{\bm{\beta}}^\omega$ on case $k$ and penalty parameter $\lambda$ is suppressed for brevity.}

\subsection{Optimality conditions}\label{opt_cond}
The regular Lasso is a special case of the case-weight adjusted Lasso when $\omega = 1$. In this section, we review the optimality conditions of the regular Lasso first and then extend them to the case-weight adjusted Lasso. The Lasso problem is defined as follows:
\begin{equation}
\begin{aligned} 
    \min_{\beta_0, \bm{\beta}}\quad \frac{1}{2}\sum_{i=1}^n (y_i - \beta_0 - \bm{x}_i^\top\bm{\beta})^2 +\lambda \sum_{j=1}^p |\beta_j|. \label{reg_Lasso}
\end{aligned}
\end{equation}

To describe the necessary and sufficient conditions for the solution to \eqref{reg_Lasso}, denoted by $\hat{\beta}_0$ and $\hat{\bm{\beta}}$, we define the derivatives of 
the residual sum of squares term first:
\begin{align*}
{d}_0 &= {d}_0(\beta_0, \bm{\beta}) =  -\bm{1}_n^\top(\bm{y} - {\beta}_0\mathbf{1}_n -X\bm{\beta})\\
\bm{d} & = \bm{d}(\beta_0, \bm{\beta}) = -X^\top (\bm{y}- {\beta}_0\bm{1}_n- X\bm{\beta}). 
\end{align*}
We define $\hat{d}_0 = {d}_0(\hat{\beta}_0,\hat{\bm{\beta}}) $ and $\hat{\bm{d}} = \bm{d}(\hat{\beta}_0,\hat{\bm{\beta}})$. Then the optimality conditions for the regular Lasso are given as follows:
\begin{align}
    &\hat{d}_0 = 0 \label{reg_d0}\\
    &\hat{\bm{d}} = - \lambda \hat{\bm{s}}\label{reg_d},
\end{align}    
{where $\hat{\bm{s}} = (\hat{s}_1, \dots, \hat{s}_p)^\top$ with $\hat{s}_j = \text{sign}(\hat{\beta}_j)$ if $\hat{\beta}_j \neq0$, and $\hat{s}_j \in [-1,1]$ if $\hat{\beta}_j=0$ for $j \in [p]$. }$\hat{s}_j$ can also be interpreted as the sub-derivative of $|\beta_j|$ at $\hat{\beta}_j$.

To express the solution corresponding to the optimality conditions,  given $\bm{\beta}$, we define $\mathcal{A} = \mathcal{A}(\bm{\beta}) = \{j \in [p]: \beta_j \neq 0\}$ as the active set. Similarly, we define $\hat{\mathcal{A}} = \mathcal{A}(\hat{\bm{\beta}})$. For any $\mathcal{A} \subset [p]$, $\backslash\mathcal{A}$ represents the complement of $\mathcal{A}$. $X_{\mathcal{A}}$ represents the sub-matrix of X containing only columns in $\mathcal{A}$, $\bm{v}_{\mathcal{A}}$ represents the sub-vector of $\bm{v}\in \mathbb{R}^p$ containing only the elements in $\mathcal{A}$.  
The regular Lasso solution is then given by
\begin{align}
    \hat{\bm{\beta}}_{\hat{\mathcal{A}}} =&\ (X_{\hat{\mathcal{A}}}^{\top}X_{\hat{\mathcal{A}}})^{-1}(X_{\hat{\mathcal{A}}}^{\top}\bm{y} - \lambda \hat{\bm{s}}_{\hat{\mathcal{A}}})\label{pseudo_beta}\\
    \hat{\bm{\beta}}_{\backslash\hat{\mathcal{A}}} =&\ \bm{0}\\
    \hat{\beta}_0 =&\ \Bar{y}\label{ybar_},
\end{align}
which imply
\begin{align}    
    \hat{\bm{d}}_{\hat{\mathcal{A}}} =& -\lambda \hat{\bm{s}}_{\hat{\mathcal{A}}}\\ 
    \hat{\bm{d}}_{\backslash\hat{\mathcal{A}}} =& -X_{\backslash\hat{\mathcal{A}}}^{\top} \left(\bm{y} - X_{\hat{\mathcal{A}}}\hat{\bm{\beta}}_{\hat{\mathcal{A}}} \right) \label{dna}\\
    \hat{\bm{y}} =&\ \hat{\beta}_0 \mathbf{1}_n + X_{\hat{\mathcal{A}}}\hat{\bm{\beta}}_{\hat{\mathcal{A}}}.\label{ycouterpart}
\end{align}

Similarly, we derive the optimality conditions and the solution for problem~\eqref{org_formula}. We use the superscript $\omega$ to emphasize the dependence of the conditions and resulting solution on $\omega$ (e.g., $\beta_0^\omega$ and $\bm{\beta}^\omega$ instead of $\beta_0$ and $\bm{\beta}$). Defining the derivatives of the residual sum of squares term analogously for the case-weight adjusted Lasso, we have
\begin{equation}
\begin{aligned}
d^\omega_0 &= -\bm{1}_n^\top(\bm{y} - \beta_0^\omega\mathbf{1}_n -X\bm{\beta}^{\omega}) + (1-\omega) (y_k - \beta_0^\omega - \bm{x}_k^\top\bm{\beta}^{\omega})\\
\bm{d}^\omega &= -X^\top(\bm{y} - \beta_0^\omega\mathbf{1}_n -X\bm{\beta}^{\omega}) + (1-\omega) \bm{x}_{k}(y_k - \beta_0^\omega - \bm{x}_k^\top\bm{\beta}^{\omega}).
\label{L_i}
\end{aligned}
\end{equation}
Then we can show that $\hat{\bm{\beta}}^\omega$ and $\hat{\beta}_0^\omega$ satisfy the following conditions: 
\begin{align}
     &\hat{d}^\omega_0 = 0\label{subgradient_0}\\
    &\hat{\bm{d}}^\omega = -\lambda \hat{\bm{s}}^\omega \label{subgradient}\\
    &\hat{s}^\omega_j = \text{sign}(\hat{\beta}_j^\omega)\quad \text{if } \hat{\beta}_j^\omega\neq0\label{gammaj}\\
    &\hat{s}^\omega_j \in [-1,1]\quad \text{if } \hat{\beta}^\omega=0.\label{gamma}
\end{align}
Note that these two sets of optimality conditions only differ in the form of $\hat{\bm{d}}$ and $\hat{d_0}$ because of the weighted case $k$, and the conditions in \eqref{subgradient_0}--\eqref{gamma} reduce to \eqref{reg_d0} and \eqref{reg_d} when the case weight $\omega$ is 1. The derivation of the conditions for the case weight adjusted Lasso is in Appendix~\ref{der_opt_cond}.

Next, we express the solution $\hat{\bm{\beta}}^\omega$, $\hat{\beta}_0^\omega$, $\hat{\bm{d}}^\omega$ and the fitted values $\hat{\bm{y}}^\omega$ based on the optimality conditions \eqref{subgradient_0}--\eqref{gamma}. 
Before presenting the results, let's first define a few auxiliary quantities.  
Letting $\hat{\mathcal{A}}^\omega=\{j \in [p]: \hat{\beta}_j^\omega \neq 0\}$ and $\widetilde{X}= (\mathbf{1}_n, X)$, we define the hat matrix $H^\omega$ as $\widetilde{X}_{\hat{\mathcal{A}}^\omega}(\widetilde{X}_{\hat{\mathcal{A}}^\omega}^{\top} \widetilde{X}_{\hat{\mathcal{A}}^\omega})^{-1}\widetilde{X}_{\hat{\mathcal{A}}^\omega}^{\top}$,
$\bm{h}^\omega_{\cdot k}$ as the $k$th column of $H^\omega$, 
and $h^\omega_{kk}$ as the $k$th diagonal entry of $H^\omega$. If $\omega=1$, we simply use $h_{kk}$ instead of $h_{kk}^{\omega=1}$. 
Then the solution for problem~\eqref{org_formula} and the fitted values are given by
\begin{align}
    \hat{\bm{\beta}}^\omega_{\hat{\mathcal{A}}^\omega} &= \hat{\bm{\beta}}_{\hat{\mathcal{A}}^\omega} - \xi^\omega(X_{\hat{\mathcal{A}}^\omega}^{\top}X_{\hat{\mathcal{A}}^\omega})^{-1}\bm{x}_{k\hat{\mathcal{A}}^\omega}(y_k-\hat{y}_k)\label{beta_e}\\
    \hat{\bm{\beta}}^\omega_{\backslash\hat{\mathcal{A}}^\omega} &= \bm{0}\label{beta_ne}\\
    \hat{\beta}_0^\omega &= \hat{\beta}_0 - \xi^\omega(y_k - \hat{y}_k)/n\label{re_beta_0}\\
    \hat{\bm{d}}^\omega_{\hat{\mathcal{A}}^\omega} &= -\lambda \hat{\bm{s}}^\omega_{\hat{\mathcal{A}}^\omega}\label{d_trivial}\\
    \hat{\bm{d}}^\omega_{\backslash\hat{\mathcal{A}}^\omega} &= \hat{\bm{d}}_{\backslash\hat{\mathcal{A}}^\omega} - \xi^\omega(X_{\backslash\hat{\mathcal{A}}^\omega}^\top \bm{h}^\omega_{\cdot k} - \bm{x}_{k\backslash\hat{\mathcal{A}}^\omega})(y_k - \hat{y}_k)\label{domega}\\
    \hat{\bm{y}}^\omega &= \hat{\bm{y}} - \xi^\omega\bm{h}^\omega_{\cdot k}(y_k - \hat{y}_k),\label{yomega}
\end{align}
where 
\begin{align}
    \xi^\omega = \frac{1-\omega}{1-(1-\omega)h_{kk}^\omega}.\label{xidef}
\end{align}
The derivation of this result is in Appendix~\ref{solution_der}. 
Note that given the active set $\hat{\mathcal{A}}^\omega$ and the sign vector $\hat{\bm{s}}^\omega$, the corresponding 
$\hat{\bm{\beta}}_{\hat{\mathcal{A}}^\omega}$, $\hat{\bm{d}}_{\backslash\hat{\mathcal{A}}^\omega}$, and $\hat{\bm{y}}$ shown in the solution take the same form as in the regular Lasso solution \eqref{pseudo_beta}--\eqref{ycouterpart}.
However, they are no longer the regular Lasso solution since $\hat{\mathcal{A}}^\omega$ and $\hat{\bm{s}}^\omega$ might be different from $\hat{\mathcal{A}}$ and $\hat{\bm{s}}$ for $\omega=1$.

\begin{figure}[htb]
    \centering
    \includegraphics[width=0.6\textwidth]{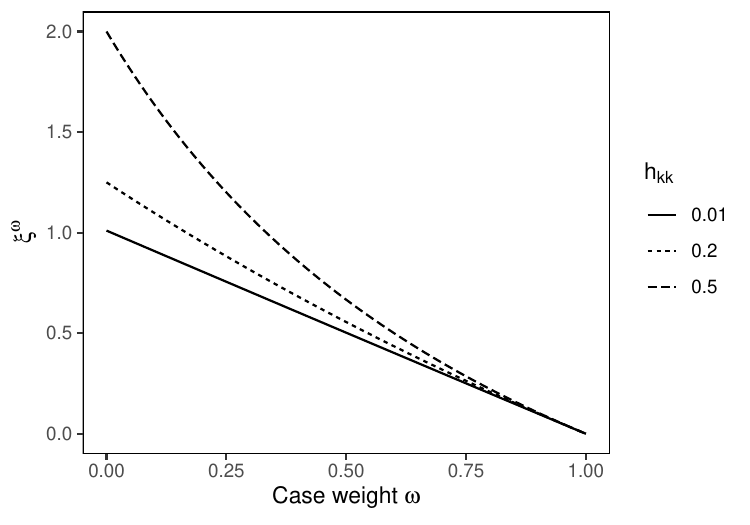}
    \caption{The relationship between the case weight $\omega$ and $\xi^\omega$ for different values of the leverage $h_{kk}^\omega$ as $\omega$ ranges from 1 to 0. Given sample size $n$, the leverage can take values from $1/n$ to 1.}
    \label{fig:xiw}
\end{figure}

As $\omega$ changes from 1 to 0 for fixed $\lambda$, $\xi^\omega$ is the only variable that changes continuously on the right-hand sides of \eqref{beta_e}--\eqref{yomega} while all others change discretely with respect to $\hat{\mathcal{A}}^\omega$.  Therefore, $\hat{\bm{\beta}}^\omega$, $\hat{\beta}_0^\omega$, $\hat{\bm{d}}^\omega$ and $\hat{\bm{y}}^\omega$ are all continuous and piecewise linear with respect to $\xi^\omega$, and their solution paths can be naturally indexed by $\xi^\omega$.
Figure~\ref{fig:xiw} shows the relationship between the case weight $\omega$ and $\xi^\omega$ for different values of the leverage $h_{kk}^\omega$. As shown in the figure, $\xi^\omega$ is a monotone decreasing function of $\omega$ (see Appendix~\ref{monot_of_xi}).
As $h_{kk}^\omega$ decreases from 1 to $1/n$, it ranges from a reciprocal function ($1/\omega-1$) to a nearly linear function (approaching $1-\omega$).

The quantities, $\hat{\bm{\beta}}^\omega$, $\hat{\beta}_0^\omega$, $\hat{\bm{d}}^\omega_{\backslash\hat{\mathcal{A}}^\omega}$ and $\hat{\bm{y}}^\omega$, all have the form of $\bm{a}+\bm{b}\xi^\omega$, where $\bm{a}$'s are their counterparts in the regular Lasso solution with active set ${\mathcal{A}}^\omega$ and $\bm{b}$'s share a common factor of $y_k-\hat{y}_k$. By \eqref{yomega}, the $k$th residual from the case-weight adjusted Lasso fit is 
\begin{align}
    y_k - \hat{y}^\omega_k= ( y_k -\hat{y}_k) + \xi^\omega h_{kk}^\omega (y_k -\hat{y}_k) = \frac{(y_k - \hat{y}_k)}{1-(1-\omega)h_{kk}^\omega}.\label{resid_form}
\end{align}
This residual becomes the regular Lasso residual when $\omega = 1$ and the leave-one-out residual when $\omega = 0$. It can be shown that the case-weight adjusted residual for OLS takes the same form as that for the Lasso but with a constant leverage $h_{kk}$. This allows us to obtain the residual with a different case weight, including the leave-one-out residual, directly from the full data fit in OLS. In fact, this is true for linear modeling procedures in general. However, for the Lasso as a nonlinear modeling procedure, we cannot obtain the case-weight adjusted residual directly from the regular residual because $h_{kk}^\omega$ here needs to be updated with $\omega$ as well. 
When there is no active set update, the residual gets smaller in magnitude as the case weight assigned to case $k$ increases.

\subsection{Determining breakpoints}\label{break_deter}
In the previous section, we have characterized the behavior of $\hat{\bm{\beta}}^\omega$, $\hat{\beta}^\omega_0$ and $\hat{\bm{d}}^\omega$ as a function of $\omega$. We now turn to the problem of locating $\omega$ values that lead to updates in the active set $\hat{\mathcal{A}}^\omega$. We call such values breakpoints.

Based on the optimality conditions \eqref{subgradient_0}--\eqref{gamma}, two kinds of changes can occur at each breakpoint:
an active variable becomes inactive or vice versa. In other words, a nonzero $\hat{\beta}_j^\omega$ becomes zero or vice versa.
By tracking the events where the active set changes as $\omega$ decreases from 1, we identify the breakpoints sequentially. Suppose that the current breakpoint is $\omega_u \in (0,1]$, and we aim to find the next breakpoint $\omega_l\in [0,\omega_u)$.
The active set $\hat{\mathcal{A}}^{\omega}$ for $\omega \in ({\omega_l}, \omega_u)$ remains the same as $\hat{\mathcal{A}}^{\omega_u}$ since there is no active set update in-between.
Therefore, regardless of the kind of change at $\omega_l$, 
$\hat{\bm{\beta}}^\omega$ and $\hat{\bm{d}}^\omega$ must satisfy the following conditions:
\begin{align*}
        &\min_{\omega\in(\omega_l, \omega_u),~j\in\hat{\mathcal{A}}^{\omega_u}} \{|\hat{\beta}_j^{\omega} |\} > 0\\
        &\max_{\omega\in({\omega_l}, \omega_u),~j\in\backslash\hat{\mathcal{A}}^{\omega_u}} \{|\hat{d}_j^\omega|\} < \lambda.
\end{align*}
 If the next update is due to a change in one variable from being active to inactive,
 $\omega_l$ should be the largest value smaller than $\omega_u$ where one $\hat{\beta}_j^\omega$ with $j\in\hat{\mathcal{A}}^{\omega_u}$ becomes 0:
\begin{align*}
    \omega_l^- = \max \left\{\omega\in [0, \omega_u)~\bigg\vert~ \prod_{j\in \hat{\mathcal{A}}^{\omega_u}} \hat{\beta}_j^{\omega} = 0\right\}.
\end{align*}
Likewise, if it is the other way around, $\omega_l$ should be the largest value smaller than $\omega_u$ where one $|\hat{d}_j^\omega|$ with $j\in\backslash\hat{\mathcal{A}}^{\omega_u}$ becomes $\lambda$:
\begin{align*}
    \omega_l^{+} = \max \left\{\omega\in [0, \omega_u)~\bigg\vert~ \prod_{j\in \backslash\hat{\mathcal{A}}^{\omega_u}} (|\hat{d}_j^{\omega}| - \lambda) = 0\right\}.
\end{align*}
Therefore, the next breakpoint $\omega_l$ is the maximum of $\omega_l^-$ and $\omega_l^{+}$:
\begin{align*}
    \omega_l =\max (\omega_l^-, \omega_l^{+}).
\end{align*}

In Section~\ref{opt_cond}, we have shown that $\hat{\bm{\beta}}^\omega$ is a linear function of $\xi^\omega$ between two adjacent breakpoints. Thus, to locate $\omega_l$, we first set $\hat{\bm{\beta}}_{\hat{\mathcal{A}}^{\omega_u}}^{\omega_u}$ to $\bm{0}$ and $\hat{\bm{d}}_{\backslash\hat{\mathcal{A}}^{\omega_u}}^{\omega_u}$ to $\pm\lambda\mathbf{1}$ to solve for all candidate values of $\xi$:
\begin{gather}
 \begin{pmatrix}
     \hat{\bm{\beta}}_{\hat{\mathcal{A}}}\\
     \hat{\bm{d}}_{\backslash\hat{\mathcal{A}}}\\
     \hat{\bm{d}}_{\backslash\hat{\mathcal{A}}}
 \end{pmatrix}
 + 
\bm{\xi}  \circ
 \begin{pmatrix}     
 (X_{\hat{\mathcal{A}}}^{\top}X_{\hat{\mathcal{A}}})^{-1}\bm{x}_{k\hat{\mathcal{A}}}\\ X_{\backslash\hat{\mathcal{A}}}^\top \bm{h}_{\cdot k}^{\omega_u} - \bm{x}_{k\backslash\hat{\mathcal{A}}}\\
 X_{\backslash\hat{\mathcal{A}}}^\top \bm{h}_{\cdot k}^{\omega_u} - \bm{x}_{k\backslash\hat{\mathcal{A}}}\end{pmatrix}
 (\hat{y}_k - y_k)
 =
  \begin{pmatrix}
  \bm{0}\\
  \lambda\mathbf{1}\\
  -\lambda\mathbf{1}
   \end{pmatrix},
   \label{matrix_eq}
\end{gather}
where $\hat{\mathcal{A}}$ represents $\hat{\mathcal{A}}^{\omega_u}$ for conciseness, $\circ$ indicates the elementwise product of vectors, and all vectors are of length $(|\hat{\mathcal{A}}^{\omega_u}| + 2(p-|\hat{\mathcal{A}}^{\omega_u}|))$. Specifically, $\bm{\xi}$ contains one candidate value $\xi$ for each index in $\hat{\mathcal{A}}^{\omega_u}$ and two for each index in $\backslash\hat{\mathcal{A}}^{\omega_u}$.  By \eqref{xidef}, we can calculate $\xi^{\omega_u}$. Since $\xi$ is monotonically decreasing in $\omega$, to locate $\omega_l$, we first find the minimum element ($\xi_{j^{'}},\ j^{'}\in[2p-|\hat{\mathcal{A}}^{\omega_u}|]$) in $\bm{\xi}$ that is greater than $\xi^{\omega_u}$ and obtain $\omega_l = \omega(\xi_{j^{'}}, h_{kk}^{\omega_u})$ by \eqref{xidef}. 
Then the active set and the sign vector, $\hat{\mathcal{A}}^{\omega_l}$ and $\hat{\bm{s}}^{\omega_l}$, can be updated accordingly. Details can be found in Section~\ref{algor}.

\subsection{Solution path algorithm}\label{algor}

Using the form of the solution to \eqref{org_formula} and the mechanism of how the breakpoints are determined, we present our solution path algorithm in Algorithm \ref{alg}.

\begin{algorithm}
\caption{Solution path algorithm}\label{alg}
\begin{algorithmic}
\State Input: $X \in \mathbb{R}^{n\times p}$, $\bm{y}\in \mathbb{R}^n$, $\lambda\in \mathbb{R}^+$, $k\in [n]$
\State Center $X$ column-wise to have mean 0.
\State Obtain the sign vector $\hat{\bm{s}}$ and the active set $\hat{\mathcal{A}}$ from the regular Lasso.  
\State Initialize $m = 0$ and $\omega_0 = 1$.

\While{$\omega_m > 0$}
    \State Calculate 
     $h_{kk}$ based on $\hat{\mathcal{A}}$, and 
     $\xi^{\omega_m}$ based on $h_{kk}$ and $\omega_m$  by (\eqref{xidef}).
    \State Calculate $\hat{\bm{\beta}}^m = (\hat{\beta}_0^{\omega_m}, \hat{\bm{\beta}}^{\omega_m})$ by  \eqref{beta_e}, (\eqref{beta_ne}, and \eqref{re_beta_0}.
    \State Calculate $\hat{\bm{\xi}} = \bm{\xi}(\hat{\bm{s}}, \hat{\mathcal{A}})$ by \eqref{matrix_eq}.
    \State Calculate $j^{'}= \argmin_{j\in [2p - |\hat{\mathcal{A}}|], \hat{\xi}_j > \xi^{\omega_m}} \hat{\xi}_j$. 
    \State Locate the covariate index $j\in[p]$ that corresponds to $j^{'}$.
    \If{$\hat{\xi}_{j^{'}}<\frac{1}{1-h_{kk}}$ ($+\infty$ if $h_{kk}=1$)}
        \State  $\omega_{m+1} = \omega(\hat{\xi}_{j^{'}}, h_{kk})$ by \eqref{xidef}
        \If{$j^{'}\leq|\hat{\mathcal{A}}|$}
            \State $\hat{\mathcal{A}} \xleftarrow{} \hat{\mathcal{A}}\backslash\{j\}$
        \Else
            \State  $\hat{\mathcal{A}} \xleftarrow{} \hat{\mathcal{A}}\cup\{j\}$
            \If{$j^{'}>p$}
                \State $\hat{\bm{s}}_j = -1$
            \Else
                \State $\hat{\bm{s}}_j = 1$
            \EndIf \EndIf
    \Else
        \State $\hat{\xi}_{j^{'}} = \frac{1}{1-h_{kk}}$ and $\omega_{m+1} = 0$
    \EndIf
    \State $\xi^{\omega_{m,2}} = \hat{\xi}_{j^{'}}$
    \State $m \gets m + 1$
\EndWhile
\State Output: $\omega_m$, and $\hat{\bm{\beta}}^m$ for $m=0,\dots, M$ (the number of breakpoints $\omega_m$)
\end{algorithmic}
\end{algorithm}

\begin{figure}[htb]
    \centering
    \includegraphics[width = 0.9\textwidth]{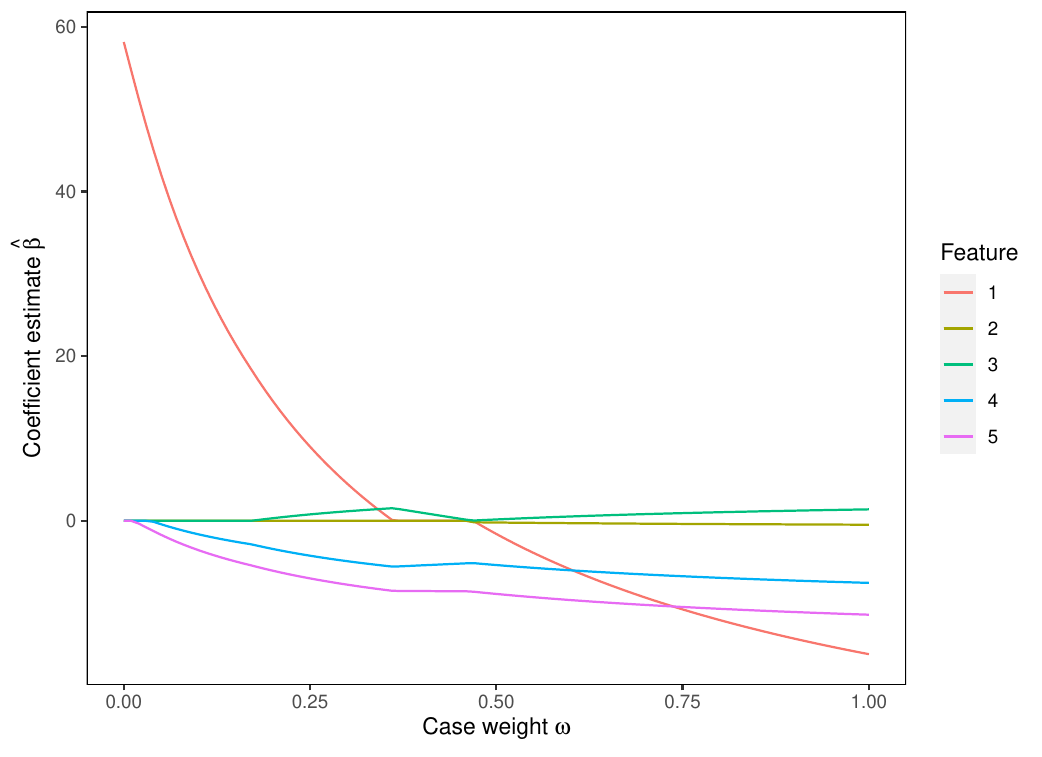}
    \caption{The case-weight adjusted Lasso coefficient paths for an influential observation in synthetic data ($n=20$, $p=5$). Each curve is the trajectory of each regression coefficient estimate indexed by case-weight $\omega$ for a fixed $\lambda$.}
    \label{fig:sp_30}
\end{figure}

Let $M$ be the number of breakpoints. Algorithm~\ref{alg} gives a sequence of $\hat{\bm{\beta}}^m$ for $m = 0,\dots,M$, breakpoints $\omega_m$ with $0 = \omega_M<\cdots<\omega_0=1$ and corresponding $(\xi^{\omega_m}, \xi^{\omega_{m,2}})$
pairs except when $m=M$, where $\xi^{\omega_m} = \xi(\omega_m, h_{kk}^{\omega_m})$ and
$\xi^{\omega_{m,2}} = \xi(\omega_{m+1}, h_{kk}^{\omega_m})$.
Once we identify the breakpoints $\omega_m$ and the corresponding case-weight adjusted Lasso estimate $\hat{\bm{\beta}}^m$, we can obtain the entire solution path through linear interpolation of the estimate with respect to $\xi^\omega$ rather than $\omega$. This is because $(\hat{\beta}_0^\omega,\hat{\bm{\beta}}^\omega)$ is piecewise linear in $\xi^\omega$ as shown in \eqref{beta_e} and \eqref{re_beta_0} while $\xi^\omega$ is a nonlinear function of $\omega$ as in \eqref{xidef}. 
Hence, for any case $k$ and weight $\omega\in[0,1]$, we can obtain the corresponding $\xi^\omega$ and the solution to problem~\eqref{org_formula} indexed by $\omega$ as follows: 
\begin{equation*}
    (\hat{\beta}_0^\omega, \hat{\bm{\beta}}^\omega) = 
    \begin{cases}
        \hat{\bm{\beta}}^{m} + (\hat{\bm{\beta}}^{m+1}- \hat{\bm{\beta}}^{m})\frac{\xi^\omega - \xi^{\omega_m}}{\xi^{\omega_{m,2}} - \xi^{\omega_m}}& \text{if  } \omega_{m+1}< \omega\leq\omega_{m}, m = 0,\dots,M-1\\
        {\hat{\bm{\beta}}^M}& \text{if  } \omega=\omega_M=0.
    \end{cases}
\end{equation*}

Figure~\ref{fig:sp_30} illustrates a solution path of $\hat{\bm{\beta}}^\omega$ for an influential point in synthetic data with five features {where only feature 1 is relevant.} It shows how the estimates of Lasso coefficients change as the case-weight $\omega$ changes from 1 to 0 when
$\lambda$ is fixed. 
It might seem more natural to use $\xi^\omega$ in the $x$-axis for indexing coefficient estimates in the figure as it offers a piecewise linear solution path. However, note that $\xi^\omega$ depends on the active set, and hence does not change continuously as the active set changes. 
For example, if $M=2$, we have $0 = \omega_2 < \omega_1 < \omega_0 =1$, 
$h_{kk}^{\omega_0}$, $h^{\omega_1}_{kk}$, 
and pairs of $(\xi^{\omega_0},\xi^{\omega_0,2})$ and $(\xi^{\omega_1},\xi^{\omega_1,2})$ associated with $\omega_0$ and $\omega_1$, respectively. $(\hat{\beta}_0^\omega,\hat{\bm{\beta}}^\omega)$ is linear in $\xi^\omega$ over $(\xi^{\omega_0}, \xi^{\omega_0,2})$ with $\hat{\mathcal{A}}^{\omega_0}$ and $(\xi^{\omega_1}, \xi^{\omega_1,2})$ with $\hat{\mathcal{A}}^{\omega_1}$. Unlike $\omega$ that ranges from 1 to 0 continuously, $\xi^\omega$ is a derived parameter that also depends on the current leverage $h_{kk}^\omega$. Therefore, $\xi^{\omega_0,2} \neq \xi^{\omega_1}$, which makes the two intervals $(\xi^{\omega_0},\xi^{\omega_0,2})$ and $(\xi^{\omega_1},\xi^{\omega_1,2})$ either overlapping or disjoint, depending on whether a variable is dropped or added in the update. Thus, $\xi^\omega$ is not an ideal index for plotting the solution path.

Algorithm~\ref{alg} can also work when $p\geq n$. Recall that the Lasso can select up to $(n-1)$ variables plus an intercept before the model saturates. Thus, it suffices to show that Algorithm~\ref{alg} works when $(n-1)$ variables are in the active set,  $\hat{\mathcal{A}}$. First, we establish that Algorithm~\ref{alg} never selects more than $(n-1)$ variables in this case. When $(n-1)$ variables are active, the matrix $(\mathbf{1}_n,X_{{\hat{\mathcal{A}}}})$ becomes full rank. Then, there exists a matrix $Q \in\mathbb{R}^{n \times (p-|{\hat{\mathcal{A}}}|)}$ such that $X_{\backslash{\hat{\mathcal{A}}}} = (\mathbf{1}_n,X_{{\hat{\mathcal{A}}}})Q$. As a result, the term in $\eqref{matrix_eq}$, $X_{\backslash\hat{\mathcal{A}}}^\top \bm{h}_{\cdot k}^{\omega_u} - \bm{x}_{k\backslash\hat{\mathcal{A}}} = Q^\top \bm{x}_{k\hat{\mathcal{A}}} - \bm{x}_{k\backslash\hat{\mathcal{A}}} = 0$, which guarantees that no additional variable can be included in ${\hat{\mathcal{A}}}$ as $\omega$ decreases. 
Moreover, whenever $(n-1)$ variables are active, the corresponding $h_{kk}^\omega$ is 1. Hence, the termination condition ($\hat{\xi}_{j^{'}}\geq \frac{1}{1-h_{kk}^\omega}$) is not met, and the algorithm proceeds to drop a variable from ${\hat{\mathcal{A}}}$. By the same condition for termination, for $\omega$ to reach 0, it is necessary that $h_{kk}^\omega<1$. This implies that at most $(n-2)$ variables can be active when the algorithm terminates, which is consistent with the fact that when $\omega=0$, the Lasso model fit on $(n-1)$ observations can include at most $(n-2)$ variables.

We have developed an R package, CaseWeightLasso, which implements Algorithm \ref{alg}. The package also provides Cook's distance for the Lasso introduced in Section~\ref{ck_Lasso}, and case influence graphs introduced in Section~\ref{cig}. 
It is available 
from the GitHub repository: \url{https://github.com/zbjiao/CaseWeightLasso}.

\section{Measuring Case Influence} \label{caseinf}
In this section, we introduce Cook's distance for the Lasso (Section~\ref{ck_Lasso}) and a measure of local influence (Section~\ref{other_measure}) that can also be useful for case influence diagnostics. 

As defined before, $\bm{y}$ is the observed response, $\hat{\bm{y}}$ is the fitted value vector in \eqref{ycouterpart} given $\lambda$ for the regular Lasso, and $\hat{\bm{y}}^\omega$ is the Lasso fitted vector with case $k$ having weight $\omega$ in \eqref{yomega}. As stated earlier, $\hat{\bm{y}}$ in \eqref{yomega} is different from the regular Lasso estimate $\hat{\bm{y}}$ in \eqref{ycouterpart} in that $\hat{\bm{y}}$ in \eqref{yomega} is a function of  $\hat{\mathcal{A}}^\omega$ and $\hat{\bm{s}}^\omega$. To differentiate these two, we write $\hat{\bm{y}}$ in \eqref{ycouterpart} as $\hat{\bm{y}}(\lambda)$ and $\hat{\bm{y}}$ in \eqref{yomega} as $\hat{\bm{y}}(\lambda,\omega)$. In addition, we define $\Tilde{\bm{y}}$ as the OLS fitted vector. We use similar notation for $\hat{\bm{\beta}}$ and $\hat{\beta_0}$. 

\subsection{Cook's distance for the Lasso}\label{ck_Lasso}
We extend the concept of Cook's distance from OLS to the Lasso.
Letting $\Tilde{y}_{i}^{(-k)}$ be the OLS fitted value for $y_i$ when observation $k$ is deleted, Cook's distance for the observation in OLS is defined as
\begin{align}
    D_k = \frac{1}{(p+1)s^2} \sum_{i=1}^n (\Tilde{y}_i - \Tilde{y}_{i}^{(-k)})^2,\label{cookdis}
\end{align}
where $s^2$ is an estimate of $\sigma^2$. Using the algebraic relation between $\hat{\bm{\beta}}$ and $\hat{\bm{\beta}}^{(-k)}$ in OLS without refitting the model, the distance can be expressed as 
\begin{align}
   D_k = \frac{r_k^2}{(p+1)s^2}\frac{h_{kk}}{(1-h_{kk})^2},\label{cookdis2}
\end{align}
 where $r_k = y_k - \Tilde{y}_k$ is the residual for observation $k$, and $h_{kk}$ is the $k$th leverage from the OLS fit.

By replacing $\Tilde{\bm{y}}^{(-k)}$ with the Lasso fitted value vector $\hat{\bm{y}}^\omega$ with case $k$ of weight $\omega$ at a given $\lambda$, we define a case influence function (also known as a case influence graph) as
\begin{align}
    D_k(\lambda, \omega) =\frac{1}{(p+1)s^2} \sum_{i=1}^n (\hat{y}_i(\lambda) - \hat{y}_{i}^\omega)^2 = \frac{1}{(p+1)s^2}\|\hat{\bm{y}}(\lambda) - \hat{\bm{y}}^\omega\|^2.\label{Dklo}
\end{align}
Then, Cook's distance for the Lasso given $\lambda$ is defined as the value of the function at $\omega = 0$, $D_k(\lambda, 0)$. 
In this definition, we keep the same factor of $(p+1)$ for normalization rather than $|\hat{\mathcal{A}}^\omega|$ in the denominator to make the case influence function continuous at breakpoints of $\omega$.
{In addition, $s^2$ as an estimate of the true error variance $\sigma^2$ doesn't depend on the choice of $\lambda$.}

The following proposition describes the monotonicity of the case influence function as the case weight changes. 
In particular, we show that as the weight $\omega$ for case $k$ increases, the derivative of its case influence viewed as a function of $\xi^\omega$, $\frac{\partial D_k}{\partial \xi^\omega}$, monotonically decreases between two consecutive breakpoints piecewise.

\begin{proposition} \label{prop2}
For each $k\in [n]$ and fixed $\lambda\in \mathbb{R}^+$, let $\{\omega_m\}_{m=0}^M$ be a sequence of breakpoints of $\omega$ with $0 = \omega_M<\cdots<\omega_0=1$. Then the derivative of $D_k(\lambda, \omega)$ with respect to $\xi^\omega$ is piecewise monotonically decreasing in $\omega\in (\omega_{m+1},\omega_m)$ for $m=0,\dots,M-1$.
\end{proposition}
The proposition implies some conditions for the monotonicity of $D_k(\lambda, \omega)$ with respect to $\omega$.
\begin{proposition} \label{coro1}
$D_k(\lambda, \omega)$ is monotonically decreasing in $\omega\in[0,1]$ for every $k\in [n]$ and fixed $\lambda\in \mathbb{R}^+$ if one of the following conditions holds: \\
i. Each active set update is a variable being added to the active set. \\
ii. Each active set update is a variable being dropped from the active set.\\ 
iii. No updates to the active set occur.
\end{proposition}
The proofs are in Appendix~\ref{app_prop2} and Appendix~\ref{app_coro1} respectively.

\subsubsection{Approximation of Cook's distance}
When there is no update on the active set $\hat{\mathcal{A}}^\omega$ as $\omega$ ranges from 1 to 0 (i.e., $M=1$), we have $\hat{\bm{y}}(\lambda,\omega) = \hat{\bm{y}}(\lambda)$. Under this scenario, by \eqref{yomega}, \eqref{xidef}, and \eqref{Dklo}, Cook's distance for the Lasso is expressed as
\begin{align}
     D_k(\lambda, 0) =&\ \frac{1}{(p+1)s^2} \sum_{i=1}^n \left(\hat{y}_i(\lambda) - \left[\hat{y}_i(\lambda,0) -  \frac{h_{ik}}{1 - h_{kk}}(y_k -\hat{y}_k(\lambda,0) )\right]\right)^2\nonumber\\
     =&\ \frac{(y_k - \hat{y}_k(\lambda))^2}{(p+1)s^2 (1-h_{kk})^2} \sum_{i=1}^n (h_{ik})^{2}\nonumber\\ 
     =&\ \frac{(y_k - \hat{y}_k(\lambda))^2}{(p+1)s^2}  \frac{h_{kk}}{(1-h_{kk})^2} \label{ocookdis}.
\end{align}
Note that Cook's distance under this scenario has the same form as that of the OLS fit in \eqref{cookdis2}, and $h_{kk}$ here is the same as the $k$th leverage based on the active set at $\lambda$ for the Lasso. This allows us to calculate Cook's distance exactly based on the full data solution. 
In addition, if the active set changes little along the solution path, the formula in \eqref{ocookdis} with the initial active set at $\omega=1$ can be used to approximate Cook's distance for the Lasso.
This approximation is compared with two alternative approximations and the exact measure numerically in Section~\ref{measure_compare}.  

From Algorithm~\ref{alg}, if no candidate in $\hat{\bm{\xi}}$ falls in $(0,\frac{1}{1-h_{kk}})$, then there will be no change in the active set. 
{From \eqref{matrix_eq}, we see that the ratios of each element in 
$\hat{\bm{\beta}}_{\hat{\mathcal{A}}}$,
    $\hat{\bm{d}}_{\backslash\hat{\mathcal{A}}}$, and 
    $\hat{\bm{d}}_{\backslash\hat{\mathcal{A}}}$, and its corresponding element in the second vector grow with $n$.} 
This indicates that a large $n$ would lead to candidate values $\xi$ of a large magnitude, and thus a change in the active set is less likely than with a small $n$. Also, a large $p$ would bring more candidates in $\bm{\xi}$, and thus it is more likely to have a change in the active set than with a small $p$. Generally speaking, this approximation of Cook's distance works well in real datasets when $n\gg p$.

\subsubsection{Identification of influential observations}\label{thresholding}
We consider identifying influential points by setting a reasonable threshold on Cook's distance for the Lasso introduced in Section~\ref{ck_Lasso}.
For Cook's distance in OLS regression, there is a heuristic for determining such a threshold based on a confidence ellipsoid for regression coefficients. For level $1-\alpha$,
\begin{align*}
    \frac{1}{{(p+1)s^2}}\left|\left|(\mathbf{1}_n, X)\left(
    \begin{pmatrix}
{\beta}_0\\
\bm{\beta}
\end{pmatrix}
     -
     \begin{pmatrix}
     \Tilde{{\beta_0}}\\
     \Tilde{\bm{\beta}}
     \end{pmatrix}
     \right)\right|\right|^2\leq F(p+1,n-p-1,1-\alpha),
\end{align*}
where $F(p+1, n-p-1, 1-\alpha)$ is the $(1-\alpha)$ quantile of the $F$ distribution with $p+1$ and $n-p-1$ degrees of freedom. Cook's distance is obtained by replacing $\begin{pmatrix}
{\beta}_0\\
\bm{\beta}
\end{pmatrix}$ with $\begin{pmatrix} \Tilde{\beta}_0^{(-k)}\\ \Tilde{\bm{\beta}}^{(-k)}\end{pmatrix}$ in the left-hand side.
Under the intuition that the estimates with the $k$th case removed would be close to the original estimates if the $k$th case is not influential, \cite{COOKDIS} proposed to make use of a quantile of the $F$ distribution as a threshold to identify influential points with a reasonable $\alpha$.  

In the Lasso setting, \cite{lee2016exact} shows that $\hat{\bm{\beta}}$ follows a truncated multivariate normal distribution. The truncated region is rectangular and centered at the origin, and its size increases with $\lambda$. 
While this makes it possible to study the exact distribution of Cook's distance for the Lasso, computation of the exact quantiles of the distribution is very challenging. 
Instead, we consider a simple approach of approximating the distribution of Cook's distance in Lasso with a scaled $\chi^2$-distribution.

\begin{figure}[htb]
    \centering
    \includegraphics[width = 0.8\textwidth]{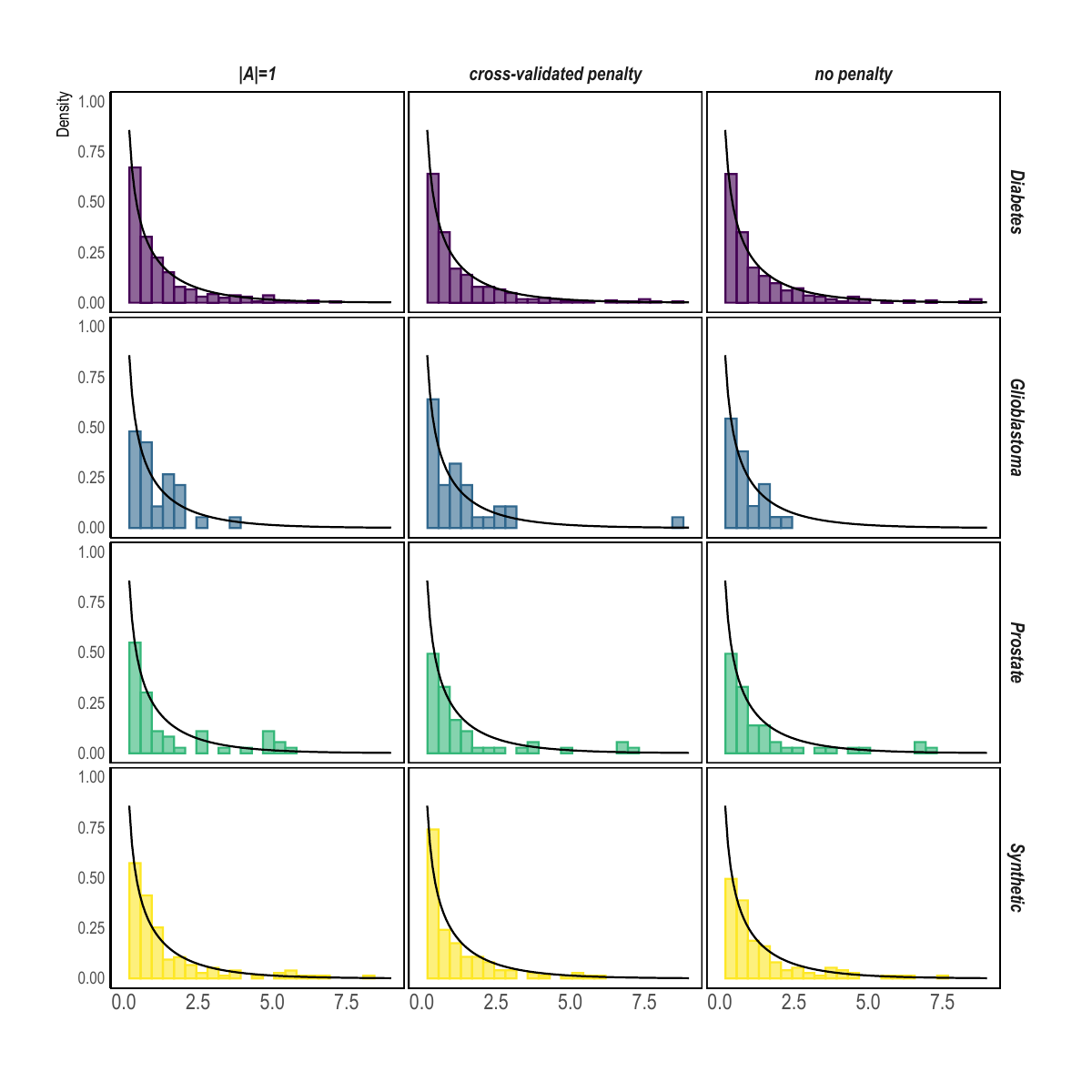}
    \caption{Histograms of $D_k(\lambda,0)/{\sqrt{\hat{\var}(D_k(\lambda,0))/2}}$ across four distinct datasets at different penalty levels. The probability density curve of $\chi^2_1$-distribution (black curve) is included in each panel for comparison.}
    \label{fig:chisq_justification}
\end{figure}

In particular, we assume that $D_k(\lambda,0)$ follows a scaled $\chi^2_1$ distribution: 
\begin{align}
\frac{D_k(\lambda,0)}{\sqrt{\var(D_k(\lambda,0))/2}}\sim \chi^2_1.\label{threshold}
\end{align}
We justify this assumption in two ways. Firstly, $D_k(\lambda,0)$ can be approximated by a residual square $\times$ a function of leverage as in \eqref{ocookdis}. If we further assume the studentized Lasso residual of $y_k$, ${(y_k - \hat{y}_k(\lambda))}/{\sqrt{1-h_{kk}}}$, follows a normal distribution for a fixed $\lambda$, the standardized Cook's distance for the Lasso in \eqref{threshold} should approximately follow a $\chi^2_1$. 
This assumption will be better met if the fitted model contains all the relevant features. 
When the model selects most of the relevant features, the residuals are composed of errors that cannot be explained by the features, and thus they are expected to be approximately normal similar to OLS residuals except for the shrinkage effect of Lasso. On the other hand, when the model is over-penalized, the residuals still contain feature information, systematically deviating from a normal distribution assumption.
As we are generally interested in candidate models that explain the data well, 
such heavily underfitting models wouldn't be of interest.
To test our assumption numerically, we evaluate Lasso residuals using both real and synthetic datasets (mentioned in the next paragraph). Appendix~\ref{resid_plot} includes the resulting distributions of studentized  Lasso residuals under varying penalty levels. The results corroborate our arguments.

Secondly, 
a $\chi^2_1$-distribution provides a good match with the histograms of standardized Cook's distance for various Lasso models empirically.
Figure~\ref{fig:chisq_justification} shows histograms of Cook's distance for the Lasso models with four distinct datasets at three different penalty levels. 
The histograms in the first row are for the diabetes data ($n=442, p=10$), introduced in Section~\ref{diabetes_example}. The second row is for the glioblastoma gene expression data ($n=50, p=3600$) introduced in Section~\ref{glio_example}. 
The third row is for the prostate cancer data ($n=97, p=8$), introduced in Section~\ref{measure_compare}. The last row is for synthetic data ($n=200, p=200$), generated following the steps provided in Section~\ref{simu_study}. The regularization parameter $\lambda$ is chosen such that only one predictor is left in the active set in the first column; is selected by 10-fold cross-validation in the second column; is set to 0 in the last column. In each histogram, a probability density curve of $\chi^2_1$ is overlaid. The distributions of Cook's distances observed in these examples closely resemble the $\chi^2_1$-distribution.

This assumption leads to a natural threshold of $T(\lambda) = \chi^2_{0.95,1}\times \sqrt{\var(D_k(\lambda,0))/2}$ given $\lambda$. In practice, the sample variance of the $n$ observed values of $D_k(\lambda,0)$ can be used as an estimate of $\var(D_k(\lambda,0))$. 
Alternatively, we can use an externally normalized variance by excluding observation $k$ for robustness.
Using the threshold, we present Algorithm~\ref{alg_detect} for identifying influential points under Lasso. To run this algorithm, one only needs to fit a full data Lasso model once, and the solution path algorithm $n$ times for each of $n$ observations. 
Note that the cross-validation step for choosing an optimal $\lambda$ can be replaced with any other model selection method.

\begin{algorithm}[htb]
\caption{Detection of influential observations}\label{alg_detect}
\begin{algorithmic}
\State Input: $X \in \mathbb{R}^{n\times p}$, $\bm{y}\in \mathbb{R}^n$
\State Center $X$ column-wise to have mean 0
\State Choose the optimal $\hat{\lambda}$ for Lasso based on cross-validation
\State Obtain the sign vector $\hat{\bm{s}}$ and the active set $\hat{\mathcal{A}}$ from the  Lasso model with $\hat{\lambda}$ 
\For{$k$ from 1 to $n$}
\State Apply Algorithm~\ref{alg} with input $X, \bm{y}, \hat{\lambda}, k$
\State Calculate $D_k(\hat{\lambda}, 0)$ based on \eqref{Dklo}
\EndFor
\State Calculate the threshold $T$ based on \eqref{threshold}
\State Output: $\{k\in[n]~\vert~ D_k(\hat{\lambda}, 0)>T \}$
\end{algorithmic}
\end{algorithm}

\subsection{Local influence}\label{other_measure}
 Local influence \citep{cook1986assessment} is  
 an alternative way to measure case influence.
 This alternative scheme considers the sensitivity of a model to an infinitesimal change in each case locally rather than its deletion.
For the Lasso, we can define local influence as the curvature of the case influence function in \eqref{Dklo}, $D_k(\lambda, \omega)$ at $\omega = 1$, taken as a function of the case weight $\omega$:
 \[\frac{1}{2} \frac{\partial^2 D_k(\lambda, \omega)}{\partial \omega^2}\Big|_{\omega=1} = \frac{1}{2(p+1)s^2}  \frac{\partial^2}{\partial \omega^2}\sum_{i=1}^n (\hat{y}_i(\lambda) - \hat{y}_{i}^\omega)^2 \Big|_{\omega=1}.\] 
 Since the fitted value from the case-weight adjusted model, $\hat{y}_{i}^\omega=\bm{x}_i^\top  \hat{\bm{\beta}}^\omega$, is the only term that depends on $\omega$, to evaluate the local influence, we need to find its derivative when the weight of case $k$ changes infinitesimally at $\omega = 1$.  
 Hence, we define the sensitivity of the fitted value to case $k$ as $\frac{\partial \hat{y}^{\omega}_i}{\partial \omega}\Big|_{\omega=1}$. 
Using the case-weight adjusted Lasso model and the fact that no active set change has yet happened at $\omega=1$, we can show that the sensitivity of each fitted value is readily available from the full data model. This allows us to evaluate the local influence without any additional computation.   
The following proposition gives explicit expressions for the sensitivity of fitted values and the local influence. The proof is in Appendix~\ref{app_prop3}.

\begin{proposition}\label{prop3}
For the Lasso model $\hat{y}=\hat{\beta}_0+\bm{x}^\top \hat{\bm{\beta}}$ with a penalty parameter $\lambda$, 
\begin{itemize}
    \item[i.] The sensitivity of the fitted value $\hat{y}_i$ to case $k$ is 
       \[ \frac{\partial \hat{y}^{\omega}_i}{\partial \omega}\Big|_{\omega=1}= h_{ik}(y_k - \hat{y}_k).\]
    \item[ii.] The local influence of case $k$ is 
    \begin{align}
        \frac{1}{2} \frac{\partial^2 D_k(\lambda, \omega)}{\partial \omega^2}\Big|_{\omega=1} = \frac{1}{(p+1)s^2} h_{kk} (y_k - \hat{y}_k)^2.\label{local_inf}
    \end{align}
\end{itemize}
Here $h_{ij}$ is the $(i,j)$th entry of the hat matrix defined with the active variables at $\lambda$, and $\hat{y}_k$ is the Lasso fitted value for case $k$.
\end{proposition}
Note that the local influence of Lasso has the same form as that of OLS. The local influence for OLS can be considered as a special case when $\lambda=0$.
It also equals $(1-h_{kk})^2 \times$ Cook's distance in \eqref{ocookdis} with no active set change.

\section{Case Influence Graph}\label{cig}

In this section, we introduce a case influence graph as a way to visualize Cook's distance for the Lasso. In the OLS setting, the case influence graph usually refers to a visualization of case influence as a function of the case-weight parameter $\omega$ \citep{cook1986assessment}. However, in this paper, we are interested in case influence in the Lasso and thus focus on its dynamics with respect to varying penalty levels while keeping $\omega$ at 0. In other words, the case influence graph we adopt is a visualization of Cook's distance for the Lasso, $D_k(\lambda,0)$, against $\lambda$ for each observation. 
To provide 
a more interpretable scale of penalty levels, we reparameterize $\lambda$ using the fraction of the $\ell_1$-norm of the full data Lasso estimate given $\lambda$ to that of the OLS estimate,
{$\|\hat{\bm{\beta}}_\lambda\|_1/\|\hat{\bm{\beta}}_{OLS}\|_1$. This fraction will be referred to as $|coef|/\max|coef|$ in short in subsequent figures.}
This reparameterization is commonly used in displaying a Lasso solution path. 
Further, for ease of reference in subsequent discussions, we term this fraction $\rho$.

In the next two subsections, we use an example to illustrate different patterns in the case influence graph and then demonstrate how it can be used for model selection. More examples of case influence graphs can be found in Section~\ref{simu}. {In the third subsection, we 
provide an analysis of the computational complexity of the proposed solution path algorithm for assessing case influence.}

\subsection{The mechanism of the case influence graph}\label{ex_mechan}

To study different modes of influence in the case influence graph, we consider a toy example with two features ($p=2$) and $n=10$. Assume that the first $(n-1)$ responses $y_i\in \mathbb{R}$ are independently generated from the following model: 
 \begin{align*}
     y = \beta_1 x_1 + \beta_2x_2 + \epsilon, 
 \end{align*}
 where $\bm{x}=(x_1,x_2)^\top \sim \mathcal{N}(\bm{0},I_2)$, $\epsilon\sim N(0,1)$, and $\beta_1 = 4\beta_2 >0$. 
To see the influence of one case, we keep the first $(n-1)$ cases fixed, assign the $n$th case, $(\bm{x}_{n},y_{n})$, four different pairs of values, and examine the resulting case influence graph. The four scenarios are described below. Different choices of $(\bm{x}_{n},y_{n})$ for the four scenarios are to demonstrate how the leverage and residual for the case play important roles in its influence and how dynamically these would change with $\lambda$.

\begin{figure}
    \centering
    \includegraphics[width = \textwidth]{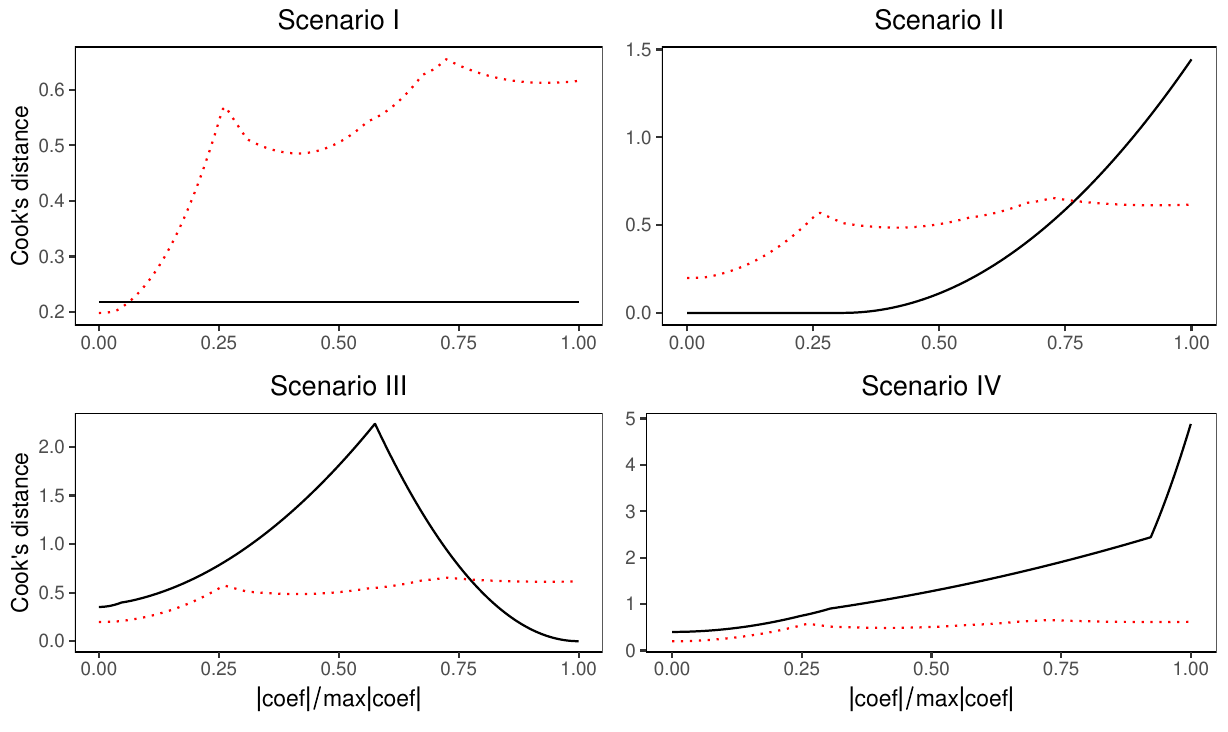}
    \caption{Case influence graphs of $(\bm{x}_{n}, y_{n})$ under four scenarios in Section~\ref{ex_mechan} as a function of $\|\hat{\bm{\beta}}_\lambda\|_1/\|\hat{\bm{\beta}}_{OLS}\|_1$. The red dotted line is the threshold proposed in Section~\ref{thresholding}.
    }
    \label{fig:mechanism}
\end{figure}

Figure~\ref{fig:mechanism} displays the case influence graph under each of the four scenarios.
In each case influence graph, the fraction $\rho = 1$ indicates the OLS coefficients, and $\rho = 0$ indicates the null coefficients that are all penalized to 0. Whenever we describe a case influence graph in this paper, we will describe it as a trajectory from the right ($\rho=1$), starting from the OLS solution, to the left ($\rho=0$). 
The threshold introduced in Section~\ref{thresholding} is also included in each graph to assist our assessment. 

\begin{itemize}
    \item Scenario I: $\bm{x}_{n} = (0,0)^\top$ and $y_{n} = c$. This scenario concerns the situation when the leverage takes its minimum value. As shown in Figure~\ref{fig:mechanism}, the case influence curve is a flat line.  This means that the penalty levels do not change Cook's distance. 
    By \eqref{re_beta_0}, when $\bm{x}_{n} = (0,0)^\top$, $\hat{\beta}_0=\bar{y}$ and $h_{nn}=1/n$, which implies $\hat{\beta}_0^{\omega=0} = \Bar{y} + \frac{1}{n-1}(\Bar{y}-y_n) = \frac{1}{n-1}\sum_{i=1}^{n-1} y_i := \Bar{y}_{-n}$. 
    {By Algorithm~\ref{alg}, no active set change happens when $\bm{x}_{n} = (0,0)^\top$, which makes \eqref{ocookdis} exact.}
    By \eqref{ocookdis},
    $D_{n}(\lambda, 0) = \frac{n}{ps^2} (\Bar{y} - \Bar{y}_{-n})^2$, which does not depend on $\lambda$. 
    
    \item Scenario II: $\bm{x}_{n} = (0,4)^\top$ and $y_{n} = \Bar{y}_{-n}$. Here $y_n$ is chosen such that the residual would become 0 when all features are dropped, and $\bm{x}_{n}$ has an extreme value in the second feature. So, its case influence is large at the start due to its large leverage but gradually becomes trivial as the second feature is dropped from the model. 
    
    \item Scenario III: $\bm{x}_{n} = (0,4)^\top$, and $y_{n}$ is chosen such that its OLS residual is 0. The influence of this case at the start, i.e., the OLS Cook's distance is 0 due to the zero residual. However, as more penalty is applied, its residual moves away from 0 while  $\hat{\beta}_2$ shrinks towards 0. 
    A larger residual means that the case is not consistent with the current parameter estimates and potentially a larger difference in the estimates if the case is deleted. On the other hand, the shrinkage of $\hat{\beta}_2$ implies less influence the $n$th case can have. These two opposite effects result in a wave-like curve as shown in the figure. The curve reaches its maximum when $\hat{\beta}_2$ is still not too close to 0 while the residual is large enough. 

    \item Scenario IV: $\bm{x}_{n} = (4,0)^\top$ and $y_{n} = -4$. Its large case influence is due to a high leverage from the first feature. Letting $y_{n} = -4$ enforces a large residual as well. 
    Both the leverage and residual remain large, making this case an influential point with its Cook's distance over the threshold indicated by the red dotted line as $\rho$ reduces from 1 to 0.
\end{itemize}
{The results from the four scenarios above show that depending on the leverage and residual of a case as a function of the penalty, the relative size of its influence can change notably, transitioning from influential to uninfluential or vice versa.  
Generally, a higher penalty results in fewer active variables and a larger residual sum of squares, which indicate a smaller leverage and larger residual, respectively. 
Understanding these dynamics is crucial for diagnosing model sensitivity to influential points in the Lasso and useful for model selection as we will see in the next section.
}

\subsection{Model selection}\label{model selection}
Cross-validation (CV) has been frequently used in model selection for the Lasso. CV selects a model based on its prediction accuracy. However, prediction accuracy may not be the sole criterion for assessing the goodness of a model. Its robustness is equally important, and CV doesn't take into account model robustness.  

Robust models should change little under small perturbations of the data. Thus, we consider a new criterion based on case influence, which can be used as an additional tool for model selection in conjunction with CV.
For measuring the robustness of a Lasso model, we
take the average Cook's distance as a criterion given $\lambda$: 
\begin{align*}
    \Bar{D}(\lambda) = \frac{1}{n}\sum_{k=1}^n D_k(\lambda, 0).
\end{align*}

We first introduce a proposition that explains how leverages behave after an active set update, which can help us explain the behavior of $\Bar{D}(\lambda)$. The proof of the proposition is in Appendix~\ref{app_prop1}.

\begin{proposition} \label{prop1}
Let $X\in\mathbb{R}^{n\times p}$ be a design matrix with $n>p$. If a feature $\bm{z}\in\mathbb{R}^{n} $ is added to the design matrix, the $k$th leverage $h_{kk}$, $k \in [n]$, will increase by $\frac{[(I-P_X)\bm{z}]_k^2}{\|(I-P_X)\bm{z}\|^2}$, where $P_X$ is the projection matrix onto the column space of $X$. 
\end{proposition}
Here $(I-P_X)\bm{z}$ is the residual vector of the new feature $\bm{z}$ after regressed on the current features ($X$).
The proposition says that when $\bm{z}$ is added, the $k$th leverage will increase by the square of the $k$th coordinate of the normalized residual vector $(I-P_X)\bm{z}$.

Next, we use an example to demonstrate 
the utility of $\Bar{D}(\lambda)$ in practice by examining a typical pattern of $\Bar{D}(\lambda)$. Consider data with three pairs of features:

\begin{align*}
    y = \bm{x}^\top\bm{\beta} +\epsilon,\quad &\bm{x}\sim N{\left(\bm{0}, \begin{pmatrix}
V & 0 & 0\\
0 & V & 0 \\
0 & 0 & V 
\end{pmatrix}\right),\quad V = \begin{pmatrix}
    1&r\\
    r&1
\end{pmatrix}},\\
&\bm{\beta} = (\beta, 0, \beta, 0, \beta, 0)^\top, \quad \epsilon \sim N(0,1),
\end{align*}
where $r=0.5$,
and features are correlated within each pair and independent between pairs. 

\begin{figure}
    \centering
    \includegraphics[width = \textwidth]{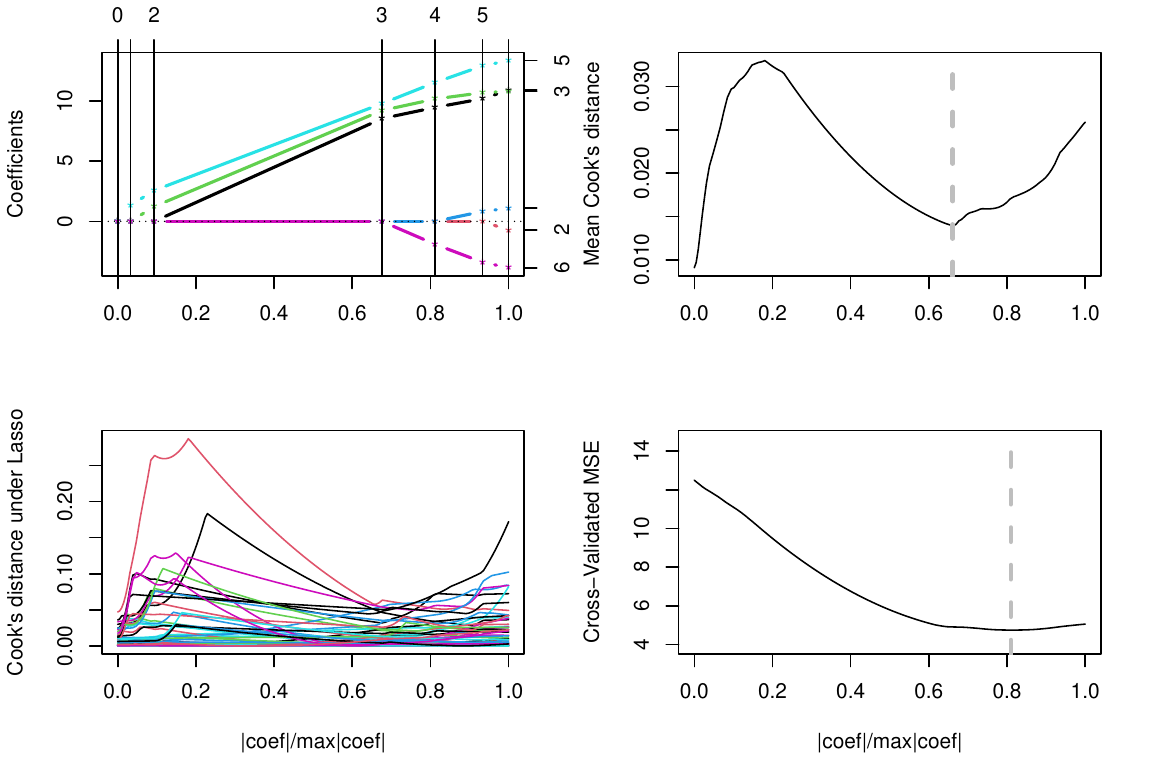}
    \caption{The top left panel is the solution path produced by the regular Lasso. 
    The bottom left panel is the corresponding case influence graph. The top right panel is the average Cook's distance, $\Bar{D}(\lambda)$, obtained by taking the average of the case influence values at $\lambda$. The dashed gray vertical line indicates its local minimum. The bottom right panel is the MSE curve of leave-one-out cross-validation. The dashed gray vertical line is located at its global minimum.}
    \label{fig:ex3}
\end{figure}

Figure~\ref{fig:ex3} shows the trajectories of estimated coefficients, case influence, average case influence, and leave-one-out cross-validated MSE, respectively, as $\rho$ changes from 1 to 0.
On the top left panel, each colored line represents a path of an estimated coefficient. The vertical lines indicate the penalty levels where features are dropped from the active set. The number on the top shows the number of features left in the active set, and the number on the right marks the index of each feature. When $\rho = 1$, all 6 features are included. The redundant features 2, 4, and 6 are dropped subsequently as $\rho$ decreases to around 0.65. An ideal model selection criterion would select the model at this stage.

The case influence graph (bottom left) and the average case influence (top right) can be examined together. We can divide the case influence graph into two stages. The first monotone decreasing stage (to the right of the vertical line) indicates overfitting, and the second unimodal stage (to the left of the line) indicates underfitting. The average Cook's distance reaches a local minimum between two stages corresponding to the optimal model fit. The reasons for this interpretation are given as follows.
 
At the beginning of the first stage, with no penalty ($\rho=1$), Cook's distances tend to be large, indicating a non-robust model. As the penalty increases, one of the coefficients in each pair shrinks to 0. Based on the true model, one non-zero coefficient in each pair can still explain $y$ well. Hence, the residual won't get too large. The leverages of all observations are guaranteed to drop by Proposition~\ref{prop1}. As a result, $\Bar{D}(\lambda)$ reaches a local minimum, which corresponds to an optimal fit.  
As the penalty increases, the remaining three features are dropped one after another, and the model can no longer explain $y$ well, causing the residuals to rise. 
Note that the leverages of some of the observations remain large.
The combination of a high leverage and a large residual of these observations induce another stage of large Cook's distances. Finally, as all features are dropped, all leverages become $\frac{1}{n}$, and $\Bar{D}(\lambda)$ reaches its minimum.

According to the leave-one-out cross-validated MSE curve in the bottom right, CV selects the model at a fraction of 0.8. As \cite{leng2006note} point out, CV is not consistent in selecting the true model in Lasso when there are redundant variables. CV lacks the capability to differentiate between overfitting models and the true model. On the other hand, the optimal fit suggested by $\Bar{D}(\lambda)$ in this example is exactly the place where the last redundant variable is dropped. 

From this example, we see that $\Bar{D}(\lambda)$ can characterize the robustness of a Lasso model as a function of $\lambda$. It offers a different perspective on model selection. However, it certainly cannot be used alone as a model selection criterion in that a model with no feature selected ($\rho=0$) would be considered robust by the criterion. 
In our ongoing work \citep{cv_cd}, we examine this issue further and consider a formal criterion that takes account of the model robustness. We show that the new criterion that combines CV with Cook's distance can outperform CV under linear models and penalized linear models while inheriting good properties of CV.

\subsection{Computational complexity analysis}
{For a fixed $\lambda$, obtaining the case-deleted lasso solution for a single case without the case-weight adjusted solution path algorithm requires refitting the Lasso model. The computational complexity of this approach is $O(np\min(n,p))$ using \texttt{lars}. 
On the contrary, obtaining the leave-one-out solution with the case-weight adjusted solution path algorithm and the full data solution takes $O(Unp) = O(np\min(p/n, n))$ where $U$ represents the number of active set updates as $\omega$ decreases from 1 to 0 and $U = O(\min(p/n, n))$ based on our simulation analysis. 
We call the former the refitting LARS approach and the latter the case-weight solution path approach.

Next, we consider a scenario with multiple values of $\lambda$. Suppose we are interested in Cook's distance for the Lasso at $L$ number of different fractions. Note that $\hat{\bm{\beta}}$ is piecewise linear with respect to $\lambda$. Calculating Cook's distance for each fraction only involves a linear interpolation of $\hat{\bm{\beta}}$ and calculating the fitted values, which takes $O(n\min(n,p))$. We need to fit Lasso models $(n+1)$ times for the full data and each case deleted. Thus, the computational complexity of the refitting LARS approach
is $O(n^2p\min(n,p) + Ln\min(n,p))$. For the case-weight solution path approach, we need to fit the full data model once, which costs $O(np\min(n,p))$. For each fraction of interest, we perform the case-weight adjusted solution path algorithm for each case starting from the full data solution. For each active set update, the dominant step is
locating $\xi$ candidates, which takes $O(np)$. 
Therefore, the computational complexity of this approach is $O(np\min(n,p) + LUn^2p) = O(np\min(n,p) + Ln^2p\min(p/n, n))$. 

For more explicit comparison, suppose we wish to assess the Lasso model with $\lambda$ cross-validated, so $L=1$. Then, the computational complexities of these two approaches are $O(n^2p\min(n,p))$ and $O(n^2p\min(p/n,n))$, respectively. Our approach only takes $O(1/n)$ of the time needed in the refitting LARS approach if $n\ge p$. This comparison remains valid whenever $L$ is a fixed number not increasing with $n$ or $p$. Table~\ref{efficiency comparison} gives a comprehensive comparison of these two approaches under different $(n,p)$ combinations. 
    
\begin{table}[htb]
    \centering
    \begin{tabular}{l|ccc}
    \hline
        & \multicolumn{3}{c}{$c$} \\ 
    $L$ & $\le 1$ & $(1,2)$ & $ \ge 2$ \\ \hline
    $O(1)$        & $O(n^{-1})$   & $O(n^{c-2})$   & $O(1)$               \\ 
    $O(\min(n,p))$ & $O(n^{c-1})$     & $O(n^{c-1})$    & $O(n)$               \\ \hline
    \end{tabular}
    \caption{
    The ratio of the computational complexity of the case-weight solution path to the refitting LARS approach for calculating Cook's distance or leave-one-out cross-validation in the Lasso for $L$ values of $\lambda$ when $p$ is allowed to change with $n$ at the rate of $p(n) = O(n^c)$. The computational complexity of the former is $O(np\min(n,p) + Ln^2p\min(p/n,n))$ while that of the latter is $O(n^2p\min(n,p) + Ln\min(n,p))$.}

    \label{efficiency comparison}
\end{table}
}

\section{Numerical Studies} \label{simu}
This section presents numerical studies with the proposed measure of case influence.
 In Section~\ref{measure_compare}, we compare Cook's distance in Lasso with other aforementioned case influence measures numerically. In Section~\ref{simu_study}, we conduct simulation studies in different settings to demonstrate the effectiveness of Cook's distance in identifying influential points. In Section~\ref{outlier_dectect}, we apply the proposed influence measure to a real-life dataset to test its usage in practice.

\begin{figure}[htb]
    \centering
    \includegraphics[width = \textwidth]{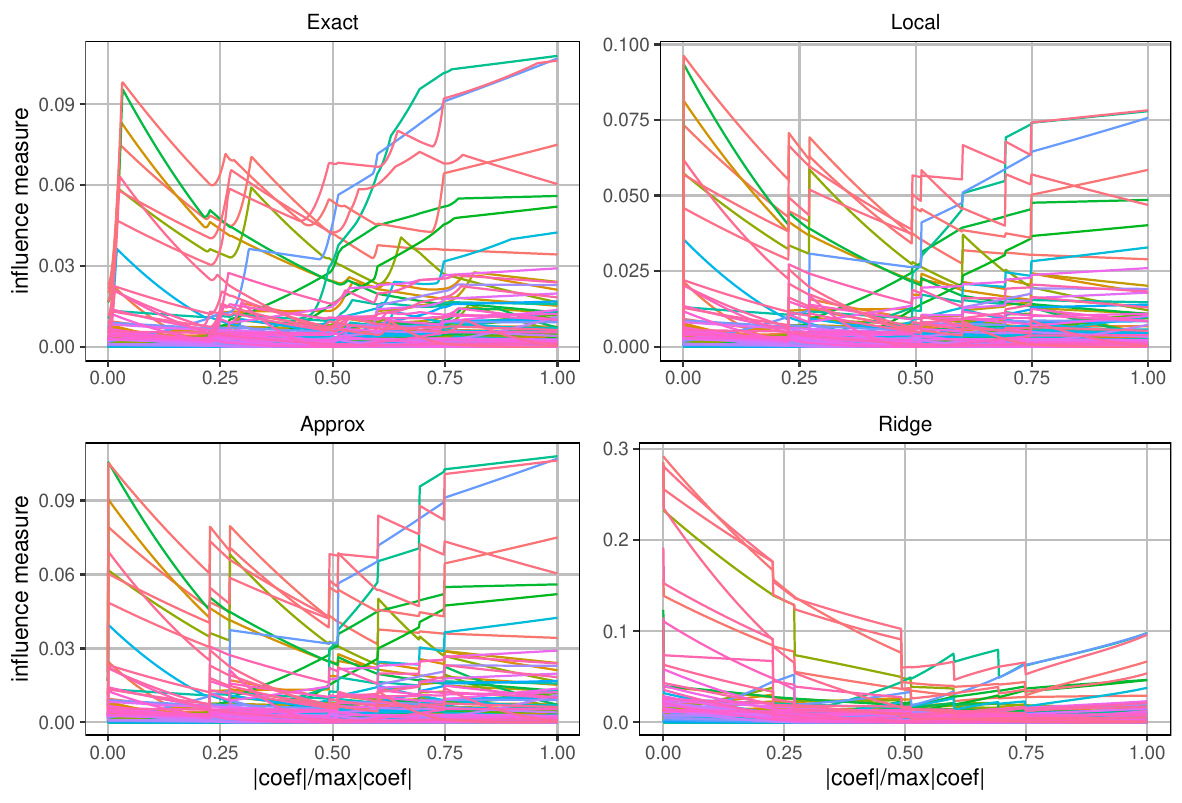}
    \caption{Case influence graphs using four proposed case influence measures on the prostate dataset. Each curve records the case influence of one observation as the fraction decreases from 1 to 0. }
    \label{fig:4compare}
\end{figure}

\subsection{Comparison of case influence measures} \label{measure_compare}
So far, we have discussed four measures that can be used for evaluating case influence in the Lasso. They are Cook's distance for the Lasso in \eqref{Dklo}, the approximate Cook's distance in \eqref{ocookdis}, the local influence in \eqref{local_inf}, and the approximate Cook's distance proposed by \cite{Kim2015}, which relies on the ridge-like estimate of $\hat{\bm{\beta}}$ suggested in \cite{tibshirani1996regression}. 
These four measures are referred to as `Exact', `Approx', `Local', and `Ridge', respectively. 
In this subsection, we compare these measures using the prostate cancer dataset \citep{stamey1989prostate}.  
The dataset consists of 97 observations with the response variable of prostate specific antigen level on the log scale and eight clinical predictors.

Figure~\ref{fig:4compare} shows the case influence graphs obtained using these four measures. Overall, `Approx' is a fairly accurate approximation of the exact measure. By definition, `Approx' and `Local' differ only by a factor of $(1-h_{kk})^2$. Proposition~\ref{prop1} guarantees that $h_{kk}$ gets smaller as more penalty is applied, and thus their difference gets smaller as well. On the other hand, `Ridge' is not an accurate approximation of `Exact'. Its values tend to inflate more as a larger penalty is applied. 
As described in \cite{Kim2015}, `Ridge' can rank observations similarly to the exact measure at a fixed fraction, but it fails to rank the magnitude of influence for the same observation across different fractions consistently with the exact measure. Moreover, `Ridge' doesn't require less amount of computation than `Approx'.

While the exact influence graph is continuous, all three approximations exhibit discontinuity in Figure~\ref{fig:4compare}. 
This discontinuity occurs at the location where there is an active set update in the regular Lasso. At a given fraction (or $\lambda$), the active set can change due to case deletion, but these approximations do not account for such active set updates. This leads to inaccurate approximations of Cook's distances around those locations. 

\begin{figure}[htb]
    \centering
    \includegraphics[width = \textwidth]{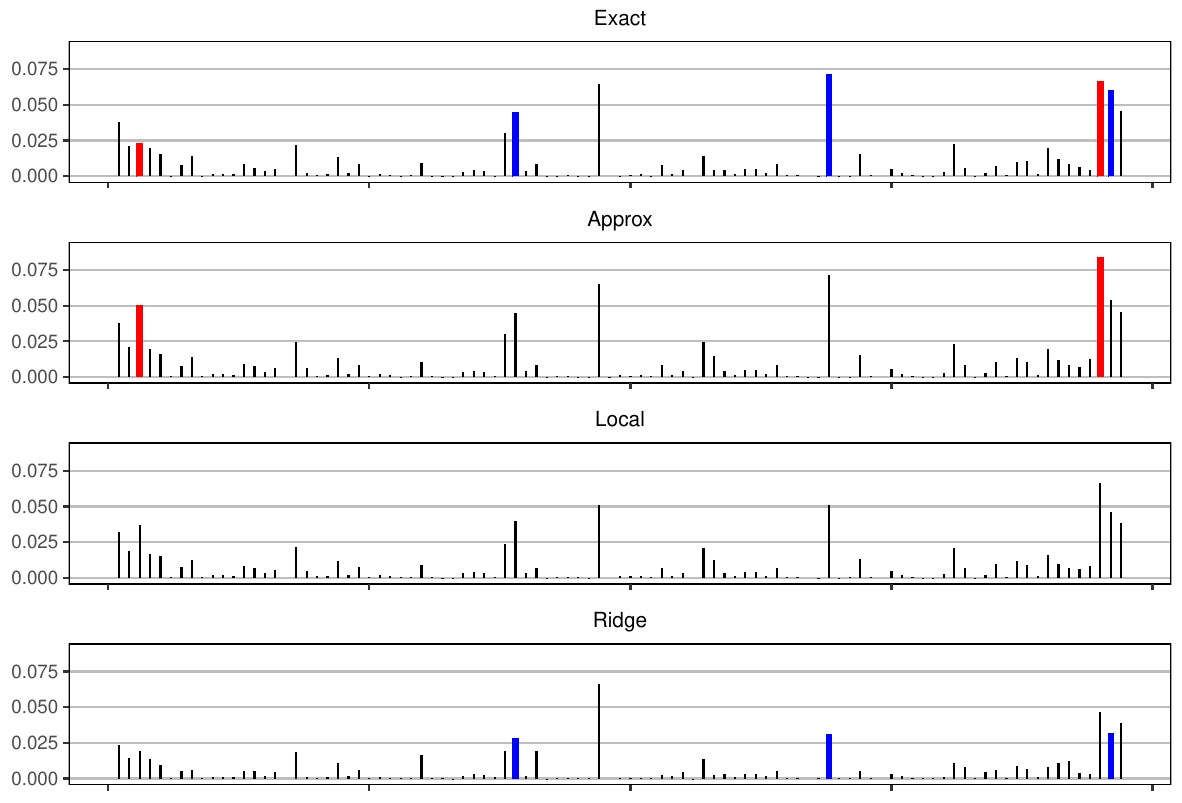}
    \caption{Influence index plots of the prostate cancer data at fraction $= 0.6$. Observations 3 and 95 are highlighted in red, and 39, 69, and 96 are in blue. }
    \label{fig:sliceplot}
\end{figure}

We take a slice (vertical cut) from Figure~\ref{fig:4compare} at the fraction ($\rho=0.6$) selected by 10-fold cross-validation and examine the four influence measures on all observations.
Figure~\ref{fig:sliceplot} displays the resulting influence index plots. Each vertical bar represents the influence of one observation.  
As observed in Figure~\ref{fig:4compare}, the `Approx' measure aligns well with the exact measure mostly. However, some discrepancies exist when the deletion of an observation causes active set updates at the given fraction. Notable examples include observations 3 and 95, highlighted in red in Figure~\ref{fig:sliceplot}.
Observations 39, 69, and 96 marked in blue are examples for which the `Ridge' measure is notably different from the exact measure. The average discrepancy of the `Ridge' measure from `Exact' is over three times larger than that of `Approx'.

\subsection{Simulation study}\label{simu_study}
We now use simulations to check how effective Algorithm~\ref{alg_detect} is in detecting influential observations. 
We consider both $n>p$ and $n<p$ cases. 
We generate data with mutually correlated features and manually assign values to the response and predictors of one observation such that it becomes influential. Ideally, the case influence measure is expected to identify the influential observation while keeping the false detection rate low.

Here is the procedure for generating data:
\begin{enumerate}
    \item Generate an $n\times p$ design matrix $X$ from a multivariate normal distribution with mean 0, variance 1, and a pairwise correlation between covariates $i$ and $j$ equal to $0.2^{|i-j|}$. 
    \item Replace the $q$th element of observation 1, $x_{1q}$, with value $a$.
    {
    \item Generate $\bm{y}\in \mathbb{R}^n$ from the following model:
     \begin{align*}
     y = \bm{x}^\top \bm{\beta} + \epsilon,
    \end{align*}
    where $\bm{\beta} = (\bm{v}^\top, 0, \dots, 0)^\top$ with $\bm{v} = (1,-1,1/2,-1/2)^\top$ and $\epsilon$ are iid with $N(0,1)$, leading to the signal-to-noise ratio of 1.872.}
    \item Replace the simulated response value, $y_1$, with $\mu(y_1) + b$, where $\mu(y_1)$ is the mean value of $y_1$ from the model in the previous step.
\end{enumerate}

We let $a$ and $b$ take values from $\{0,2,3,5\}$ to simulate varying levels of leverage and residual for observation 1 and consider four different combinations of $(n,p)$: $(50,10)$, $(500,10)$, $(50,500)$, and $(200,200)$. 
We also vary the parameter $q$ from $\{1, 10\}$ to control the leverage of the observation due to a valid feature ($q=1$) or a redundant feature ($q=10$).
For each combination of $(a,b,n,p,q)$, we generate $1000$ replicates of data and record the proportion of times each observation is flagged as influential by Algorithm~\ref{alg_detect}. {To account for different signs of the coefficients for the four features and different signal-to-noise ratios, we also consider $\bm{v}$ taking value from $\{(1,1,1/2,1/2)^\top, (1,1,1,0)^\top, (4,3,2,1)^\top\}$. The full simulation results are in Appendix~\ref{simu_sum}.}

\begin{table}
\centering
\begin{tabular}{cccccc|cccccc}
\hline
$a$ & $b$ & $n$  & $p$ &  1 & 2--$n$ & $a$ & $b$ & $n$ & $p$ & 1 & 2--$n$ \\\hline
0 & 2 & 50 & 10  & 0.26 & 0.05 & 0 & 2 & 200 & 200 & 0.32 & 0.05 \\
0 & 2 & 50 & 500 & 0.11 & 0.05 & 0 & 2 & 500 & 10  & 0.31 & 0.04 \\\hline
0 & 3 & 50 & 10  & 0.67 & 0.04 & 0 & 3 & 200 & 200 & 0.91 & 0.04 \\
0 & 3 & 50 & 500 & 0.31 & 0.04 & 0 & 3 & 500 & 10  & 0.86 & 0.04 \\\hline
0 & 5 & 50 & 10  & 0.97 & 0.02 & 0 & 5 & 200 & 200 & 1.00 & 0.02 \\
0 & 5 & 50 & 500 & 0.65 & 0.03 & 0 & 5 & 500 & 10  & 1.00 & 0.03 \\\hline
2 & 0 & 50 & 10  & 0.01 & 0.06 & 2 & 0 & 200 & 200 & 0.00 & 0.05 \\
2 & 0 & 50 & 500 & 0.02 & 0.05 & 2 & 0 & 500 & 10  & 0.00 & 0.04 \\\hline
2 & 2 & 50 & 10  & 0.38 & 0.05 & 2 & 2 & 200 & 200 & 0.33 & 0.05 \\
2 & 2 & 50 & 500 & 0.16 & 0.04 & 2 & 2 & 500 & 10  & 0.52 & 0.04 \\\hline
3 & 0 & 50 & 10  & 0.02 & 0.06 & 3 & 0 & 200 & 200 & 0.00 & 0.05 \\
3 & 0 & 50 & 500 & 0.02 & 0.05 & 3 & 0 & 500 & 10  & 0.00 & 0.04 \\\hline
3 & 3 & 50 & 10  & 0.85 & 0.04 & 3 & 3 & 200 & 200 & 0.91 & 0.04 \\
3 & 3 & 50 & 500 & 0.27 & 0.04 & 3 & 3 & 500 & 10  & 0.94 & 0.04 \\\hline
5 & 0 & 50 & 10  & 0.07 & 0.06 & 5 & 0 & 200 & 200 & 0.00 & 0.05 \\
5 & 0 & 50 & 500 & 0.03 & 0.05 & 5 & 0 & 500 & 10  & 0.00 & 0.04 \\\hline
5 & 5 & 50 & 10  & 0.99 & 0.01 & 5 & 5 & 200 & 200 & 1.00 & 0.02 \\
5 & 5 & 50 & 500 & 0.63 & 0.03 & 5 & 5 & 500 & 10  & 1.00 & 0.01 \\\hline
\end{tabular}
\caption{Proportion of the times that Algorithm~\ref{alg_detect} detects observation 1 or observations 2--$n$ as an influential point from simulated data of size $n$ and dimension $p$ with different parameters controlling the leverage ($a$) and residual ($b$) of observation 1 due to a redundant predictor ($q=10$). }
\label{simu_table_partial}
\end{table}

Table~\ref{simu_table_partial} summarizes the results when $q=10$. 
When $b=5$, observation 1 is identified as influential over $97\%$ of the time except when $p\gg n$. When $b=3$, the percentage can reach as high as $94\%$.
The percentage of observation 1 being identified as influential decreases as $p/n$ increases. This is because an increase in the average leverage ($(p+1)/n$) with $p$ largely dilutes
the overall influence observation 1 can have on the model. When $b=0$, no matter how extreme $a$ is, the algorithm seldom considers observation 1 as influential. 
This is reasonable because $X_{10}$ is a redundant variable and unlikely to be included in the active set at $\hat{\lambda}$ chosen by CV. Thus, the leverage of observation 1 is never large, and with $b=0$, the residual would be small as well. Hence the observation would not be influential.

We turn our attention to the scenarios with $q=1$ where the value of a relevant predictor is modified. 
Table~\ref{simu_table_partial2} includes the results that can show the difference from $q=10$. We observe a much higher detection rate simply 
because covariate $1$ as a relevant feature is very likely to be included in the active set at $\hat{\lambda}$.
For example, the detection rate increases from $27\%$ to $71\%$ when $a=3,~b=3,~n=50$, and $p=500$ and from $33\%$ to $69\%$ when
$a=2,~b=2,~n=200$, and $p=200$.
The effect of parameter $a$ on the detection rate depends heavily on which covariate is modified.
When the covariate is important ($q=1$), 
$a$ is more likely to affect the leverage of observation 1 in the fitted model at $\hat{\lambda}$ and thus its detection rate. Conversely, if $q=10$, the detection rate is less influenced by $a$ and primarily determined by $b$. 

\begin{table}[htb]
    \centering
    \begin{tabular}{cccccc|cccccc}
    \hline
$a$ & $b$ & $n$  & $p$ & 1 & 2--$n$ & $a$ & $b$ & $n$ & $p$ & 1 & 2--$n$ \\\hline
2 & 2 & 50 & 10  & 0.58 & 0.05 & 2 & 2 & 200 & 200 & 0.69 & 0.05 \\
2 & 2 & 50 & 500 & 0.44 & 0.04 & 2 & 2 & 500 & 10  & 0.73 & 0.04 \\\hline
3 & 3 & 50 & 10  & 0.97 & 0.02 & 3 & 3 & 200 & 200 & 1.00 & 0.03 \\
3 & 3 & 50 & 500 & 0.71 & 0.02 & 3 & 3 & 500 & 10  & 1.00 & 0.03 \\\hline
5 & 5 & 50 & 10  & 1.00 & 0.00 & 5 & 5 & 200 & 200 & 1.00 & 0.00 \\
5 & 5 & 50 & 500 & 0.73 & 0.01 & 5 & 5 & 500 & 10  & 1.00 & 0.00 \\\hline
    \end{tabular}
\caption{Proportion of the times that Algorithm~\ref{alg_detect} detects observation 1 or observations 2--$n$ as an influential point. The results are obtained in the same way as in Table~\ref{simu_table_partial} but for settings with a crucial predictor ($q=1$). }
    \label{simu_table_partial2}
\end{table}
\begin{table}[htb]
    \centering
    \begin{tabular}{cccc|cccc|cccc}
    \hline
$a$ & $b$ & 1 & $1_{df}$ & $a$ & $b$ & 1 & $1_{df}$ & $a$ & $b$ & 1 & $1_{df}$\\\hline
10 & 10 & 1.00 & 1.00 & 5 & 5 & 0.92 & 0.96 & 2 & 2 & 0.23 & 0.11 \\\hline
10 & 0 & 0.02 & 0.00 & 5 & 0 & 0.02 & 0.00 &2 & 0 & 0.01 & 0.00 \\\hline
0 & 10 & 1.00 & 1.00 & 0 & 5 & 0.92 & 0.96 &0 & 2 & 0.19 & 0.09 \\\hline
    \end{tabular}
\caption{A comparison of the detection rate of an influential point between Cook's distance for the Lasso (Column `1') and df-cvpath in \cite{influencediag} (Column `$1_{df}$') when $n=50$, $p=1000$, and $q=100$. }
    \label{simu_table_partial3}
\end{table}

Additionally, for observations 2--$n$, when $b$ does not take extreme values, the false detection rate is at around $5\%$, which 
corresponds to the choice of $95\%$ quantile in \eqref{threshold} as a threshold. When $b\geq3$, the false detection rate gets smaller due to 
the extreme influence of observation 1, which raises the threshold.
Thus, replacing the sample variance with an externally normalized variance--where the variance is computed without the observation of interest--can effectively prevent an extremely influential point from masking itself and other points.

\cite{influencediag} introduced four specific measures for the detection of influential observations when $p\gg n$. We replicate the same simulation setting in their paper ($\bm{v}=(1,2,3,4,5)^\top,~n=50,~p=1000,~q=100$) and compare our measure to the best-performing one (df-cvpath) of theirs. The results of this comparison are presented in Table~\ref{simu_table_partial3}.  Our measure is more sensitive to the outlyingness of observation 1 in the response than df-cvpath given a moderately extreme value $b=2$, while its detection rate is slightly less than that of df-cvpath given a more extreme value $b=5$. 
However, note that the implementation of df-cvpath requires fitting a Lasso model $(n+1)$ times for each $\lambda$ value, while our measure only requires model fitting just once and applying the solution path algorithm in Algorithm~\ref{alg} $n$ times. In this example, using an Apple M2 pro chip, the average run times required to compute our measure and df-cvpath are 0.38s and 2.26s, respectively. 
Note that the speed of our algorithm could be significantly improved if it were implemented in C++, similar to $glmnet$ \citep{friedman2010regularization}, which is the primary tool used in df-cvpth.
Overall, we conclude that Algorithm~\ref{alg_detect} can effectively and efficiently identify influential points in both $p<n$ and $p>n$ cases. 

\subsection{Real data analysis} \label{outlier_dectect}
{In this section, we demonstrate the applications of our procedures on three real-world datasets. We evaluate the effectiveness of our procedures by comparing the Lasso models before and after the removal of influential observations.}

\subsubsection{Diabetes data}\label{diabetes_example}
The diabetes data first analyzed in \cite{lars} comprises 442 patients with diabetes ($n=442$). For each patient, ten variables ($p=10$) are measured: age, sex, body mass index ($bmi$), average blood pressure ($bp$), and six blood serum measurements: total cholesterol ($tc$), low-density lipoprotein ($ldl$), high-density lipoprotein ($hdl$), total cholesterol to high-density lipoprotein ratio ($tch$), log of triglycerides ($ltg$) and glucose ($glu$).
The response variable quantitatively measures disease progression one year after baseline. All variables are standardized to have mean 0 and variance $1/n$ except for the response.

We compare the Lasso models with $\lambda$ selected by CV before and after influential points are identified and removed. 
Specifically, we take two approaches. 
In Approach 1, we apply a 10-fold CV to Lasso, choose the optimal $\hat{\lambda}$ that offers the smallest MSE, and record the Lasso estimate of $\bm{\beta}$. In Approach 2, we remove influential observations with Algorithm~\ref{alg_detect} first and then take Approach 1 for refitting a model.

\begin{figure}
    \centering
     \includegraphics[width = \textwidth]{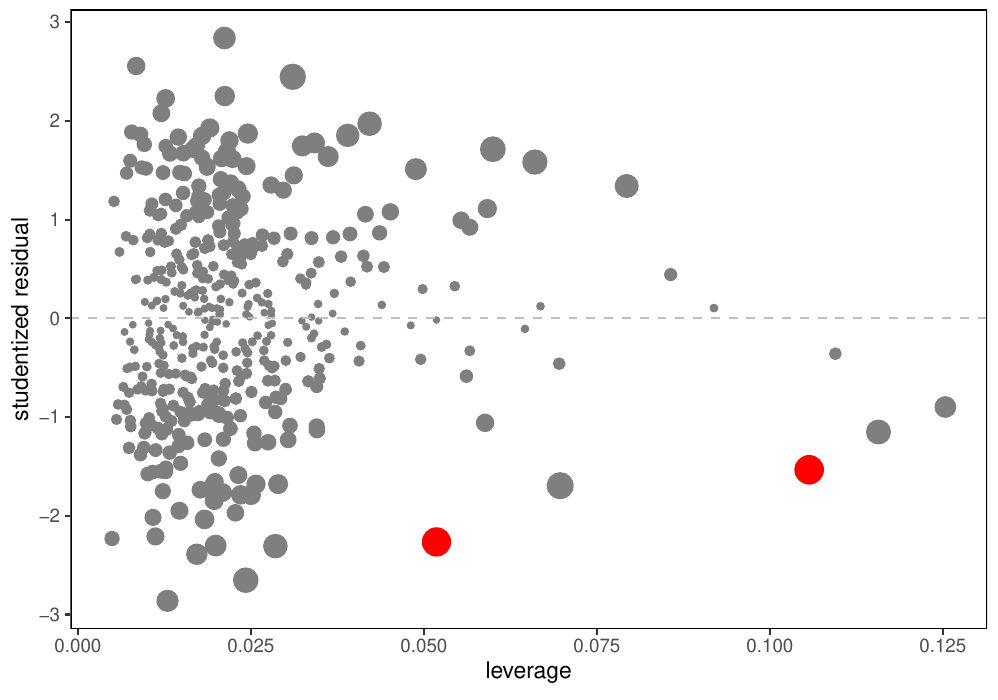}
    \caption{A plot of the studentized residuals versus leverages of all 442 observations from the Lasso model for the diabetes data when $\lambda=3$ (chosen by CV) and the fraction is 0.76. The size of each point represents the magnitude of its influence. The points marked in red are observations 170 (right) and 383 (left), which are the two most influential points.}
    \label{fig:res_lev}
\end{figure}

Figure~\ref{fig:res_lev} shows Cook's distances of all observations for the Lasso model with the penalty level ($\lambda=3$) determined by CV. Each point in the figure is an observation with the $x$-axis being its leverage under the active set chosen by CV and the $y$-axis being its studentized residual from the full data Lasso model. Even though we no longer have the explicit form of Cook's distance as a function of the residual and leverage from the full data fit as in OLS, Cook's distance for the Lasso still reflects the overall size of the residual and leverage to a large extent. Observations 170 (right) and 383 (left), marked in red, are the two most influential points. They both have large leverages and residuals simultaneously. 

\begin{table}[htb]
\centering
\begin{tabular}{c|rr|rr|rr}
\hline
& \multicolumn{2}{c|}{Mean} & \multicolumn{2}{c|}{Standard Dev} & \multicolumn{2}{c}{Inclusion$\%$}  \\
&$m(A1)$&$m(A2)$&$sd(A1)$&$sd(A2)$&In$(A1)$&In$(A2)$\\\hline
$age$ & -2.30   & -2.87   & 3.33   & 4.46  & 32.7            & 46.0  \\
$sex$ & -210.03 & -231.39 & 20.89  & 12.21 & 100.0           & 100.0   \\
$bmi$ & 522.00  & 518.04  & 1.50   & 5.56  & 100.0           & 100.0   \\
$bp$ & 305.13  & 302.13  & 12.67  & 7.57  & 100.0            & 100.0    \\
$tc$  & -269.66 & -59.45  & 229.48 & 72.31  &100.0        & 51.6     \\
$ldl$ & 105.88  & -63.81  & 153.20 & 48.75 & 32.7        & 65.8    \\
$hdl$ & -139.34 & -246.97  & 105.39 & 31.95 & 79.9           & 100.0     \\
$tch$ & 58.15   & 2.00    & 65.36  & 12.89 & 53.0           & 3.2     \\
$ltg$ & 567.78  & 576.59  & 78.80  & 22.99 & 100.0           & 100.0    \\
$glu$ & 58.09   & 27.25   & 7.26   & 8.58  & 100.0           & 100.0     \\\hline
$\hat{\lambda}$ & 14.35 & 16.36 & 10.33 & 4.13 &     &      \\
$|\hat{\mathcal{A}}|$ & 7.98 & 7.67 & 1.08 & 0.74 &  &      \\\hline
\end{tabular}
\caption{Summary statistics of coefficient estimates, $\lambda$ value chosen by CV, and the active set size $|\hat{\mathcal{A}}|$
from Approach 1  and Approach 2 applied to the diabetes data. 
For example, $m(A1)$, $sd(A1)$, and In$(A1)$ represent the mean and standard deviation of estimates, and the percentage of inclusion of each variable over 1000 experiments with Approach 1, respectively.
}
\label{rda_sum}
\end{table}

Since there is randomness in the choice of $\lambda$ due to CV, we repeat the experiment 1000 times with different data splits for CV. 
Table~\ref{rda_sum} contains summary statistics of the estimates of coefficients, $\lambda$ values chosen by CV, and the cardinality of active variables obtained from the two approaches.
Examining the summary of $\hat{\lambda}$, we see that after influential points are removed, the result from CV becomes more consistent in that the standard deviation of $\hat{\lambda}$ is only $40\%$ of that before. As a consequence, the estimates of coefficients also become more consistent after the removal of influential points. 
For example, variable $hdl$ is included by Approach 1 for $80\%$ of the times, $87\%$ being negative and $13\%$ being positive, while Approach 2 includes it for $100\%$ of the times with a negative estimate. 
Variable $tch$ is rarely included in the model by Approach 2 ($3.2\%$ of inclusion rate) while it is chosen by Approach 1 with around $50\%$ of chance. Note that $hdl$ and $tch$ are highly negatively correlated with correlation of $-0.74$. This leads to their inconsistent selection under Approach 1. 
With larger $\lambda$ values chosen after removing highly influential observations, the model is more likely to include $hdl$ and exclude $tch$.
Appendix~\ref{lasso_sol_path_app} includes the Lasso coefficient paths and elaborates on the reason for this pattern further.
As for $tc$ and $ldl$, the standard deviations of their estimates are greatly reduced, which also indicates an increased level of consistency. 
The increase in the standard deviation of $bmi$ coefficient is trivial compared to the scale of its estimate. We conclude that removing influential points identified by our measure can help select models more consistently. 

Next, we compare the model's prediction performance before and after removing influential points.  
The dataset is randomly split into training (354 observations) and testing (88 observations) sets. Using each training set, we implement Approach 1 and Approach 2. For each approach, 

\begin{enumerate}
    \item Obtain the Lasso estimate.
    \item Calculate the OLS estimate based on the active set identified.
    \item Compute MSE on the testing set for both the Lasso and OLS estimates.
\end{enumerate}

This process yields four distinct versions of MSE: MSE from the Lasso estimate and OLS estimate for both Approach 1 and Approach 2, denoted as $MSE_1(Lasso)$, $MSE_1(OLS)$, $MSE_2(Lasso)$, and $MSE_2(OLS)$, respectively. For comparison, we calculate the ratios of $MSE_1(Lasso)$ to $MSE_2(Lasso)$  and $MSE_1(OLS)$ to $MSE_2(OLS)$ for 1000 random splits of training and testing sets. The average and standard deviation of the ratios for the Lasso estimates and the OLS estimates are $0.999$ $(0.022)$ and $0.995$ $(0.023)$, respectively, suggesting no significant difference in the prediction performance before and after removal of influential points. Notably, Approach 2 estimates $\bm{\beta}$ with $5\%$ less data than Approach 1, yet it maintains nearly identical performance. Our analysis reveals that the removal of influential observations, as identified by our measure, enables a more consistent selection of models without sacrificing the prediction accuracy.

\begin{figure}[htb]
    \centering
    \includegraphics[width=\linewidth]{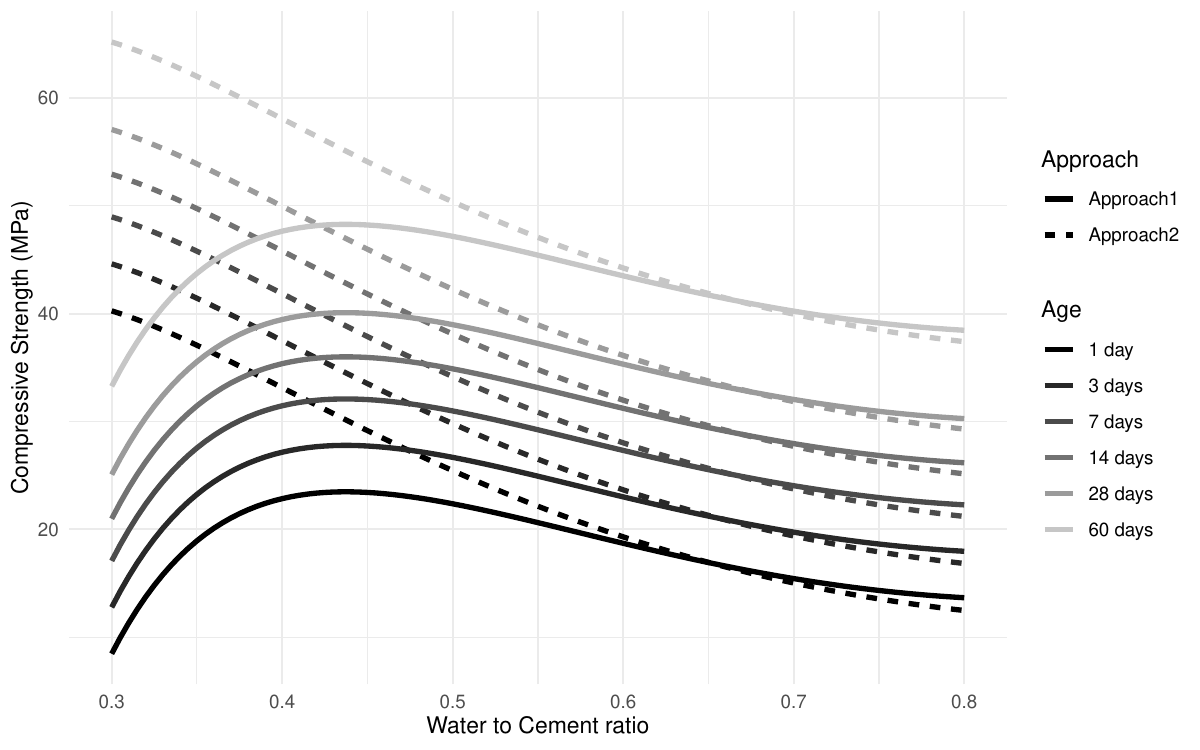}
    \caption{Compressive strength estimates as a function of water-to-cement ratio across different curing ages via two approaches. All other ingredient densities are fixed at typical values: cement at 350, blast furnace slag at 10, fly ash at 10, superplasticizer at 5, coarse aggregate at 1100, and fine aggregate at 750 (in kg/$m^3$).}
    \label{fig:concrete_example}
\end{figure}

{\subsubsection{Concrete compressive strength data}

The concrete compressive strength dataset, introduced by \citet{Yeh1998ModelingOS}, contains 1,030 concrete samples from 17 different laboratories and provides information on various ingredients and factors influencing the compressive strength of concrete. The dataset includes 8 input variables including the densities (in kg/$m^3$) of cement content, blast furnace slag, fly ash, water content, superplasticizer, coarse aggregate, fine aggregate, and the age of the concrete (in days). The response variable is the compressive strength, measured in megapascals (MPa), which is a critical factor in construction, determining the load-bearing capacity of structures.

Since concrete strength is a nonlinear function of its ingredients and age, we consider variable transformations to capture these complex relationships. Specifically, we add the reciprocals of water and cement density as two new features and transform the age variable into two new features: (1)  \text{min(age, 28)}, which accounts for early strength development, and (2)  $\max(0,  \mathrm{age} - 28)$, which models the maturity effects after 28 days when strength growth slows \citep{taylor2013curing}.
Additionally, all two-way interaction terms among the ingredients and their log transformations are included in the model. After adding those new features and interactive terms, the dataset comprises 45 features ($p=45$), allowing for a more detailed investigation of how the proportions of ingredients and factors such as cement-to-water ratio and curing age impact the compressive strength of concrete.


Next, we repeat the procedures outlined in the previous example. Across 1,000 replicates resulting from different splits of the data into 10 folds, the total variance (the sum of individual variances) of the inclusion probability of each covariate in Approach 2 is $78.6\%$ of that in Approach 1. 
The variance of the fractions, $\|\hat{\bm{\beta}}_\lambda\|_1/\|\hat{\bm{\beta}}_{OLS}\|_1$, also decreases by $38.2\%$.
Both of these indicate that removing influential points the algorithm identifies makes model selection decisions more consistent across different data splits. Regarding prediction accuracy, the average and standard deviation of the out-of-sample $R^2$ before and after the removal are 0.874 (0.001) and 0.881 (0.002), respectively. 


To demonstrate the difference in the fitted models between the two approaches, we present a plot of compressive strength estimates versus the water-to-cement (w/c) ratio, one of the most critical factors determining the compressive strength across different curing ages for the two approaches in Figure~\ref{fig:concrete_example} while keeping the densities of all other ingredients fixed at typical values. The relationship from Approach 2 indicates that a lower w/c ratio results in a higher compressive strength over a typical range of the w/c ratio, which aligns with established knowledge \citep{SINGH201594, WANG2022128833}.  In contrast, Approach 1 fails to capture this monotonic relationship. This difference is due to the presence of influential observations (381-385) identified by our algorithm and removed in Approach 2. These observations have high cement densities and low w/c ratios but have significantly lower compressive strengths than predicted by the model. Consequently, they pulled the estimated curve downward when the w/c ratio is small.
The original study noted that some samples were removed due to their large aggregate size, abnormal cement types, or special curing conditions that could potentially affect the compressive strength without specifying which ones. Given the consecutive numbering of the influential observations, we suspect that they may have come from the same source and suffer from the issues listed. } 

\begin{table}
\centering
\begin{tabular}{@{}l ccc l ccc@{}}
\toprule
\multicolumn{1}{l}{\textbf{Gene}} & \multicolumn{3}{c}{\textbf{Inclusion Prob}} & \multicolumn{1}{l}{\textbf{Gene}} & \multicolumn{3}{c}{\textbf{Inclusion Prob}} \\
\cmidrule(lr){2-4} \cmidrule(lr){6-8}
& \textbf{A2} & \textbf{IL} & \textbf{A1} & & \textbf{A2} & \textbf{IL} & \textbf{A1} \\ \midrule
CNN3      & 0.97 & 0.73 & 0.00 & GTSE1     & 0.90 & 0.57 & 0.00 \\
FLJ12443  & 0.97 & 0.68 & 0.00 & IRF3      & 0.86 & 0.54 & 0.32 \\
PTEN      & 0.97 & 0.73 & 0.14 & JIK       & 0.24 & 0.04 & 0.22 \\
KCNC1     & 0.97 & 0.85 & 0.03 & PPAP2C    & 0.05 & 0.02 & 0.89 \\
SYNJ2     & 0.97 & 0.71 & 0.00 & HS3ST2    & 0.00 & 0.07 & 0.89 \\
CGI-115   & 0.97 & 0.69 & 0.00 & TRPM2     & 0.00 & 0.01 & 0.01 \\
PTGDS     & 0.91 & 0.59 & 0.03 & ARHGAP15  & 0.00 & 0.01 & 0.03 \\
ADIPOR1   & 0.91 & 0.54 & 0.00 & RNF44     & 0.00 & 0.00 & 0.32 \\
\bottomrule
\end{tabular}
\caption{Genes with top-10 inclusion probability under Approaches 1, 2, and the Influence-Lasso (IL) procedure proposed in \citet{influencediag}. The genes are ranked based on the inclusion probability under Approach 2.}
\label{Table Glioblastoma}
\end{table}

{\subsubsection{Glioblastoma gene expression data} \label{glio_example}
Lastly, we test our algorithm on a high-dimensional dataset from a glioblastoma microarray gene expression study \citep{horvath2006analysis}. 
\citet{influencediag} also used this dataset to test their Lasso case influence measure in high-dimensional settings. Following the same preprocessing procedures in the paper, the dataset contains 3600 gene expressions and the survival time (in days) from 50 patients. See Section 2.2.2 of \citet{influencediag} for details. We regress the survival times on gene expressions and use the inclusion probability of each gene across 100 different data splits for 10-fold CV to identify genes potentially related to patients' survival using Approaches 1 and 2. 



Table~\ref{Table Glioblastoma} presents the top-10 genes based on the results of the two approaches and the influence-lasso procedure proposed by \citet{influencediag} in terms of the inclusion probability.
The top-10 genes using Approach 2 are consistent with the top-10 genes using influence-lasso but only with much higher inclusion probabilities, indicating a more consistent selection of genes. Regarding these selected genes, 
\citet{influencediag} provides a discussion of the relevant literature 
on the biological relevance of each gene to glioblastoma. 

Observation 29 is the only observation flagged by our algorithm and flagged 100$\%$ of the time across all data splits. The influence-lasso procedure flagged observations 27 and 29, $98\%$ and $100\%$ of the time, respectively, and other 19 observations at least once. 
Observation 29 is an outlier in that this patient is among the youngest and has an extremely shorter survival time than other patients, suggesting a potentially different pathological cause. Table~\ref{Table Glioblastoma} shows that the genes, PPAP2C and HS3ST2, are among the top 2 by Approach 1 with observation 29, but not anymore after observation 29 is removed.
This indicates that the importance of the two genes is only due to observation 29, and this conclusion may not be reliable. However, further excluding observation 27 does not alter the top-10 genes but leads to lower inclusion probabilities. In our analysis, Cook's distance for observation 27 is not even close to the threshold. 
Overall, observation 27 has a high leverage but small residual in the model determined by cross-validated $\lambda$ across all experiments. 
Note that the influence-lasso procedure determines influential points based on four separate measures, two of which are measures integrated over $\lambda$. This raises a concern that an observation like 27 that can be helpful for gene selection at some $\lambda$ value might get flagged due to its behavior at other $\lambda$ values. The unnecessary removal of such observations will waste valuable data and can impair the predictive power of the model.
}


\section{Conclusions and Future Directions}\label{conclusion}
 In this paper, we have introduced a case-weight adjusted Lasso model given a fixed penalty and derived its optimality conditions. From this, we have shown that the estimated coefficients change piecewise linearly with respect to a simple function of the weight parameter. 
 This property allows us to obtain the leave-one-out Lasso models from the full data model without refitting the model.
 Taking advantage of this, we have extended Cook's distance to the Lasso as a case influence measure and proposed a criterion to identify influential observations.
Furthermore, we have examined Cook's distance as a function of the penalty parameter $\lambda$ through a case influence graph,
which can highlight influential points at different penalty levels and also help with model selection. 
Through simulation studies, we have numerically validated the effectiveness of Cook's distance for the Lasso in identifying influential points in both $p<n$ and $p>n$ cases.
Our definition of an influential point as an observation that significantly alters estimates highlights that these points are essentially heterogeneous compared to other observations. 
Through our experiments on real data, we have shown that the exclusion of those heterogeneous observations can lead to more robust and reliable estimates without sacrificing the prediction accuracy.

There are several directions we can pursue in the future.
Firstly, this work only considers 
the influence of a single case on the Lasso model. 
Since the influence of one case can be easily masked by others, an extension of this work to assess the cumulative influence of multiple cases is a meaningful direction.
The solution path algorithm for calculating Cook's distance without refitting the model can be extended to the scenario when multiple cases are removed \citep{cv_cd}. So can be the case influence graph in this scenario. Secondly, as pointed out in Section~\ref{thresholding}, 
the Lasso coefficient estimates in the linear model setting with normal errors are shown to follow a truncated multivariate normal distribution. 
We may examine and utilize such distributional properties of the Lasso to refine the threshold for the identification of influential observations.
Last but not least, it is well known that the sensitivity of a model to data perturbation is inherently related to the notion of model complexity \citep{stein1981estimation, ye1998measuring, efron2004estimation}, and it is worth investigating the implications of the case-weight adjusted model on the complexity of a Lasso model. 
This will result in an alternative argument for the degrees of freedom of a Lasso model examined in \cite{zou2007degrees}. 

\clearpage
\bibliography{ref} 

\clearpage
\appendix
\section{Derivations for Section~\ref{cwasp}} \label{app_cwasp}
\subsection{Optimality conditions} \label{der_opt_cond}
To handle $|\beta_j|$ for each $j\in [p]$, we introduce upper bounds $a^+_j \geq 0 $ and $a^-_j \geq 0$ for the positive ($\beta_j^+$) and negative ($\beta_j^-$) parts of $\beta_j$ such that $-a^-_j\leq\beta_j\leq a^+_j$. Then, letting $\bm{a}^+=(a^+_1,\dots,a^+_p)^\top$ and $\bm{a}^-=(a^-_1,\dots,a^-_p)^\top$, we can rewrite Problem~\eqref{org_formula} as

\begin{equation}
\begin{aligned} 
    \min_{\bm{\beta},\beta_0,\bm{a}^+, \bm{a}^-} \quad &\frac{1}{2}\sum_{i\neq k} (y_i - \beta_0 - \bm{x}_i^\top\bm{\beta})^2 + \frac{1}{2}\omega(y_k - \beta_0 - \bm{x}_k^\top\bm{\beta})^2 +\lambda \sum_{j=1}^p (a^+_j + a^-_j)\\
    &\text{subject to}\quad -a^-_j\leq\beta_j\leq a^+_j, \\
    &\ \  \quad \quad \quad \quad \quad a^+_j \geq 0,\ a^-_j \geq 0\ \text{for} \ j \in [p].
\label{formula_lee}
\end{aligned}
\end{equation}

Next, by introducing Lagrange multipliers $\gamma_j^+\ge 0$ and $\gamma_j^-\ge 0$ for the inequality constraints $\beta_j \leq a_j^+$ and $-\beta_j \leq a_j^-$, and $\theta^+_j \ge 0$ and $\theta^-_j \ge 0$ for $a_j^+\geq0$ and $a_j^-\geq0$, we get the following Lagrangian function:
\begin{align*} 
   L(\bm{\beta},\beta_0,\bm{a}^+, \bm{a}^-, &\bm{\gamma}^+, \bm{\gamma}^-, \bm{\theta}^+,\bm{\theta}^-) 
   = \frac{1}{2}RSS(\bm{\beta},\beta_0) +\lambda \mathbf{1}_p^\top (\bm{a}^+ + \bm{a}^-) \\
    &+ (\bm{\gamma}^{+})^{\top} (\bm{\beta} -\bm{a}^+) - (\bm{\gamma}^{-})^{\top} (\bm{\beta} + \bm{a}^-) - \left((\bm{\theta}^{+})^{\top} \bm{a}^+ + (\bm{\theta}^{-})^{\top} \bm{a}^- \right),
\end{align*}
where 
$\bm{\gamma}^+, \bm{\gamma}^-, \bm{\theta}^+$, and $\bm{\theta}^-$ refer to the vectors with $\gamma^+_j$, $\gamma^-_j$, $\theta^+_j$, and $\theta^-_j$ for $j\in [p]$ as elements, respectively, and
\begin{align*}
    RSS(\bm{\beta},\beta_0)&=\sum_{i\neq k} (y_i - \beta_0 - \bm{x}_i^\top\bm{\beta})^2 + \omega(y_k - \beta_0 - \bm{x}_k^\top\bm{\beta})^2\\
    &= (\bm{y}-\beta_0\bm{1}_n- X\bm{\beta})^\top(\bm{y}-\beta_0\bm{1}_n- X\bm{\beta}) -(1-\omega)(y_k - \beta_0 - \bm{x}_k^\top\bm{\beta})^2. 
\end{align*}

According to the KKT conditions for a convex-constrained optimization problem, the gradients of $L$ with respect to the primal variables should vanish. Those conditions for our problem are stated as 
\begin{equation}
\begin{aligned} 
&\frac{\partial L}{\partial \bm{\beta}} = 
\bm{d}^\omega + (\bm{\gamma}^+ - \bm{\gamma}^-) = \bm{0},\\
&\frac{\partial L}{\partial \beta_0} = d_0^\omega = 0,\\
&\frac{\partial L}{\partial \bm{a}^+} = \lambda \mathbf{1}_p - \bm{\gamma}^+ - \bm{\theta}^+ = \bm{0}, \\
&\frac{\partial L}{\partial \bm{a}^-} = \lambda \mathbf{1}_p - \bm{\gamma}^{-} - \bm{\theta}^{-} = \bm{0},
\label{formula_lee3}
\end{aligned}
\end{equation}
where
\begin{align*}
\bm{d}^\omega &= \frac{1}{2}\frac{\partial RSS}{\partial \bm{\beta}} = 
-X^\top(\bm{y} - \beta_0\mathbf{1}_n -X\bm{\beta}) + (1-\omega) \bm{x}_{k}(y_k - \beta_0 - \bm{x}_k^\top\bm{\beta}),\\
d^\omega_0 &= \frac{1}{2}\frac{\partial RSS}{\partial \beta_0} = -\bm{1}_n^\top(\bm{y} - \beta_0\mathbf{1}_n -X\bm{\beta}) + (1-\omega) (y_k - \beta_0 - \bm{x}_k^\top\bm{\beta}).
\end{align*}
In addition, the complementarity conditions for the constraint functions and the corresponding dual variables should hold, which are stated as 

\begin{equation}
\begin{aligned}
&\bm{\gamma}^+ \circ(\bm{\beta} - \bm{a}^+) = \bm{0},\\
&\bm{\gamma}^- \circ(\bm{\beta} + \bm{a}^-) = \bm{0},\\
&\bm{\theta}^+ \circ \bm{a}^+ = \bm{0},\\
&\bm{\theta}^- \circ \bm{a}^- = \bm{0},
\label{formula_lee4}
\end{aligned}
\end{equation}
where $\circ$ indicates the elementwise product of vectors.
To emphasize the dependence of the optimal 
$\bm{\beta}$ and $\bm{\beta}_0$ on the case weight parameter $\omega$, we will add the superscript $\omega$ to $\bm{\beta}$ and $\bm{\beta}_0$ in the remaining derivations.

Using \eqref{formula_lee3} and \eqref{formula_lee4}, we discuss how the dual variables ($\bm{\gamma}^+$, $\bm{\gamma}^-$, $\bm{\theta}^+$, and $\bm{\theta}^-$) are related to $\bm{\beta}^\omega$, which further simplifies the KKT conditions. Given $\bm{\beta}^\omega$, for $j\in[p]$,
\begin{itemize}
    \item If $\beta_j^\omega = 0$, the values of $a_j^+$ and $a_j^-$ that minimize the objective function \eqref{formula_lee} are 0.
    \item If $\beta_j^\omega > 0$, then $a_j^+ >0$ since $\beta_j^\omega\leq a^+_j$ from \eqref{formula_lee}, which implies $\theta^+_j = 0$ by the complementarity condition in \eqref{formula_lee4}.
     This further implies $\gamma_j^+ = \lambda$ and $a^+_j = \beta_j^\omega$ by the conditions $\bm{\gamma}^+ =\lambda \mathbf{1}_p - \bm{\theta}^+$ in \eqref{formula_lee3} and $\bm{\gamma}^+ \circ(\bm{\beta}^\omega - \bm{a}^+) = \bm{0}$ in \eqref{formula_lee4}, respectively.  As for $a_j^-$ and $\gamma_j^-$ in this case, they both take value 0 due to \eqref{formula_lee} and  $\bm{\gamma}^- \circ(\bm{\beta}^\omega + \bm{a}^-) = \bm{0}$ in \eqref{formula_lee4}. 
    \item If $\beta_j^\omega < 0$, similarly, we have $\theta_j^-=0$, $\gamma_j^- = \lambda$, $\gamma_j^+ = 0$, and $a_j^-= -\beta_j^\omega$.
\end{itemize}

Next, we characterize $\bm{d}^\omega$ based on $\bm{\gamma}^+$ and $\bm{\gamma}^-$ using $\bm{d}^\omega = -\bm{\gamma}^+ + \bm{\gamma}^-$ 
in \eqref{formula_lee3}. In general, from the last two conditions in \eqref{formula_lee3} and the fact that $\bm{\theta}^+\ge \bm{0}$ and $\bm{\theta}^-\ge \bm{0}$,  we have 
    $\bm{\theta}^+ = \lambda \mathbf{1}_p - \bm{\gamma}^+ \ge \bm{0}$ and 
$\bm{\theta}^{-} =\lambda \mathbf{1}_p - \bm{\gamma}^{-} \ge \bm{0}$, which implies 
 $-\lambda \mathbf{1}_p \leq\bm{\gamma}^+ - \bm{\gamma}^-\leq\lambda\mathbf{1}_p$. Thus, $d_j^\omega \in [-\lambda, \lambda]$ for $j\in[p]$.
\begin{itemize}
    \item If $\beta_j^\omega = 0$, the value of $d_j$ cannot be determined explicitly. $d_j^\omega \in [-\lambda, \lambda]$.
    \item If $\beta_j^\omega > 0$, $d_j^\omega = -\gamma^+_j + \gamma^-_j = -\lambda +0 = -\lambda$. 
    \item If $\beta_j^\omega < 0$, $d_j^\omega = -\gamma^+_j + \gamma^-_j = 0 + \lambda = \lambda$. 
\end{itemize}
Letting $s_j^\omega = -\frac{1}{\lambda} d_j^\omega$, we arrive at the following optimality conditions for the primal variables: 
\begin{align}
     &d_0^\omega = 0\tag*{\eqref{subgradient_0}},\\
    &\bm{d}^\omega = -\lambda \bm{s}^\omega, \tag*{\eqref{subgradient}}\\
   \mbox{where  }   &s_j^\omega = \text{sign}(\beta_j^\omega)\quad \text{if } \beta_j^\omega\neq0,\tag*{\eqref{gammaj}}\\
    &s_j^\omega \in [-1,1]\quad \text{if } \beta_j=0.\tag*{\eqref{gamma}}
\end{align}
These conditions indicate that once we know the signs of $\bm{\beta}^\omega$, we can identify $\bm{\beta}^\omega$ directly.

\subsection{Solving for \texorpdfstring{$\bm{\beta}$}{beta}, \texorpdfstring{$\beta_0$}{beta0}, and \texorpdfstring{$\mathbf{d}$}{d} as a function of \texorpdfstring{$\omega$}{omega}} \label{solution_der}

For given $\bm{\beta}$, we define $\mathcal{A} = \{j\in[p]~|~\beta_j\neq0\}$ as the active set. Similarly, we define $\hat{\mathcal{A}}^\omega = \mathcal{A}(\hat{\bm{\beta}}^\omega)$. For any $\mathcal{A}\subset[p]$, $\backslash\mathcal{A}$ is the complement of $\mathcal{A}$.
We use $X_{\mathcal{A}}$ to denote the submatrix of $X$ containing only the columns in $\mathcal{A}$. For $\bm{v} \in \mathbb{R}^p$,    $\bm{v}_{\mathcal{A}}$ denotes the subvector of  $\bm{v}$ containing only the elements in $\mathcal{A}$. 
Using \eqref{L_i}--\eqref{gamma} and solving for $\bm{\beta}$ and $\beta_0$, we have
\begin{align}
   &\hat{\bm{\beta}}^\omega_{\backslash {\hat{\mathcal{A}}^{\omega}}} = \bm{0}, \notag \tag*{\eqref{beta_ne}}\\
   & \hat{\bm{\beta}}^\omega_{{\hat{\mathcal{A}}^{\omega}}} = \left( X_{{\hat{\mathcal{A}}^{\omega}}}^{\top}X_{{\hat{\mathcal{A}}^{\omega}}} - \frac{n (1-\omega)}{n-1+\omega} \bm{x}_{k{\hat{\mathcal{A}}^{\omega}}}\bm{x}_{k{\hat{\mathcal{A}}^{\omega}}}^{\top}\right)^{-1}\nonumber\\
   &\quad \quad \quad\quad\quad\quad \quad \left(X_{{\hat{\mathcal{A}}^{\omega}}}^{\top}\bm{y}  - \lambda \hat{\bm{s}}^\omega_{\hat{\mathcal{A}}^{\omega}} - \frac{n(1-\omega)  (y_k - \Bar{y})}{n-1+\omega} \bm{x}_{k{\hat{\mathcal{A}}^{\omega}}}\right),\label{beta}\\
   &\hat{\beta}_0^\omega = \Bar{y} + \frac{1-\omega}{n-1+\omega}(\Bar{y}+\bm{x}_{k{\hat{\mathcal{A}}^{\omega}}}^\top \hat{\bm{\beta}}^{\omega}_{{\hat{\mathcal{A}}^{\omega}}}-y_k), \label{beta_0}
\end{align}
where $\Bar{y}$ is the sample mean of the response vector $\bm{y}$.
To split out the second term depending on $\omega$ in the inverse in \eqref{beta}, we introduce the following Lemma~\ref{lemma1} from \cite{miller1981inverse}. 

\begin{lemma}
For matrices $A$ and $B$, if $A$ and $A+B$ are invertible and $B$ has rank 1, then $\tr(BA^{-1}) \neq -1$ and
\begin{align*}
    (A+B)^{-1} = A^{-1} - \frac{1}{1+\tr(BA^{-1})}A^{-1}BA^{-1}.
\end{align*}
\label{lemma1}
\end{lemma}
Applying Lemma~\ref{lemma1} to the inverse in \eqref{beta} with $A = X_{{\hat{\mathcal{A}}^{\omega}}}^{\top}X_{{\hat{\mathcal{A}}^{\omega}}}$ and $B=-\frac{n (1-\omega)}{n-1+\omega} \bm{x}_{k{\hat{\mathcal{A}}^{\omega}}}\bm{x}_{k{\hat{\mathcal{A}}^{\omega}}}^{\top}$, we have

\begin{align*}
(X_{{\hat{\mathcal{A}}^{\omega}}}^{\top}X_{{\hat{\mathcal{A}}^{\omega}}})^{-1}  +\frac{n (1-\omega)(X_{{\hat{\mathcal{A}}^{\omega}}}^{\top}X_{{\hat{\mathcal{A}}^{\omega}}})^{-1} \bm{x}_{k{\hat{\mathcal{A}}^{\omega}}}\bm{x}_{k{\hat{\mathcal{A}}^{\omega}}}^{\top}(X_{{\hat{\mathcal{A}}^{\omega}}}^{\top}X_{{\hat{\mathcal{A}}^{\omega}}})^{-1}}{({n-1+\omega}) -n(1-\omega)\bm{x}_{k{\hat{\mathcal{A}}^{\omega}}}^{\top} (X_{{\hat{\mathcal{A}}^{\omega}}}^{\top}X_{{\hat{\mathcal{A}}^{\omega}}})^{-1}\bm{x}_{k{\hat{\mathcal{A}}^{\omega}}}}.
\end{align*}
We can simplify $\bm{x}_{k{\hat{\mathcal{A}}^{\omega}}}^{\top} (X_{{\hat{\mathcal{A}}^{\omega}}}^{\top}X_{{\hat{\mathcal{A}}^{\omega}}})^{-1}\bm{x}_{k{\hat{\mathcal{A}}^{\omega}}}$ by noting the fact that
the hat matrix for the least squares regression with the design matrix, $\widetilde{X} = (\mathbf{1}_n, X)$, is given as 
$H^\omega = \widetilde{X}_{\hat{\mathcal{A}}^{\omega}}(\widetilde{X}^{\top}_{\hat{\mathcal{A}}^{\omega}}\widetilde{X}_{\hat{\mathcal{A}}^{\omega}})^{-1}\widetilde{X}_{\hat{\mathcal{A}}^{\omega}}^\top$, and 
when all features in $X$ are centered to 0, each entry of $H^\omega$ is $h_{ij}^\omega = 1/n +  \bm{x}_{i{\hat{\mathcal{A}}^{\omega}}}^{\top} (X_{{\hat{\mathcal{A}}^{\omega}}}^{\top}X_{{\hat{\mathcal{A}}^{\omega}}})^{-1}\bm{x}_{j{\hat{\mathcal{A}}^{\omega}}}$. Replacing $\bm{x}_{k{\hat{\mathcal{A}}^{\omega}}}^{\top} (X_{{\hat{\mathcal{A}}^{\omega}}}^{\top}X_{{\hat{\mathcal{A}}^{\omega}}})^{-1}\bm{x}_{k{\hat{\mathcal{A}}^{\omega}}}$ with $(h_{kk}^\omega-1/n)$, we have the following: 
\begin{align}
(X_{{\hat{\mathcal{A}}^{\omega}}}^{\top}X_{{\hat{\mathcal{A}}^{\omega}}})^{-1}  +\frac{(1-\omega)}{1 -(1-\omega)h_{kk}^\omega}(X_{{\hat{\mathcal{A}}^{\omega}}}^{\top}X_{{\hat{\mathcal{A}}^{\omega}}})^{-1} \bm{x}_{k{\hat{\mathcal{A}}^{\omega}}}\bm{x}_{k{\hat{\mathcal{A}}^{\omega}}}^{\top}(X_{{\hat{\mathcal{A}}^{\omega}}}^{\top}X_{{\hat{\mathcal{A}}^{\omega}}})^{-1}.\label{split_inv}
\end{align}
Letting $\xi^\omega = \frac{1-\omega}{1-(1-\omega)h_{kk}^\omega}$, and 
replacing the inverse in $\hat{\bm{\beta}}^\omega_{{\hat{\mathcal{A}}^{\omega}}}$ \eqref{beta} with \eqref{split_inv}, we have
\begin{align*}
\hat{\bm{\beta}}^\omega_{{\hat{\mathcal{A}}^{\omega}}} =&\ (X_{{\hat{\mathcal{A}}^{\omega}}}^{\top}X_{{\hat{\mathcal{A}}^{\omega}}})^{-1} (X_{{\hat{\mathcal{A}}^{\omega}}}^{\top}\bm{y} - \lambda \hat{\bm{s}}^\omega_{\hat{\mathcal{A}}^{\omega}}) \\
&+\xi^\omega(X_{{\hat{\mathcal{A}}^{\omega}}}^{\top}X_{{\hat{\mathcal{A}}^{\omega}}})^{-1} \bm{x}_{k{\hat{\mathcal{A}}^{\omega}}}\bm{x}_{k{\hat{\mathcal{A}}^{\omega}}}^{\top} (X_{{\hat{\mathcal{A}}^{\omega}}}^{\top}X_{{\hat{\mathcal{A}}^{\omega}}})^{-1}(X_{{\hat{\mathcal{A}}^{\omega}}}^{\top}\bm{y} - \lambda \hat{\bm{s}}^\omega_{\hat{\mathcal{A}}^{\omega}})\\
&- \frac{n(1-\omega)  (y_k - \Bar{y})}{n-1+\omega}(X_{{\hat{\mathcal{A}}^{\omega}}}^{\top}X_{{\hat{\mathcal{A}}^{\omega}}})^{-1}  \bm{x}_{k{\hat{\mathcal{A}}^{\omega}}}\\
&-\frac{n(1-\omega)  (y_k - \Bar{y})}{n-1+\omega}\xi^\omega (X_{{\hat{\mathcal{A}}^{\omega}}}^{\top}X_{{\hat{\mathcal{A}}^{\omega}}})^{-1} \bm{x}_{k{\hat{\mathcal{A}}^{\omega}}}
\left[\bm{x}_{k{\hat{\mathcal{A}}^{\omega}}}^{\top}(X_{{\hat{\mathcal{A}}^{\omega}}}^{\top}X_{{\hat{\mathcal{A}}^{\omega}}})^{-1} \bm{x}_{k{\hat{\mathcal{A}}^{\omega}}}\right]\\
=&\ (X_{{\hat{\mathcal{A}}^{\omega}}}^{\top}X_{{\hat{\mathcal{A}}^{\omega}}})^{-1} (X_{{\hat{\mathcal{A}}^{\omega}}}^{\top}\bm{y} - \lambda \hat{\bm{s}}^\omega_{\hat{\mathcal{A}}^{\omega}}) \\
&+\xi^\omega(X_{{\hat{\mathcal{A}}^{\omega}}}^{\top}X_{{\hat{\mathcal{A}}^{\omega}}})^{-1} \bm{x}_{k{\hat{\mathcal{A}}^{\omega}}}\bm{x}_{k{\hat{\mathcal{A}}^{\omega}}}^{\top} (X_{{\hat{\mathcal{A}}^{\omega}}}^{\top}X_{{\hat{\mathcal{A}}^{\omega}}})^{-1}(X_{{\hat{\mathcal{A}}^{\omega}}}^{\top}\bm{y} - \lambda \hat{\bm{s}}^\omega_{\hat{\mathcal{A}}^{\omega}})\\
&- \frac{n(1-\omega)  (y_k - \Bar{y})}{n-1+\omega} \left(1+\xi^\omega \left(h_{kk}^\omega-\frac{1}{n}\right)\right)(X_{{\hat{\mathcal{A}}^{\omega}}}^{\top}X_{{\hat{\mathcal{A}}^{\omega}}})^{-1}\bm{x}_{k{\hat{\mathcal{A}}^{\omega}}}. 
\end{align*}
For the third term,
\begin{align*}
    \frac{n(1-\omega)  (y_k - \Bar{y})}{n-1+\omega} \left(1+\xi^\omega \left(h_{kk}^\omega-\frac{1}{n}\right)\right) =&\ \frac{n(1-\omega)  (y_k - \Bar{y})}{n-1+\omega}  \frac{1-(1-\omega)/n}{1-(1-\omega)h_{kk}^\omega}\\
    =& \frac{1-\omega}{1-(1-\omega)h_{kk}^\omega}(y_k-\Bar{y}) \\
    =&\ \xi^\omega (y_k-\Bar{y}),
\end{align*}
which leads to the following expression for $\hat{\bm{\beta}}^\omega_{{\hat{\mathcal{A}}^{\omega}}}$:  
\begin{align*}
\hat{\bm{\beta}}^\omega_{{\hat{\mathcal{A}}^{\omega}}} =& \ (X_{{\hat{\mathcal{A}}^{\omega}}}^{\top}X_{{\hat{\mathcal{A}}^{\omega}}})^{-1} (X_{{\hat{\mathcal{A}}^{\omega}}}^{\top}\bm{y} - \lambda \hat{\bm{s}}^\omega_{\hat{\mathcal{A}}^{\omega}}) \nonumber\\
&+\xi^\omega(X_{{\hat{\mathcal{A}}^{\omega}}}^{\top}X_{{\hat{\mathcal{A}}^{\omega}}})^{-1} \bm{x}_{k{\hat{\mathcal{A}}^{\omega}}}\left(\bm{x}_{k{\hat{\mathcal{A}}^{\omega}}}^{\top} (X_{{\hat{\mathcal{A}}^{\omega}}}^{\top}X_{{\hat{\mathcal{A}}^{\omega}}})^{-1}(X_{{\hat{\mathcal{A}}^{\omega}}}^{\top}\bm{y} - \lambda \hat{\bm{s}}^\omega_{\hat{\mathcal{A}}^{\omega}}) + \Bar{y} - y_k)\right) .
\end{align*}
This seemingly complicated formula can be simplified if we adopt some of the results for regular Lasso corresponding to $\omega=1$:
\begin{align}
    &\hat{\bm{\beta}}_{\hat{\mathcal{A}}} = (X_{{\hat{\mathcal{A}}}}^{\top}X_{{\hat{\mathcal{A}}}})^{-1}(X_{{\hat{\mathcal{A}}}}^{\top}\bm{y} - \lambda \hat{\bm{s}}_{\hat{\mathcal{A}}}),\tag*{\eqref{pseudo_beta}}\\
    &\hat{\beta}_0 = \Bar{y}, \tag*{\eqref{ybar_}}\\
     & \hat{\bm{d}}_{\backslash{\hat{\mathcal{A}}}} = -X_{\backslash{\hat{\mathcal{A}}}}^{\top} \left(\bm{y} - X_{{\hat{\mathcal{A}}}}\hat{\bm{\beta}}_{\hat{\mathcal{A}}} \right), \tag*{\eqref{dna}}\\
    &\hat{\bm{y}} =  \hat{\beta}_0 \mathbf{1}_n + X_{{\hat{\mathcal{A}}}}\hat{\bm{\beta}}_{\hat{\mathcal{A}}}, \tag*{\eqref{ycouterpart}}
\end{align}
where $\hat{\mathcal{A}} = \mathcal{A}(\hat{\bm{\beta}})$.
Using \eqref{ycouterpart}, $\hat{y}_k = \hat{\beta}_0 + \bm{x}_{k{\hat{\mathcal{A}}}}^\top\hat{\bm{\beta}}_{\hat{\mathcal{A}}} = \Bar{y} + \bm{x}_{k{\hat{\mathcal{A}}}}^{\top} (X_{{\hat{\mathcal{A}}}}^{\top}X_{{\hat{\mathcal{A}}}})^{-1}(X_{{\hat{\mathcal{A}}}}^{\top}\bm{y} - \lambda \hat{\bm{s}}_{\hat{\mathcal{A}}})$. Thus, $\hat{\bm{\beta}}^\omega_{{\hat{\mathcal{A}}^\omega}}$ can be rewritten as
\begin{align}
         \hat{\bm{\beta}}^\omega_{\hat{\mathcal{A}}^\omega} &= \hat{\bm{\beta}}_{\hat{\mathcal{A}}} - \xi^\omega(X_{{\hat{\mathcal{A}}^\omega}}^{\top}X_{{\hat{\mathcal{A}}^\omega}})^{-1}\bm{x}_{k{\hat{\mathcal{A}}^\omega}}(y_k-\hat{y}_k)\notag\tag*{\eqref{beta_e}}.
\end{align}
Subsequently, by plugging \eqref{beta_e} into \eqref{beta_0}, we get 
\begin{align*}
   \hat{\beta}_0^\omega = \hat{\beta}_0 - \xi^\omega(y_k - \hat{y}_k )/n \tag*{\eqref{re_beta_0}}.
\end{align*}
Once we have $\hat{\bm{\beta}}^\omega$ and $\hat{\beta}_0^\omega$, the case-weighted fitted value $\hat{\bm{y}}^\omega$  can be calculated:
\begin{align}
    \hat{\bm{y}}^\omega =&\ \hat{\beta}_0^\omega\mathbf{1}_n + X_{{\hat{\mathcal{A}}^\omega}}\hat{\bm{\beta}}^\omega_{\hat{\mathcal{A}}^\omega}\nonumber\\
    =&\ \left(\hat{\beta}_0 - \xi^\omega(y_k - \hat{y}_k)/n\right) \mathbf{1}_n + X_{{\hat{\mathcal{A}}^\omega}}\left(\hat{\bm{\beta}}_{\hat{\mathcal{A}}} + \xi^\omega(X_{{\hat{\mathcal{A}}^\omega}}^{\top}X_{{\hat{\mathcal{A}}^\omega}})^{-1}\bm{x}_{k{\hat{\mathcal{A}}^\omega}}(\hat{y}_k - y_k)\right)\nonumber\\
    =&\ \hat{\bm{y}} - \xi^\omega \bm{h}_{\cdot k}^\omega(y_k - \hat{y}_k),  \tag*{\eqref{yomega}}
\end{align}
where $\bm{h}_{\cdot k}^\omega$ is the $k$th column of $H^\omega$.

Next, we consider $\bm{d}^\omega$ defined in \eqref{L_i}.
By the optimality conditions \eqref{subgradient} and \eqref{gammaj},
\begin{align}
    \hat{\bm{d}}^\omega_{{\hat{\mathcal{A}}^{\omega}}} &= -\lambda ~\text{sign}(\hat{\bm{\beta}}^\omega_{\hat{\mathcal{A}}^{\omega}}) =-\lambda \hat{\bm{s}}^\omega_{\hat{\mathcal{A}}^{\omega}} \tag*{\ref{d_trivial}}.
\end{align}
As for $\hat{\bm{d}}^\omega_{\backslash{\hat{\mathcal{A}}^{\omega}}}$, by $\hat{\bm{y}}^\omega$ in \eqref{yomega}, we have 
\begin{align*}
    \bm{d}^\omega_{\backslash{\hat{\mathcal{A}}^{\omega}}} =&\ -X_{\backslash{\hat{\mathcal{A}}^{\omega}}}^\top (\bm{y}-\beta_0^\omega\bm{1}_n- X_{{\hat{\mathcal{A}}^{\omega}}}\bm{\beta}^\omega_{\hat{\mathcal{A}}^{\omega}})+(1-\omega)\bm{x}_{k\backslash{\hat{\mathcal{A}}^{\omega}}} (y_k - \hat{\beta}_0^\omega - \bm{x}_{k{\hat{\mathcal{A}}^{\omega}}}^\top\hat{\bm{\beta}}^\omega_{\hat{\mathcal{A}}^{\omega}})\\
    =&\ -X_{\backslash{\hat{\mathcal{A}}^{\omega}}}^\top (\bm{y}-\hat{\bm{y}}^\omega)+(1-\omega)\bm{x}_{k\backslash{\hat{\mathcal{A}}^{\omega}}} (y_k - \hat{y}_k^\omega)\\
    =&\ -X_{\backslash{\hat{\mathcal{A}}^{\omega}}}^\top\left(\bm{y}-\bm{\hat{y}}+\xi^\omega \bm{h}_{\cdot k}^\omega(y_k - \hat{y}_k) \right)+(1-\omega)\bm{x}_{k\backslash{\hat{\mathcal{A}}^{\omega}}} \left(y_k - \hat{y}_k+\xi^\omega h_{kk}^\omega(y_k -\hat{y}_k)\right).
\end{align*}
Further, using $\hat{\bm{d}}^\omega_{\backslash{\hat{\mathcal{A}}^{\omega}}}$ in \eqref{dna}, we have the following relation between $\hat{\bm{d}}^\omega_{\backslash{\hat{\mathcal{A}}^{\omega}}}$ and $\hat{\bm{d}}_{\backslash{\hat{\mathcal{A}}}}$:
\begin{align}
\hat{\bm{d}}^\omega_{\backslash{\hat{\mathcal{A}}^{\omega}}}  =&\ \hat{\bm{d}}_{\backslash{\hat{\mathcal{A}}^\omega}} - \xi^\omega
X_{\backslash{\hat{\mathcal{A}}^{\omega}}}^\top \bm{h}^\omega_{\cdot k}(y_k - \hat{y}_k) + (1-\omega)(1+\xi^\omega h_{kk}^\omega)\bm{x}_{k\backslash{\hat{\mathcal{A}}^{\omega}}}(y_k - \hat{y}_k)\nonumber\\
    =&\ \hat{\bm{d}}_{\backslash{\hat{\mathcal{A}}^\omega}} - \xi^\omega X_{\backslash{\hat{\mathcal{A}}^{\omega}}}^\top \bm{h}^\omega_{\cdot k}(y_k - \hat{y}_k) + \xi^\omega\bm{x}_{k\backslash{\hat{\mathcal{A}}^{\omega}}}(y_k - \hat{y}_k)\nonumber\\
    =&\ \hat{\bm{d}}_{\backslash{\hat{\mathcal{A}}^\omega}} - \xi^\omega (X_{\backslash{\hat{\mathcal{A}}^{\omega}}}^\top \bm{h}^\omega_{\cdot k} - \bm{x}_{k\backslash{\hat{\mathcal{A}}^{\omega}}})(y_k - \hat{y}_k) \tag*{\eqref{domega}}.
\end{align}

\subsection{Monotonicity of \texorpdfstring{$\xi^\omega$}{xi}} \label{monot_of_xi}
We verify that
\begin{align}
   \frac{\partial \xi^\omega}{\partial \omega} = -\frac{1}{(1-(1-\omega){h_{kk}^\omega)}^2} < 0. \label{dev_xi}
\end{align}
Thus, $\xi^\omega$ 
is a monotone decreasing function of $\omega$ and hence has a one-to-one relationship with $\omega$ between two adjacent breakpoints. 

\section{Proofs of Propositions}
\subsection{Proof of Proposition~\ref{prop2}} \label{app_prop2}
Restatement of the proposition: For each $k\in [n]$ and fixed $\lambda\in \mathbb{R}^+$, let $\{\omega_m\}_{m=0}^M$ be a sequence of breakpoints of $\omega$ with $0 = \omega_M<\cdots<\omega_0=1$. Then the derivative of $D_k(\lambda, \omega)$ with respect to $\xi^\omega$ is piecewise monotonically decreasing in $\omega\in (\omega_{m+1},\omega_m)$ for $m=0,\dots,M-1$.
\begin{proof}
For any $k \in [n]$, $\lambda>0$, and $\omega\in(\omega_{m+1},\omega_m)$, by \eqref{yomega}, 
\begin{align*}
    \frac{\partial}{\partial \omega}\left(\frac{\partial D_k(\lambda, \omega)}{\partial \xi^\omega} \right)\propto
    & ~\frac{\partial}{\partial \omega}\left(\sum_{i=1}^n (\hat{y}^\omega_i - \hat{y}_i(\lambda))\frac{\partial \hat{y}^\omega_i}{\partial \xi^\omega}\right) \nonumber\\
    =& ~\frac{\partial}{\partial \omega}\left((\hat{\bm{y}}^\omega - \hat{\bm{y}}(\lambda))^\top\frac{\partial\hat{\bm{y}}^\omega}{\partial\xi^\omega}\right) \nonumber\\
    =& ~\frac{\partial}{\partial \omega}(\hat{\bm{y}}^\omega - \hat{\bm{y}}(\lambda))^\top\bm{h}^{\omega_m}_{\cdot k} (\hat{y}_k(\lambda, \omega_m) - y_k)\nonumber\\
    =& ~h_{kk}^{\omega_m}(\hat{y}_k(\lambda, \omega_m) - y_k)^2\frac{\partial \xi^\omega}{\partial\omega}.
\end{align*}
By \eqref{dev_xi},   $\frac{\partial\xi^\omega}{\partial\omega}<0$, and thus we have $\frac{\partial}{\partial \omega}\left(\frac{\partial D_k(\lambda, \omega)}{\partial \xi^\omega} \right)\le0$.
\end{proof}

\subsection{Proof of Proposition~\ref{coro1}} \label{app_coro1}
Restatement of the proposition: $D_k(\lambda, \omega)$ is monotonically decreasing in $\omega\in[0,1]$ for every $k\in [n]$ and fixed $\lambda\in \mathbb{R}^+$ if one of the following conditions holds: \\
i. Each active set update is a variable being added to the active set. \\
ii. Each active set update is a variable being dropped from the active set.\\ 
iii. No updates to the active set occur.
\begin{proof}
For any $k \in [n]$ and $\lambda>0$,
\begin{align*}
    \frac{\partial D_k(\lambda, \omega)}{\partial \omega} =& \frac{\partial D_k(\lambda, \omega)}{\partial \xi^\omega}\frac{\partial \xi^\omega}{\partial \omega}.
\end{align*}
By Proposition~\ref{prop2} and \eqref{dev_xi}, to show $\frac{\partial D_k(\lambda, \omega)}{\partial \omega} \le 0$, we verify that
$\frac{\partial D_k(\lambda, \omega)}{\partial \xi^\omega} \ge 0$ at the right boundary of each $(\omega_{m+1},\omega_m)$, i.e.,
\begin{align}
   f(m) = \lim_{\omega \uparrow \omega_m}\frac{\partial D_k(\lambda, \omega)}{\partial \xi^\omega} = (\hat{\bm{y}}^{\omega_m} - \hat{\bm{y}}(\lambda))^\top\bm{h}^{\omega_m}_{\cdot k} (\hat{y}_k(\lambda, \omega_m) - y_k) \ge 0, \quad m = 0, \dots, M-1.\label{target}
\end{align}  
If $m=0$, $\hat{\bm{y}}^{\omega_m} = \hat{\bm{y}}(\lambda)$, and \eqref{target} holds obviously. Under condition iii ($M=1$), $m$ can only be 0, and the monotonicity holds. 
If $m>0$, $\hat{\bm{y}}^{\omega_m} -  \hat{\bm{y}}(\lambda)$ can be rewritten as
\begin{align*}
     \hat{\bm{y}}^{\omega_m} -  \hat{\bm{y}}(\lambda) =&\ \hat{\bm{y}}^{\omega_m} -  \hat{\bm{y}}^{\omega_0}\\
    =&\ \sum_{i=0}^{m-1} (\hat{\bm{y}}^{\omega_{i+1}} -  \hat{\bm{y}}^{\omega_i})\\
    =&\ \sum_{i=0}^{m-1} (\xi^{\omega_{i,2}} - \xi^{\omega_i})\bm{h}_{\cdot k}^{\omega_i}(\hat{y}_k(\lambda, \omega_i) - y_k).
\end{align*}
For the third step, since both $\hat{\bm{y}}^{\omega_{i+1}}$ and $\hat{\bm{y}}^{\omega_{i}}$ can be evaluated under the same active set, using \eqref{yomega}, we have
\begin{align*}
    \hat{\bm{y}}^{\omega_{i+1}} - \hat{\bm{y}}^{\omega_i} =& \Big(\hat{\bm{y}}(\lambda,\omega_i) - \xi^{\omega_{i,2}}\bm{h}^{\omega_i}_{\cdot k}(y_k - \hat{y}_k(\lambda, \omega_i))\Big) - \Big(\hat{\bm{y}}(\lambda,\omega_i) - \xi^{\omega_{i}}\bm{h}^{\omega_i}_{\cdot k}(y_k - \hat{y}_k(\lambda, \omega_i))\Big)\\
    =& (\xi^{\omega_{i,2}} - \xi^{\omega_i})\bm{h}_{\cdot k}^{\omega_i}(\hat{y}_k(\lambda, \omega_i) - y_k).
\end{align*}
Then $f(m)$ becomes
\begin{align*}
    f(m) &= \left(\sum_{i=0}^{m-1} (\xi^{\omega_{i,2}} - \xi^{\omega_i})\bm{h}_{\cdot k}^{\omega_i}(\hat{y}_k(\lambda,\omega_i) - y_k)\right)^\top\bm{h}_{\cdot k}^{\omega_m} (\hat{y}_k(\lambda, \omega_m) - y_k) \\ 
    &= \sum_{i=0}^{m-1} (\xi^{\omega_{i,2}} - \xi^{\omega_i})(\hat{y}_k(\lambda,\omega_i) - y_k)(\hat{y}_k(\lambda,\omega_m) - y_k)(\bm{h}_{\cdot k}^{\omega_i})^\top\bm{h}_{\cdot k}^{\omega_m},
\end{align*}
which is a weighted sum of $(\hat{y}_k(\lambda,\omega_i) - y_k)(\hat{y}_k(\lambda,\omega_m) - y_k)(\bm{h}_{\cdot k}^{\omega_i})^\top\bm{h}_{\cdot k}^{\omega_m}$ for $i=0,\ldots,m-1$.

To prove $(\hat{y}_k(\lambda,\omega_i) - y_k)(\hat{y}_k(\lambda,\omega_m) - y_k)>0$, we show that the sign of $(\hat{y}_k(\lambda,\omega_i)- y_k)$ stays the same for $i = 0,\dots,m$.
For each breakpoint $\omega_m$, $\hat{y}^{\omega_m}_k$ can be evaluated in two ways by considering it at the right end of the interval $[\omega_{m+1}, \omega_{m}]$ and at the left end of the interval $[\omega_{m}, \omega_{m-1}]$. By \eqref{resid_form}, we have
\begin{align*}
    \hat{y}^{\omega_m}_k - y_k &=(1+\xi^{\omega_m} h_{kk}^{\omega_m}) (\hat{y}_k(\lambda,\omega_m) - y_k)\\
                           &=(1+\xi^{\omega_{m-1,2}} h_{kk}^{\omega_{m-1}})(\hat{y}_k(\lambda, \omega_{m-1}) - y_k),
\end{align*}
which results in  
\begin{align*}
    \hat{y}_k(\lambda,\omega_m) - y_k &=\frac{1+\xi^{\omega_{m-1,2}} h_{kk}^{\omega_{m-1}}}{1+\xi^{\omega_m} h_{kk}^{\omega_m}}(\hat{y}_k(\lambda, \omega_{m-1}) - y_k).
\end{align*}
Therefore, 
$(\hat{y}_k(\lambda, \omega_{i}) - y_k)$, $i=0,\dots,m$, all have the same sign,
and $(\hat{y}_k(\lambda,\omega_i) - y_k)(\hat{y}_k(\lambda,\omega_m) - y_k)>0$. The last term $(\bm{h}_{\cdot k}^{\omega_i})^\top\bm{h}_{\cdot k}^{\omega_m}$ is greater than 0 when $\hat{\mathcal{A}}^{\omega_i} \subset \hat{\mathcal{A}}^{\omega_m}$ or $\hat{\mathcal{A}}^{\omega_m} \subset \hat{\mathcal{A}}^{\omega_i}$. 
For example, when $\hat{\mathcal{A}}^{\omega_i} \subset \hat{\mathcal{A}}^{\omega_m}$, $(\bm{h}_{\cdot k}^{\omega_i})^\top\bm{h}_{\cdot k}^{\omega_m} = (H^{\omega_i}H^{\omega_m})_{kk} = (H^{\omega_i})_{kk} = h^{\omega_i}_{kk} >0$. 
$\hat{\mathcal{A}}^{\omega_i} \subset \hat{\mathcal{A}}^{\omega_m}$ or $\hat{\mathcal{A}}^{\omega_m} \subset \hat{\mathcal{A}}^{\omega_i}$ for all $i<m$ corresponds to the first two conditions stated in the proposition. 
\end{proof}

\subsection{Proof of Proposition~\ref{prop3}}\label{app_prop3}
Restatement of the proposition: For the Lasso model $\hat{y}=\hat{\beta}_0+\bm{x}^\top \hat{\bm{\beta}}$ with a penalty parameter $\lambda$, 
\begin{itemize}
    \item[i.] The sensitivity of the fitted value $\hat{y}_i$ to case $k$ is 
       \[ \frac{\partial \hat{y}^{\omega}_i}{\partial \omega}\Big|_{\omega=1}= h_{ik}(y_k - \hat{y}_k).\]
    \item[ii.] The local influence of case $k$ is 
    \[\frac{1}{2} \frac{\partial^2 D_k(\lambda, \omega)}{\partial \omega^2}\Big|_{\omega=1} = \frac{1}{(p+1)s^2} h_{kk} (y_k - \hat{y}_k)^2.\]
\end{itemize}
Here $h_{ij}$ is the $(i,j)$th entry of the hat matrix defined with the active variables at $\lambda$, and $\hat{y}_k$ is the Lasso fitted value for case $k$.

\begin{proof}
By \eqref{yomega} and \eqref{dev_xi}, the sensitivity of the fitted value $\hat{y}_i$ to case $k$ is 
\begin{align*}
    \frac{\partial \hat{y}^{\omega}_i}{\partial \omega}\Big|_{\omega=1} &= 
    \frac{\partial}{\partial \omega}\Big(\hat{y}_i(\lambda) - \xi^\omega h^\omega_{ik}(y_k - \hat{y}_k(\lambda))\Big)\Big|_{\omega=1} \\
    &= - \frac{\partial \xi^\omega}{\partial \omega}\Big|_{\omega=1}\cdot h_{ik}(y_k - \hat{y}_k(\lambda))\\
    &= h_{ik}(y_k - \hat{y}_k(\lambda)),
\end{align*}
where $h^{\omega=1}_{ik} = h_{ik}$, and it is constant with respect to $\omega$ in the left neighborhood of $\omega=1$. 
As for the local influence of case $k$, from \eqref{Dklo}, we have
\begin{align*}
    \frac{(p+1)s^2}{2}\cdot \frac{\partial^2 D_k(\lambda, \omega)}{\partial \omega^2}\Bigg|_{\omega=1} 
    =&
    \left(\frac{\partial}{\partial \omega}\sum_{i=1}^n (\hat{y}_{i}^\omega - \hat{y}_i(\lambda) )\frac{\partial \hat{y}^{\omega}_i}{\partial \omega}\right)\Bigg|_{\omega=1}\nonumber\\
    =& 
    \sum_{i=1}^n\left( (\hat{y}^{\omega}_i - \hat{y}_i(\lambda))\frac{\partial^2 \hat{y}^{\omega}_i}{\partial \omega^2} + \left(\frac{\partial \hat{y}^{\omega}_i}{\partial \omega}\right)^2\right)\Bigg|_{\omega=1}
    \nonumber\\
    =&\sum_{i=1}^n \left(\frac{\partial \hat{y}^{\omega}_i}{\partial \omega}\right)^2\Big|_{\omega=1} \nonumber\\
    =& ~(y_k - \hat{y}_k(\lambda) )^2 \sum_{i=1}^n (h_{ik})^2 \nonumber\\
    =&  ~h_{kk} (y_k - \hat{y}_k(\lambda))^2 
    .
\end{align*}
In the third step, $\hat{y}^{\omega}_i|_{\omega=1} =\hat{y}_i(\lambda)=\hat{y}_i$. As a result, the local influence is represented as a sum of case sensitivity squares. Then we can plug in the result of the sensitivity of fitted values in the fourth step. Finally, $\sum_{i=1}^n (h_{ik})^2 = h_{kk}$ by the idempotent property of the hat matrix.  
\end{proof}

\subsection{Proof of Proposition~\ref{prop1}}\label{app_prop1}

Restatement of the proposition: Let $X\in\mathbb{R}^{n\times p}$ be a design matrix with $n>p$. If a feature $\bm{z}\in\mathbb{R}^{n} $ is added to the design matrix, the $k$th leverage $h_{kk}$, $k \in [n]$, will increase by $\frac{[(I-P_X)\bm{z}]_k^2}{\|(I-P_X)\bm{z}\|^2}$, where $P_X$ is the projection matrix onto the column space of $X$. 
\begin{proof}
    Consider the new design matrix $X^* = (X, \bm{z})$ and the new $k$th observation $\bm{x}^*_k = (\bm{x}^\top_k, z_k)^\top$. Then, the $k$th leverage before $\bm{z}$ has been added is $h_{kk} = \bm{x}_k^\top (X^\top X)^{-1}\bm{x}_k$ and after is $h_{kk}^{*} = (\bm{x}^{*}_k)^\top ({X^*}^\top X^*)^{-1}\bm{x}_k^{*}$.
    By applying the formula for block matrix inversion, $({X^*}^\top X^*)^{-1}$ can be written as the sum of  $\begin{pmatrix}
            (X^\top X)^{-1}& \bm{0}\\
            \bm{0}^\top & 0
        \end{pmatrix}$ and 
    \begin{align*}
        \frac{1}{\bm{z}^\top (I-P_X)\bm{z}}\begin{pmatrix}
            (X^\top X)^{-1}X^\top \bm{z}  \bm{z}^\top X (X^\top X)^{-1} & -(X^\top X)^{-1}X^\top \bm{z}\\
            -\bm{z}^\top X(X^\top X)^{-1} & 1
        \end{pmatrix},
    \end{align*}
where $P_X = X(X^\top X)^{-1} X^\top$. Consequently, $h_{kk}^*$ is the sum of $h_{kk}$ and \begin{align*}
     \frac{1}{\bm{z}^\top (I-P_X)\bm{z}}(\bm{x}^\top_k, z_k)\begin{pmatrix}
            (X^\top X)^{-1}X^\top \bm{z}  \bm{z}^\top X (X^\top X)^{-1} & -(X^\top X)^{-1}X^\top \bm{z}\\
            -\bm{z}^\top X(X^\top X)^{-1} & 1
        \end{pmatrix}\begin{pmatrix}
            \bm{x}_k\\ z_k
        \end{pmatrix}.
\end{align*}
Therefore, 
\begin{align*}
h_{kk}^* &= h_{kk} + \frac{1}{\bm{z}^\top (I-P_X)\bm{z}}\left(\bm{h}_k^\top\bm{z}  \bm{z}^\top \bm{h}_k - 2\bm{h}_k^\top \bm{z}z_k + z_k^2\right)    \\
&=  h_{kk} + \frac{(z_k - \bm{h}_k^\top \bm{z} )^2}{\bm{z}^\top (I-P_X)\bm{z}} = h_{kk} + \frac{[(I-P_X)\bm{z}]_k^2}{\|(I-P_X)\bm{z}\|^2}.
\end{align*}

\end{proof}

This proposition can be easily generalized to the update of the $(i,j)$th entry and the $H$ itself:
\begin{align*}
    h_{ij}^* &=  h_{ij} + \frac{(z_i - \bm{h}_i^\top \bm{z} )(z_j - \bm{h}_j^\top \bm{z} )}{\bm{z}^\top (I-P_X)\bm{z}} = h_{ij} + \frac{[(I-P_X)\bm{z}]_i [(I-P_X)\bm{z}]_j}{\|(I-P_X)\bm{z}\|^2},\nonumber\\
    H^* &=  H + \frac{[(I-H)\bm{z}] [(I-H)\bm{z}]^\top}{\|(I-H)\bm{z}\|^2}.
\end{align*}

\section{Additional Results}
\subsection{Distributions of studentized Lasso residuals}\label{resid_plot}
\begin{figure}[htb]
    \centering
    \includegraphics[width = 0.8\textwidth]{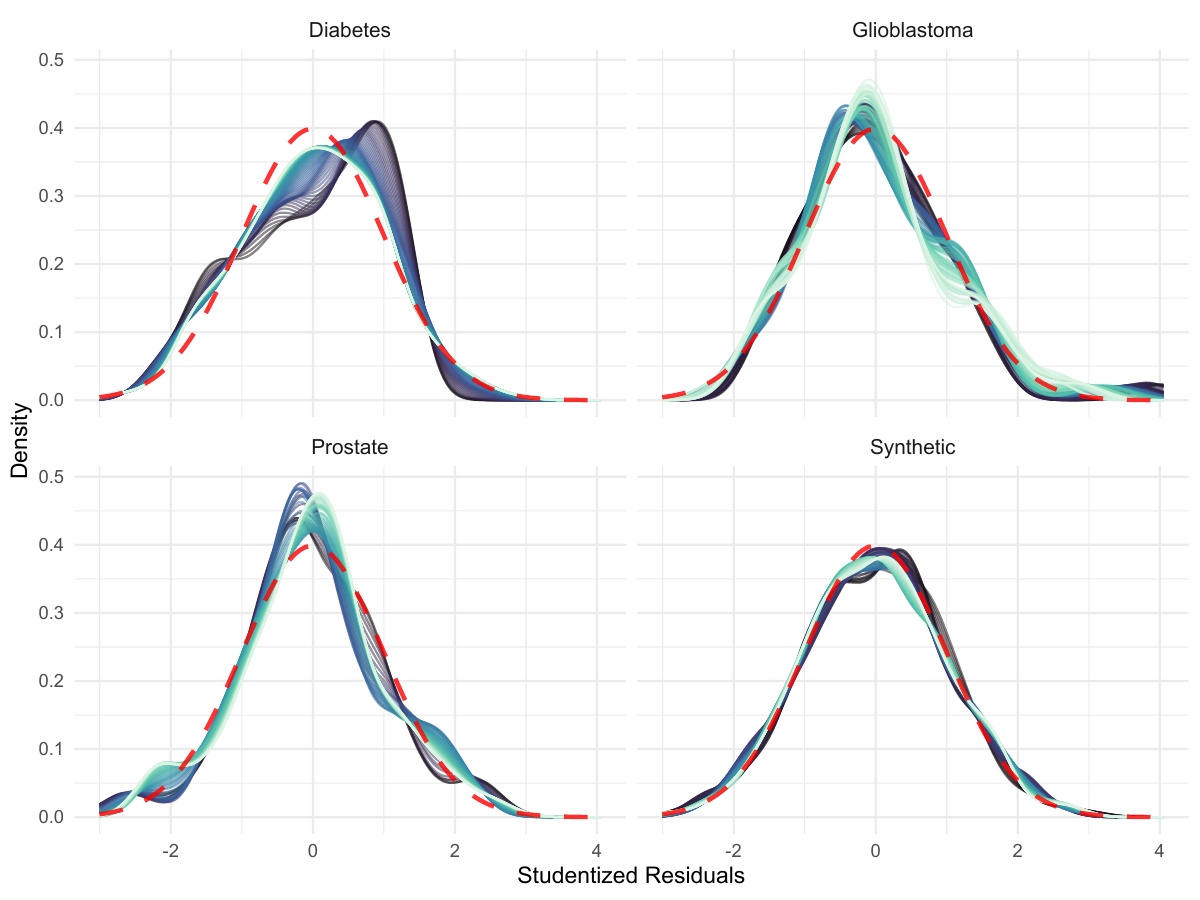}
    \caption{Probability density estimates of studentized Lasso residuals for four datasets. Each panel represents one dataset. Each curve represents the distribution of studentized residuals for a different value of $\lambda$. Darker colors indicate residuals coming from models with higher penalties. The red dashed curve in each panel is the probability density function of the standard normal distribution, serving as a reference for assessing the normality of the residuals.}
    \label{fig:resid_plot}
\end{figure}

Figure~\ref{fig:resid_plot} shows the distributions of studentized  Lasso residuals from four distinct datasets described in Section~\ref{thresholding} when we vary the penalty parameter $\lambda$. In each panel, each curve represents the distribution of studentized residuals for a different value of $\lambda$. A darker color indicates a higher penalty. When the penalty is relatively small, studentized residuals roughly follow a standard normal distribution. When a larger penalty is applied to the model, the residuals gradually deviate from the normal distribution assumption. 

\subsection{Summary of simulation results}\label{simu_sum}
Tables~\ref{simu_table_beta1} through \ref{simu_table_beta8} present additional results of the simulation study in Section~\ref{simu_study} for different values of $\bm{\beta}$ and $q$.

    \begin{table}[htb]
        \centering
    \begin{tabular}{cccccc|cccccc}
    \hline
    $a$ & $b$ & $n$  & $p$ &  1 & 2--$n$ & $a$ & $b$ & $n$ & $p$ & 1 & 2--$n$ \\\hline
    0 & 2 & 50 & 10  & 0.23 & 0.05 & 0 & 2 & 200 & 200 & 0.28 & 0.05 \\
    0 & 2 & 50 & 500 & 0.15 & 0.05 & 0 & 2 & 500 & 10  & 0.24 & 0.04 \\\hline
    0 & 3 & 50 & 10  & 0.66 & 0.05 & 0 & 3 & 200 & 200 & 0.88 & 0.04 \\
    0 & 3 & 50 & 500 & 0.33 & 0.04 & 0 & 3 & 500 & 10  & 0.80 & 0.04 \\\hline
    0 & 5 & 50 & 10  & 0.98 & 0.02 & 0 & 5 & 200 & 200 & 1.00 & 0.02 \\
    0 & 5 & 50 & 500 & 0.71 & 0.03 & 0 & 5 & 500 & 10  & 1.00 & 0.03 \\\hline
    2 & 0 & 50 & 10  & 0.01 & 0.05 & 2 & 0 & 200 & 200 & 0.00 & 0.05 \\
    2 & 0 & 50 & 500 & 0.02 & 0.05 & 2 & 0 & 500 & 10  & 0.00 & 0.04 \\\hline
    2 & 2 & 50 & 10  & 0.58 & 0.05 & 2 & 2 & 200 & 200 & 0.71 & 0.04 \\
    2 & 2 & 50 & 500 & 0.39 & 0.04 & 2 & 2 & 500 & 10  & 0.73 & 0.04 \\\hline
    3 & 0 & 50 & 10  & 0.04 & 0.05 & 3 & 0 & 200 & 200 & 0.00 & 0.05 \\
    3 & 0 & 50 & 500 & 0.09 & 0.05 & 3 & 0 & 500 & 10  & 0.00 & 0.04 \\\hline
    3 & 3 & 50 & 10  & 0.97 & 0.02 & 3 & 3 & 200 & 200 & 1.00 & 0.03 \\
    3 & 3 & 50 & 500 & 0.71 & 0.02 & 3 & 3 & 500 & 10  & 1.00 & 0.03 \\\hline
    5 & 0 & 50 & 10  & 0.25 & 0.05 & 5 & 0 & 200 & 200 & 0.11 & 0.05 \\
    5 & 0 & 50 & 500 & 0.35 & 0.04 & 5 & 0 & 500 & 10  & 0.00 & 0.04 \\\hline
    5 & 5 & 50 & 10  & 1.00 & 0.00 & 5 & 5 & 200 & 200 & 1.00 & 0.00 \\
    5 & 5 & 50 & 500 & 0.74 & 0.01 & 5 & 5 & 500 & 10  & 1.00 & 0.00 \\\hline
    \end{tabular}
        \caption{Proportion of the times that the proposed Cook's distance detects an influential observation. Columns $a$ and $b$ are the parameter values that were used to generate observation 1 with the value of a relevant predictor ($q=1$) modified.  Column $p$ is the dimension of the data. Column `1' is for the proportion of the time the measure flagged observation 1 as influential and Column `2--$n$' for the average proportion of observations 2--$n$ being flagged. The results for each quadruple $(a,b,n,p)$ are based on 1000 simulations when $\bm{\beta} = (1,1,1/2,1/2,0,\dots,0)^\top$ and SNR = 3.288.}
        \label{simu_table_beta1}
    \end{table}
    
    \begin{table}[htb]
    \centering
    \begin{tabular}{cccccc|cccccc}
    \hline
    $a$ & $b$ & $n$  & $p$ &  1 & 2--$n$ & $a$ & $b$ & $n$ & $p$ & 1 & 2--$n$ \\\hline
    0 & 2 & 50 & 10  & 0.22 & 0.05 & 0 & 2 & 200 & 200 & 0.29 & 0.05 \\
    0 & 2 & 50 & 500 & 0.09 & 0.05 & 0 & 2 & 500 & 10  & 0.24 & 0.04 \\\hline
    0 & 3 & 50 & 10  & 0.67 & 0.05 & 0 & 3 & 200 & 200 & 0.88 & 0.04 \\
    0 & 3 & 50 & 500 & 0.27 & 0.04 & 0 & 3 & 500 & 10  & 0.84 & 0.04 \\\hline
    0 & 5 & 50 & 10  & 0.96 & 0.02 & 0 & 5 & 200 & 200 & 1.00 & 0.02 \\
    0 & 5 & 50 & 500 & 0.64 & 0.03 & 0 & 5 & 500 & 10  & 1.00 & 0.03 \\\hline
    2 & 0 & 50 & 10  & 0.02 & 0.06 & 2 & 0 & 200 & 200 & 0.00 & 0.05 \\
    2 & 0 & 50 & 500 & 0.07 & 0.05 & 2 & 0 & 500 & 10  & 0.00 & 0.04 \\\hline
    2 & 2 & 50 & 10  & 0.58 & 0.05 & 2 & 2 & 200 & 200 & 0.69 & 0.05 \\
    2 & 2 & 50 & 500 & 0.44 & 0.04 & 2 & 2 & 500 & 10  & 0.73 & 0.04 \\\hline
    3 & 0 & 50 & 10  & 0.04 & 0.06 & 3 & 0 & 200 & 200 & 0.00 & 0.05 \\
    3 & 0 & 50 & 500 & 0.22 & 0.02 & 3 & 0 & 500 & 10  & 0.00 & 0.03 \\\hline
    3 & 3 & 50 & 10  & 0.97 & 0.02 & 3 & 3 & 200 & 200 & 1.00 & 0.03 \\
    3 & 3 & 50 & 500 & 0.71 & 0.02 & 3 & 3 & 500 & 10  & 1.00 & 0.03 \\\hline
    5 & 0 & 50 & 10  & 0.23 & 0.05 & 5 & 0 & 200 & 200 & 0.11 & 0.05 \\
    5 & 0 & 50 & 500 & 0.55 & 0.03 & 5 & 0 & 500 & 10  & 0.00 & 0.04 \\\hline
    5 & 5 & 50 & 10  & 1.00 & 0.00 & 5 & 5 & 200 & 200 & 1.00 & 0.00 \\
    5 & 5 & 50 & 500 & 0.73 & 0.01 & 5 & 5 & 500 & 10  & 1.00 & 0.00 \\\hline
    \end{tabular}
        \caption{Proportion of the times that the proposed Cook's distance detects an influential observation. Columns $a$ and $b$ are the parameter values that were used to generate observation 1 with the value of a relevant predictor ($q=1$) modified.  Column $p$ is the dimension of the data. Column `1' is for the proportion of the time the measure flagged observation 1 as influential and Column `2--$n$' for the average proportion of observations 2--$n$ being flagged. The results for each quadruple $(a,b,n,p)$ are based on 1000 simulations when $\bm{\beta} = (1,-1,1/2,-1/2,0,\dots,0)^\top$ and SNR = 1.872.}
        \label{simu_table_beta2}
    \end{table}
        
    \begin{table}[htb]
    \centering
    \begin{tabular}{cccccc|cccccc}
    \hline
    $a$ & $b$ & $n$  & $p$ &  1 & 2--$n$ & $a$ & $b$ & $n$ & $p$ & 1 & 2--$n$ \\\hline
    0 & 2 & 50 & 10  & 0.21 & 0.05 & 0 & 2 & 200 & 200 & 0.33 & 0.05 \\
    0 & 2 & 50 & 500 & 0.14 & 0.05 & 0 & 2 & 500 & 10  & 0.22 & 0.04 \\\hline
    0 & 3 & 50 & 10  & 0.64 & 0.04 & 0 & 3 & 200 & 200 & 0.80 & 0.04 \\
    0 & 3 & 50 & 500 & 0.34 & 0.04 & 0 & 3 & 500 & 10  & 0.72 & 0.04 \\\hline
    0 & 5 & 50 & 10  & 0.98 & 0.02 & 0 & 5 & 200 & 200 & 0.99 & 0.03 \\
    0 & 5 & 50 & 500 & 0.68 & 0.03 & 0 & 5 & 500 & 10  & 1.00 & 0.03 \\\hline
    2 & 0 & 50 & 10  & 0.01 & 0.06 & 2 & 0 & 200 & 200 & 0.00 & 0.05 \\
    2 & 0 & 50 & 500 & 0.03 & 0.05 & 2 & 0 & 500 & 10  & 0.00 & 0.04 \\\hline
    2 & 2 & 50 & 10  & 0.68 & 0.04 & 2 & 2 & 200 & 200 & 0.72 & 0.05 \\
    2 & 2 & 50 & 500 & 0.44 & 0.04 & 2 & 2 & 500 & 10  & 0.81 & 0.04 \\\hline
    3 & 0 & 50 & 10  & 0.05 & 0.06 & 3 & 0 & 200 & 200 & 0.00 & 0.05 \\
    3 & 0 & 50 & 500 & 0.10 & 0.05 & 3 & 0 & 500 & 10  & 0.00 & 0.04 \\\hline
    3 & 3 & 50 & 10  & 0.99 & 0.01 & 3 & 3 & 200 & 200 & 1.00 & 0.03 \\
    3 & 3 & 50 & 500 & 0.75 & 0.02 & 3 & 3 & 500 & 10  & 1.00 & 0.03 \\\hline
    5 & 0 & 50 & 10  & 0.30 & 0.05 & 5 & 0 & 200 & 200 & 0.11 & 0.05 \\
    5 & 0 & 50 & 500 & 0.43 & 0.04 & 5 & 0 & 500 & 10  & 0.00 & 0.04 \\\hline
    5 & 5 & 50 & 10  & 1.00 & 0.00 & 5 & 5 & 200 & 200 & 1.00 & 0.00 \\
    5 & 5 & 50 & 500 & 0.77 & 0.01 & 5 & 5 & 500 & 10  & 1.00 & 0.00 \\\hline
    \end{tabular}
        \caption{Proportion of the times that the proposed Cook's distance detects an influential observation. Columns $a$ and $b$ are the parameter values that were used to generate observation 1 with the value of a relevant predictor ($q=1$) modified.  Column $p$ is the dimension of the data. Column `1' is for the proportion of the time the measure flagged observation 1 as influential and Column `2--$n$' for the average proportion of observations 2--$n$ being flagged. The results for each quadruple $(a,b,n,p)$ are based on 1000 simulations when $\bm{\beta} = (1,1,1,0,\dots,0)^\top$ and SNR = 3.880.}
        \label{simu_table_beta3}
    \end{table}

    \begin{table}[htb]
    \centering
    \begin{tabular}{cccccc|cccccc}
    \hline
    $a$ & $b$ & $n$  & $p$ &  1 & 2--$n$ & $a$ & $b$ & $n$ & $p$ & 1 & 2--$n$ \\\hline
    0 & 2 & 50 & 10  & 0.20 & 0.05 & 0 & 2 & 200 & 200 & 0.31 & 0.05 \\
    0 & 2 & 50 & 500 & 0.13 & 0.05 & 0 & 2 & 500 & 10  & 0.24 & 0.04 \\\hline
    0 & 3 & 50 & 10  & 0.69 & 0.05 & 0 & 3 & 200 & 200 & 0.86 & 0.04 \\
    0 & 3 & 50 & 500 & 0.29 & 0.04 & 0 & 3 & 500 & 10  & 0.81 & 0.04 \\\hline
    0 & 5 & 50 & 10  & 0.98 & 0.02 & 0 & 5 & 200 & 200 & 0.99 & 0.02 \\
    0 & 5 & 50 & 500 & 0.64 & 0.03 & 0 & 5 & 500 & 10  & 1.00 & 0.03 \\\hline
    2 & 0 & 50 & 10  & 0.00 & 0.05 & 2 & 0 & 200 & 200 & 0.00 & 0.05 \\
    2 & 0 & 50 & 500 & 0.04 & 0.04 & 2 & 0 & 500 & 10  & 0.00 & 0.04 \\\hline
    2 & 2 & 50 & 10  & 0.62 & 0.05 & 2 & 2 & 200 & 200 & 0.72 & 0.05 \\
    2 & 2 & 50 & 500 & 0.36 & 0.04 & 2 & 2 & 500 & 10  & 0.72 & 0.04 \\\hline
    3 & 0 & 50 & 10  & 0.04 & 0.06 & 3 & 0 & 200 & 200 & 0.00 & 0.05 \\
    3 & 0 & 50 & 500 & 0.06 & 0.05 & 3 & 0 & 500 & 10  & 0.00 & 0.04 \\\hline
    3 & 3 & 50 & 10  & 0.98 & 0.02 & 3 & 3 & 200 & 200 & 1.00 & 0.02 \\
    3 & 3 & 50 & 500 & 0.63 & 0.03 & 3 & 3 & 500 & 10  & 1.00 & 0.03 \\\hline
    5 & 0 & 50 & 10  & 0.23 & 0.05 & 5 & 0 & 200 & 200 & 0.07 & 0.05 \\
    5 & 0 & 50 & 500 & 0.32 & 0.04 & 5 & 0 & 500 & 10  & 0.00 & 0.04 \\\hline
    5 & 5 & 50 & 10  & 1.00 & 0.00 & 5 & 5 & 200 & 200 & 1.00 & 0.00 \\
    5 & 5 & 50 & 500 & 0.71 & 0.01 & 5 & 5 & 500 & 10  & 1.00 & 0.00 \\\hline
    \end{tabular}
        \caption{Proportion of the times that the proposed Cook's distance detects an influential observation. Columns $a$ and $b$ are the parameter values that were used to generate observation 1 with the value of a relevant predictor ($q=1$) modified.  Column $p$ is the dimension of the data. Column `1' is for the proportion of the time the measure flagged observation 1 as influential and Column `2--$n$' for the average proportion of observations 2--$n$ being flagged. The results for each quadruple $(a,b,n,p)$ are based on 1000 simulations when $\bm{\beta} = (4,3,2,1,0,\dots,0)^\top$ and SNR = 38.944.}
        \label{simu_table_beta4}
    \end{table}

    \begin{table}[htb]
    \centering
    \begin{tabular}{cccccc|cccccc}
    \hline
    $a$ & $b$ & $n$  & $p$ &  1 & 2--$n$ & $a$ & $b$ & $n$ & $p$ & 1 & 2--$n$ \\\hline
    0 & 2 & 50 & 10  & 0.30 & 0.05 & 0 & 2 & 200 & 200 & 0.31 & 0.05 \\
    0 & 2 & 50 & 500 & 0.14 & 0.05 & 0 & 2 & 500 & 10  & 0.30 & 0.04 \\\hline
    0 & 3 & 50 & 10  & 0.71 & 0.04 & 0 & 3 & 200 & 200 & 0.90 & 0.04 \\
    0 & 3 & 50 & 500 & 0.36 & 0.04 & 0 & 3 & 500 & 10  & 0.83 & 0.04 \\\hline
    0 & 5 & 50 & 10  & 0.99 & 0.02 & 0 & 5 & 200 & 200 & 1.00 & 0.02 \\
    0 & 5 & 50 & 500 & 0.74 & 0.03 & 0 & 5 & 500 & 10  & 1.00 & 0.03 \\\hline
    2 & 0 & 50 & 10  & 0.01 & 0.06 & 2 & 0 & 200 & 200 & 0.00 & 0.05 \\
    2 & 0 & 50 & 500 & 0.02 & 0.05 & 2 & 0 & 500 & 10  & 0.00 & 0.04 \\\hline
    2 & 2 & 50 & 10  & 0.32 & 0.05 & 2 & 2 & 200 & 200 & 0.32 & 0.05 \\
    2 & 2 & 50 & 500 & 0.16 & 0.04 & 2 & 2 & 500 & 10  & 0.46 & 0.04 \\\hline
    3 & 0 & 50 & 10  & 0.01 & 0.06 & 3 & 0 & 200 & 200 & 0.00 & 0.05 \\
    3 & 0 & 50 & 500 & 0.02 & 0.05 & 3 & 0 & 500 & 10  & 0.00 & 0.04 \\\hline
    3 & 3 & 50 & 10  & 0.86 & 0.03 & 3 & 3 & 200 & 200 & 0.89 & 0.04 \\
    3 & 3 & 50 & 500 & 0.36 & 0.04 & 3 & 3 & 500 & 10  & 0.93 & 0.04 \\\hline
    5 & 0 & 50 & 10  & 0.04 & 0.06 & 5 & 0 & 200 & 200 & 0.00 & 0.05 \\
    5 & 0 & 50 & 500 & 0.01 & 0.05 & 5 & 0 & 500 & 10  & 0.00 & 0.04 \\\hline
    5 & 5 & 50 & 10  & 0.99 & 0.00 & 5 & 5 & 200 & 200 & 1.00 & 0.02 \\
    5 & 5 & 50 & 500 & 0.72 & 0.03 & 5 & 5 & 500 & 10  & 1.00 & 0.02 \\\hline
    \end{tabular}
        \caption{Proportion of the times that the proposed Cook's distance detects an influential observation. Columns $a$ and $b$ are the parameter values that were used to generate observation 1 with the value of a relevant predictor ($q=10$) modified.  Column $p$ is the dimension of the data. Column `1' is for the proportion of the time the measure flagged observation 1 as influential and Column `2--$n$' for the average proportion of observations 2--$n$ being flagged. The results for each quadruple $(a,b,n,p)$ are based on 1000 simulations when $\bm{\beta} = (1,1,1/2,1/2,0,\dots,0)^\top$ and SNR = 3.288.}
        \label{simu_table_beta5}
    \end{table}

    \begin{table}[htb]
    \centering
    \begin{tabular}{cccccc|cccccc}
    \hline
    $a$ & $b$ & $n$  & $p$ &  1 & 2--$n$ & $a$ & $b$ & $n$ & $p$ & 1 & 2--$n$ \\\hline
    0 & 2 & 50 & 10  & 0.26 & 0.05 & 0 & 2 & 200 & 200 & 0.33 & 0.05 \\
    0 & 2 & 50 & 500 & 0.17 & 0.05 & 0 & 2 & 500 & 10  & 0.28 & 0.04 \\\hline
    0 & 3 & 50 & 10  & 0.70 & 0.04 & 0 & 3 & 200 & 200 & 0.86 & 0.04 \\
    0 & 3 & 50 & 500 & 0.41 & 0.04 & 0 & 3 & 500 & 10  & 0.82 & 0.04 \\\hline
    0 & 5 & 50 & 10  & 0.99 & 0.02 & 0 & 5 & 200 & 200 & 1.00 & 0.02 \\
    0 & 5 & 50 & 500 & 0.72 & 0.03 & 0 & 5 & 500 & 10  & 1.00 & 0.03 \\\hline
    2 & 0 & 50 & 10  & 0.00 & 0.06 & 2 & 0 & 200 & 200 & 0.00 & 0.05 \\
    2 & 0 & 50 & 500 & 0.03 & 0.05 & 2 & 0 & 500 & 10  & 0.00 & 0.04 \\\hline
    2 & 2 & 50 & 10  & 0.32 & 0.05 & 2 & 2 & 200 & 200 & 0.33 & 0.05 \\
    2 & 2 & 50 & 500 & 0.16 & 0.05 & 2 & 2 & 500 & 10  & 0.46 & 0.04 \\\hline
    3 & 0 & 50 & 10  & 0.01 & 0.06 & 3 & 0 & 200 & 200 & 0.00 & 0.05 \\
    3 & 0 & 50 & 500 & 0.01 & 0.05 & 3 & 0 & 500 & 10  & 0.00 & 0.04 \\\hline
    3 & 3 & 50 & 10  & 0.81 & 0.03 & 3 & 3 & 200 & 200 & 0.87 & 0.04 \\
    3 & 3 & 50 & 500 & 0.37 & 0.04 & 3 & 3 & 500 & 10  & 0.91 & 0.04 \\\hline
    5 & 0 & 50 & 10  & 0.04 & 0.06 & 5 & 0 & 200 & 200 & 0.00 & 0.05 \\
    5 & 0 & 50 & 500 & 0.03 & 0.05 & 5 & 0 & 500 & 10  & 0.00 & 0.04 \\\hline
    5 & 5 & 50 & 10  & 1.00 & 0.01 & 5 & 5 & 200 & 200 & 1.00 & 0.02 \\
    5 & 5 & 50 & 500 & 0.71 & 0.02 & 5 & 5 & 500 & 10  & 1.00 & 0.02 \\\hline
    \end{tabular}
        \caption{Proportion of times that the proposed Cook's distance detects an influential observation. Columns $a$ and $b$ are the parameter values that were used to generate observation 1 with the value of a relevant predictor ($q=10$) modified.  Column $p$ is the dimension of the data. Column `1' is for the proportion of the time the measure flagged observation 1 as influential and Column `2--$n$' for the average proportion of observations 2--$n$ being flagged. The results for each quadruple $(a,b,n,p)$ are based on 1000 simulations when $\bm{\beta} = (1,1,1,0,\dots,0)^\top$ and SNR = 3.880.}
        \label{simu_table_beta7}
    \end{table}

    \begin{table}[htb]
    \centering
    \begin{tabular}{cccccc|cccccc}
    \hline
    $a$ & $b$ & $n$  & $p$ &  1 & 2--$n$ & $a$ & $b$ & $n$ & $p$ & 1 & 2--$n$ \\\hline
    0 & 2 & 50 & 10  & 0.23 & 0.05 & 0 & 2 & 200 & 200 & 0.28 & 0.05 \\
    0 & 2 & 50 & 500 & 0.14 & 0.05 & 0 & 2 & 500 & 10  & 0.29 & 0.04 \\\hline
    0 & 3 & 50 & 10  & 0.70 & 0.04 & 0 & 3 & 200 & 200 & 0.89 & 0.04 \\
    0 & 3 & 50 & 500 & 0.32 & 0.04 & 0 & 3 & 500 & 10  & 0.83 & 0.04 \\\hline
    0 & 5 & 50 & 10  & 0.99 & 0.02 & 0 & 5 & 200 & 200 & 1.00 & 0.02 \\
    0 & 5 & 50 & 500 & 0.61 & 0.03 & 0 & 5 & 500 & 10  & 1.00 & 0.03 \\\hline
    2 & 0 & 50 & 10  & 0.00 & 0.06 & 2 & 0 & 200 & 200 & 0.00 & 0.05 \\
    2 & 0 & 50 & 500 & 0.02 & 0.05 & 2 & 0 & 500 & 10  & 0.00 & 0.04 \\\hline
    2 & 2 & 50 & 10  & 0.36 & 0.05 & 2 & 2 & 200 & 200 & 0.33 & 0.05 \\
    2 & 2 & 50 & 500 & 0.14 & 0.04 & 2 & 2 & 500 & 10  & 0.45 & 0.04 \\\hline
    3 & 0 & 50 & 10  & 0.01 & 0.06 & 3 & 0 & 200 & 200 & 0.00 & 0.05 \\
    3 & 0 & 50 & 500 & 0.01 & 0.05 & 3 & 0 & 500 & 10  & 0.00 & 0.04 \\\hline
    3 & 3 & 50 & 10  & 0.82 & 0.04 & 3 & 3 & 200 & 200 & 0.89 & 0.04 \\
    3 & 3 & 50 & 500 & 0.33 & 0.04 & 3 & 3 & 500 & 10  & 0.93 & 0.04 \\\hline
    5 & 0 & 50 & 10  & 0.04 & 0.06 & 5 & 0 & 200 & 200 & 0.00 & 0.05 \\
    5 & 0 & 50 & 500 & 0.02 & 0.05 & 5 & 0 & 500 & 10  & 0.00 & 0.04 \\\hline
    5 & 5 & 50 & 10  & 1.00 & 0.01 & 5 & 5 & 200 & 200 & 1.00 & 0.02 \\
    5 & 5 & 50 & 500 & 0.67 & 0.03 & 5 & 5 & 500 & 10  & 1.00 & 0.02 \\\hline
    \end{tabular}
        \caption{Proportion of the times that the proposed Cook's distance detects an influential observation. Columns $a$ and $b$ are the parameter values that were used to generate observation 1 with the value of a relevant predictor ($q=10$) modified.  Column $p$ is the dimension of the data. Column `1' is for the proportion of the time the measure flagged observation 1 as influential and Column `2--$n$' for the average proportion of observations 2--$n$ being flagged. The results for each quadruple $(a,b,n,p)$ are based on 1000 simulations when $\bm{\beta} = (4,3,2,1,0,\dots,0)^\top$ and SNR = 38.944.}
        \label{simu_table_beta8}
    \end{table}

\clearpage
\subsection{Lasso coefficient paths for diabetes data}\label{lasso_sol_path_app}
\begin{figure}[htb]
    \centering
    \includegraphics[width = \textwidth]{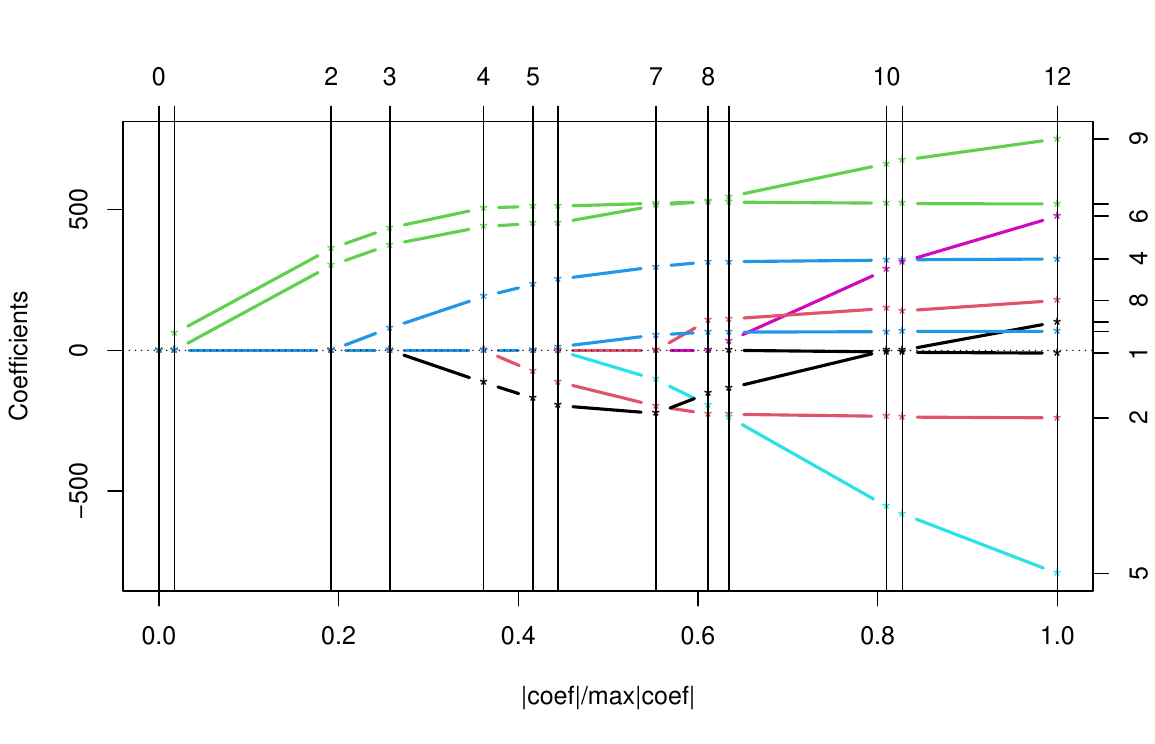}
    \caption{The Lasso solution path for the diabetes data obtained by LARS. The numbers on the top represent the order of active set changes. The numbers on the right-hand side represent predictors. Predictor 7 corresponds to $hdl$ and the tick right above predictor 7 is $tch$.}
    \label{fig:reg_sol_path}
\end{figure}
Figure~\ref{fig:reg_sol_path} shows the Lasso coefficient paths for 10 variables in the diabetes data using LARS.
Due to the high negative correlation ($-0.74$) of $hdl$ and $tch$, their coefficient paths move in the same direction at the beginning. 
The $\lambda$ values selected by CV in 1000 experiments mainly fall around the fraction of 0.7, making $hdl$ oscillate among positive, zero, and negative. After removing influential points, a smaller fraction tends to be selected, which renders the $hdl$ coefficient consistently negative, while reducing the $tch$ coefficient to zero for almost all experiments.
\end{document}